
\input phyzzx
\input tables

%

%
\papers
\def\Lagr{{\cal L}}
\def\dslash{\not{\hbox{\kern-2pt $\partial$}}}
\def\Dslash{\not{\hbox{\kern-4pt $D$}}}
\def\pslash{\not{\hbox{\kern-2.3pt $p$}}}
 \newtoks\slashfraction
 \slashfraction={.13}
 \def\slash#1{\setbox0\hbox{$ #1 $}
 \setbox0\hbox to \the\slashfraction\wd0{\hss \box0}/\box0 }
\def\bigtstrut{\vrule height 20pt depth 6pt width 0pt}

\def\scrscr{\scriptscriptstyle}
\def\scrtwo{\scrscr{1\over 2}}
\def\hc{{\rm h.c.}}

\def\tfrac#1#2{{\textstyle{#1\over #2}}}
\def\Re{{\rm Re}}
\def\sgn{{\rm sgn}}
\def\dag{\dagger}
\def\ttwo{{\textstyle{1\over 2}}}
\def\tthree{{\textstyle {1\over 3}}}
\def\tfour{{\textstyle{1\over 4}}}
\def\ovl{\overline}

\def\ag{\alpha_{\scrscr G}}
\def\Mv{M_{{\scrscr V}}}
\def\Msig{M_{{\scrscr\Sigma}}}
\def\Mhs{M_{{\scrscr H_3}}}
\def\Ms{M_{{\scrscr SUSY}}}
\def\Mx{M_{{\scrscr GUT}}}
\def\Mpl{M_{\scrscr Planck}}
\def\mst{m_{\tilde{t}_i}}
\def\msti{m_{\tilde{t}_i}}
\def\mstj{m_{\tilde{t}_j}}
\def\mstk{m_{\tilde{t}_k}}
\def\mgl{m_{\tilde{g}}}
\def\msb{m_{\tilde{b}_i}}
\def\msbi{m_{\tilde{b}_i}}
\def\msbj{m_{\tilde{b}_j}}
\def\msbk{m_{\tilde{b}_k}}
\def\mstau{m_{\tilde{\tau}_i}}
\def\mstaui{m_{\tilde{\tau}_i}}
\def\mstauj{m_{\tilde{\tau}_j}}

\def\mchg{m_{\chi^{\pm}_j}}
\def\mneut{m_{\chi^0_j}}
\def\msnutau{m_{\tilde{\nu}_{\tau}}}
\def\mchgi{m_{\chi^{\pm}_i}}
\def\mneuti{m_{\chi^0_i}}
\def\mchgj{m_{\chi^{\pm}_j}}
\def\mneutj{m_{\chi^0_j}}
\def\mchgk{m_{\chi^{\pm}_k}}
\def\mneutk{m_{\chi^0_k}}
\def\mhino{m_{\tilde{H}^{\pm}}}
\def\mh0ino{m_{\tilde{H}^0}}
\def\mbino{m_{\tilde{B}}}
\def\mwino{m_{\tilde{W}}}
\def\mw3ino{m_{\tilde{W}^3}}
\def\mbh{m_{H_0}}
\def\mhpm{m_{H^{\pm}}}
\def\mlh{m_{h_0}}
\def\muh{\mu_{\scrscr H}}
\hyphenation{char-gino}

\def\Wisc{\address{Physics Department, University of Wisconsin\break
Madison, WI 53706, USA}}

\hskip.25in \raise.5in\hbox{{\fourteenbf University of Wisconsin - Madison}}
\vskip-1in\hfill\hbox{\vtop{\hbox{\tenbf MAD/PH/812}\vskip-2mm
\hbox{\tenrm March 1994}}}
\nopubblock
\vskip.5in
\titlepage
\title{Yukawa Coupling Thresholds: Application to the MSSM
and the Minimal Supersymmetric SU(5) GUT}
\bigskip
\author{Brian~D.~Wright\footnote{\dag}{E-mail address:
wright@phenot.physics.wisc.edu}}
\Wisc
\bigskip
\abstract{{\singlespace\Tenpoint
We consider a particular class of threshold corrections to
Yukawa couplings and mass relations in the MSSM and
supersymmetric grand unified models. We give a complete
treatment of Yukawa coupling thresholds at the
unification scale $\Mx$ and the effective supersymmetry scale
$\Ms$ and apply them
to corrections to the tree-level prediction $y_b(\Mx) = y_{\tau}(\Mx)$
in minimal supersymmetric SU(5). We apply both gauge and Yukawa coupling
thresholds to gauge unification and the above Yukawa unification condition to
find predictions for the top quark mass, $M_t$, the superheavy vector boson
mass $\Mv$ and the colored Higgs triplet mass $\Mhs$.
We discuss the dependencies of $\Mx$ and $\Mhs$ on $\alpha_s(M_Z)$,
$M_t$ and the sparticle spectrum as well as those of $M_t$ on
$\tan\beta$, $\alpha_s(M_Z)$ and the bottom quark mass, $M_b$.
The effect of the Yukawa coupling thresholds on $M_t$ are given
for representative sparticle spectra. We describe the
quantitative differences between these effects for low and high $\tan\beta$.
We also give new bounds on superheavy masses,
incorporating proton decay as well as unification constraints, the
former leading to a lower bound on $\alpha_s$.}}

\submit{Physical Review D}
\endpage

\def\PRL#1&#2&#3&{\sl\ Phys. Rev. Lett.\ \bf #1\rm\ (19#3)\ #2}
\def\PRD#1&#2&#3&{\sl\ Phys. Rev.\ \bf D#1\rm\ (19#3)\ #2}
\def\ZPC#1&#2&#3&{\sl\ Z. Phys.\ \bf C#1\rm\ (19#3)\ #2}
\def\NPB#1&#2&#3&{\sl\ Nucl. Phys.\ \bf B#1\rm\ (19#3)\ #2}
\def\TMP#1&#2&#3&{\sl\ Theor. Math. Phys.\ \bf #1\rm\ (19#3)\ #2}
\def\PTP#1&#2&#3&{\sl\ Prog. Theor. Phys.\ \bf B#1\rm\ (19#3)\ #2}
\def\PLB#1&#2&#3&{\sl\ Phys. Lett.\ \bf #1B\rm\ (19#3)\ #2}
\def\CNPP#1&#2&#3&{\sl\ Comm. Nucl. Part. Phys.\ \bf#1\rm\ (19#3)\ #2}
\def\PREP#1&#2&#3&{\sl\ Phys. Reports\ \bf#1\rm\ (19#3)\ #2}
\def\NC#1&#2&#3&{\sl\ Nuovo Cim.\ \bf #1\rm\ (19#3)\ #2}
\def\NCL#1&#2&#3&{\sl\ Nuovo Cim. Lett.\ \bf#1\rm\ (19#3)\ #2}
\def\CMP#1&#2&#3&{\sl\ Commun. Math. Phys.\ \bf#1\rm\ (19#3)\ #2}
\def\AnnPhys#1&#2&#3&{\sl\ Ann. Phys.\ \bf #1\rm\ (19#3)\ #2}
\def\AnnMath#1&#2&#3&{\sl\ Ann. Math.\ \bf #1\rm\ (19#3)\ #2}
\def\TransMath#1&#2&#3&{\sl\ Trans. Amer. Math. Soc. \ \bf #1\rm\ (19#3)\ #2}
\def\SJNP#1&#2&#3&{\sl\ Sov. J. Nucl. Phys.\ \bf #1\rm\ (19#3)\ #2}
\def\JMP#1&#2&#3&{\sl\ J. Math. Phys.\ \bf #1\rm\ (19#3)\ #2}
\def\ApJ#1&#2&#3&{\sl\ Astrophys. J.\ \bf #1\rm\ (19#3)\ #2}

\REF{\arnnath}{R.~Arnowitt and P.~Nath, \PRL 69&725&92&; \PLB 289&368&92&.}
\REF{\rossrob}{G.~G.~Ross and R.~G.~Roberts, \NPB 377&571&92&.}
\REF{\nano}{S.~Kelley, J.~L.~Lopez, D.~V.~Nanopoulos, H.~Pois
and K.~Yuan \NPB 398&3&93&.}
\REF{\OP}{M.~Olechowski and S.~Pokorski, \NPB 404&590&93&.}
\REF{\ramond}{P.~Ramond, Talk presented at the International Workshop on Recent
Advances in the Superworld, Woodlands, TX, Institute for Fundamental Theory
preprint
UFIFT-HEP-93-13 (Apr. 1993);
D.~J.~Casta\~no, E.~J.~Piard and P.~Ramond, UFIFT-HEP-93-18 (Aug. 1993).}
\REF\BBOspectra{V.~Barger, M.~S.~Berger and P.~Ohmann,
U. Wisconsin preprint\brk MAD/PH/801 (Nov. 1993).}
\REF\Kanesusy{G.~L.~Kane, C.~Kolda, L.Roszkowski and J.~D.~Wells,
U. Michigan preprint UM-TH-93-24 (hep-ph/9312272).}
\REF\LangA{P.~Langacker and N.~Polonsky, University of Pennsylvania
preprint, UPR-0556T, May 1993.}
\REF\HRS{L.~Hall, R.~Rattazzi and U.~Sarid, Lawrence Berkeley Lab
preprint, LBL-333997 (hep-ph/9306309) (June 1993).}
\REF\Hempthr{R.~Hempfling, DESY preprint, DESY-93-092 (Jul. 1993).}
\REF\CarenaSOten{W.~A.~Bardeen, M.~Carena, S.~Pokorski and
C.~E.~M.~Wagner, Max Planck Institute preprint, MPI-PH-93-103 (Feb. 1994).}
\REF\Hempfling{H.~E.~Haber and R.~Hempfling, \PRD 48&4280&93&;
R.~Hempfling, DESY preprint, DESY-93-012 (Feb. 1993).}
\REF\template{H.~Arason, D.~Casta\~no, B.~Keszthelyi, S.~Mikaelian, E.~Piard,
P.~Ramond
and B.~D.~Wright, \PRD 46&3945&92&.}
\REF{\FO}{J.~Oliensis and M.~Fishler, \PRD 28&194&83&.}
\REF\DHR{S.~Dimopoulos, L.~J.~Hall and S.~Raby, \PRL 68&1984&92&;
\PRD 45&4195&92&.}
\REF\BBHZ{V.~Barger, M.~S.~Berger, T.~Han and M.~Zralek, \PRL 68&3394&92&.}
\REF\CPR{H.~Arason, D.~J.~Casta\~no, E.~J.~Piard and P.~Ramond, \PRD
47&232&93&.}
\REF\quilt{P.~Ramond, R.~G.~Roberts and G.~G.~Ross, \NPB 406&19&93&.}
\REF\ADHRS{G.~Anderson, S.~Dimopoulos, L.~J.~Hall, S.~Raby and
G.~D.~Starkman, Lawrence Berkeley Lab preprint, LBL-33531 (Aug. 1993).}
\REF{\missdoublet}{A.~Masiero, D.~V.~Nanopoulos, K.~Tamvakis
and T.~Yanagida, \PLB 115&380&82&.}
\REF{\apple}{T.~W.~Appelquist and J.~Carrazzone,\PRD 11&2856&75&.}
\REF{\weinberg}{S.~Weinberg,\PLB 91&51&80&.}
\REF{\hall}{L.~Hall,\NPB 178&75&81&.}
\REF\Schnitzrut{B.~A.Ovrut and H.~J.~Schnitzer,\PRD 21&3369&80&;\brk
\PRD 22&2518&80&;
\NPB 179&381&81&;\brk
\PLB 100&403&81&;\NPB 184&109&81&.}
\REF{\inprep}{B.~Wright and B.~Keszthelyi, in preparation.}
\REF\us{H.~Arason, D.~Casta\~no, B.~Keszthelyi, S.~Mikaelian, E.~Piard,
P.~Ramond
and B.~D.~Wright, \PRL 67&2933&91&.}
\REF\KLN{S.~Kelly, J.~L.~Lopez and D.~V.~Nanopoulos, \PLB 274&387&92&.}
\REF\ALS{B.~Ananthanarayan, G.~Lazarides and Q.~Shafi, \PRD 44&1613&91&.}
\REF\GHS{A.~Giveon, L.~Hall and U.~Sarid, \PLB 271&138&91&.}
\REF\EJ{M.~B.~Einhorn and D.~R.~T.~Jones, \NPB 196&475&82&.}
\REF\IKT{I.~Antoniadis, C.~Kounnas and K.~Tamvakis, \PLB 119&377&82&.}
\REF\HY{K.~Hagiwara and Y.~Yamada, \PRL 70&709&93&; Y.~Yamada, \ZPC 60&83&93&.}
\REF\HMY{J.~Hisano, H.~Murayama and T.~Yanagida, \NPB 402&46&93&; \PRL
69&1014&92&.}
\REF\Rxi{B.~Ovrut and J.~Wess, \PRD 25&409&82&.}
\REF\SRxi{P.~Bin\'etruy, P.~Sorba and R.~Stora, \PLB 129&85&83&.}
\REF\GSR{M.~T.~Grisaru, W.~Siegel and M.~Ro\v{c}ek, \NPB 159&429&79&.}
\REF\DR{W.~Siegel, \PLB 84&193&79&, \PLB 94&37&80&,\brk
D.~M.~Capper, D.~R.~T.~Jones and P.~van~Nieuwenhuizen, \NPB 167&479&80&.}
\REF\nonrenorm{J.~Wess and B.~Zumino, \PLB 49&52&74&;
J.~Iliopoulos and B.~Zumino, \NPB 76&310&74&; S.~Ferrara,
J.~Ilioupoulos and B.~Zumino, \NPB 77&413&74&; B.~Zumino, \NPB 89&535&75&;
S.~Ferrara and O.~Piguet, \NPB 93&261&75&.}
\REF\NRO{L.~J.~Hall and U.~Sarid, \PRL 70&2673&93&.}
\REF\DFS{G.~Degrassi, S.~Fanchiotti and A.~Sirlin, \NPB 351&49&91&.}
\REF\LangB{P.~Langacker and N.~Polonsky, \PRD 47&4028&93&.}
\REF\LangC{P.~Langacker and N.~Polonsky, University of Pennsylvania
preprint, UPR-0594T (Feb. 1994).}
\REF{\gray}{N.~Gray, D.~J.~Broadhurst, W.~Grafe, and K.~Schilcher,\ZPC
48&673&90&.}
\REF{\BW}{W.~Bernreuther and W.~Wetzel, \NPB 197&228&82&.}
\REF{\Marsiske}{H.~Marsiske, Talk presented at the Second Workshop on
Tau Lepton Physics, Ohio State University, Sept 1992,
SLAC preprint SLAC-PUB-5977 (Oct. 1992)}
\REF\glexp{F.~M.~Borzumati, DESY preprint 93-090 (1993).}
\REF\LEPupdate{The LEP Collaborations ALEPH, DELPHI, L3, OPAL
and the LEP Electroweak Working Group, CERN preprint, CERN/PPE/93-157
(Aug. 1993).}
\REF\Carena{M.~Carena, S.~Pokorski and C.~E.~M.~Wagner, \NPB 406&59&93&.}
\REF\Dzerotop{The D0\llap/ Collaboration, S.~Abachi \etal, FERMILAB
preprint, FERMILAB Pub-94/004-E (Jan. 1994).}
\REF\inpreptwo{V.~Barger, M.~S.~Berger, P.~Ohmann and B.~Wright, in
preparation.}
\REF\Naculich{S.~G.~Naculich, \PRD 48&5293&93&.}
\REF\BBOfixpt{V.~Barger, M.~S.~Berger and P.~Ohmann, \PRD 47&2038&93&;
V.~Barger, M.~S.~Berger, P.~Ohmann and R.~J.~Phillips, \PLB 314&351&93&.}
\REF\CarenaIRone{M.~Carena, M.~Olechowski, S.~Pokorski and
C.~E.~M.~Wagner, \PLB 320&110&94&.}
\REF\pdecayorig{S.~Weinberg, \PRD 26&287&82&; N.~Sakai and T.~Yanagida, \NPB
197&533&82&;
S.~Dimopoulos, S.~Raby and F.~Wilczek, \PLB 112&133&82&; J.~Ellis,
D.~V.~Nanopoulos and S.~Rudaz,
\NPB 202&43&82&; S.~Chadha and M.~Daniels, \NPB 229&105&83&; B.~A.~Campbell,
J.~Ellis
and D.~V.~Nanopoulos, \PLB 141&229&84&; R.~Arnowitt, A.~H.~Chamseddine and
P.~Nath, \PLB 156&215&85&,
 \PRD 32&2348&85&.}
\REF\Kam{Kamiokande Collaboration, K.~S.~Hirata, \etal, \PLB 220&308&89&.}
\REF{\MV}{M.~E.~Machacek and M.~T.~Vaughn,\NPB 222&83&83&;\brk
\NPB 236&221&84&;\NPB 249&70&85&.}
\REF\BBORGE{V.~Barger, M.~S.~Berger and P.~Ohmann, \PRD 47&1093&1993&.}
\REF\MY{T.~Moroi and T.~Yanagida, Tohoku University preprint, TU-455
(Mar. 1994).}
\REF\Polonyi{G.~D.~Coughlan, N.~Fischler, E.~W.~Kolb, S.~Raby and
G.~G.~Ross, \PLB 131&59&83&.}
\REF\Higgshunt{J.~F.~Gunion, H.~E.~Haber, G.~L.~Kane and S.~Dawson,
``The Higgs Hunter's Guide,'' Addison-Wesley (1990).}
\REF\HaberKane{H.~Haber and G.~Kane, \PREP C117&75&85&.}
\REF\Rosiek{J.~Rosiek, \PRD 41&3464&90&.}
\REF\HaberGun{J.~F.~Gunion and H.~E.~Haber, \NPB 272&1&86&.}
\REF\CarenaIRtwo{M.~Carena, M.~Olechowski, S.~Pokorski and
C.~E.~M.~Wagner, CERN preprint, CERN-TH.7060/93 (Oct. 1993).}

\chapter{Introduction}

In light of the current renewal of interest in renormalization
group (RG) constraints arising from supersymmetric
grand unified theories (SUSY-GUTs),
it has become increasingly important to quantify corrections to
GUT scale predictions. Recently several two loop
renormalization group analyses of SUSY-GUTs with soft supersymmetry-breaking
induced via minimal N=1 supergravity have been
performed\rlap.\refmark{\arnnath - \Kanesusy}
In certain scenarios, these analyses have made predictions for
the sparticle spectrum and low energy parameters arising from
GUT scale constraints. Emphasis should be placed on determining
the uncertainties in these predictions
and their dependence on the unknown heavy mass spectrum as well as
the details of supersymmetry breaking.
However, only some of these analyses treat
the one loop threshold matching conditions which are necessary
for a consistent two loop analysis of the gauge coupling unification
and few consider thresholds
occuring in Yukawa couplings and mass parameters in
general. Usually threshold effects are
considered only for the gauge couplings.\footnote{1}{
For a discussion of this indirect influence of gauge coupling
thresholds on the prediction of $m_b(M_Z)$ from Yukawa
unification, see Ref.~{\LangA}.}
We find that the Yukawa coupling thresholds
are typically at least as important as the gauge coupling
threshold effects.

The purpose of this article is to present a complete treatment of
Yukawa threshold corrections in the Standard Model (SM), the
Minimal Supersymmetric Standard Model (MSSM) and
the minimal supersymmetric SU(5) GUT.
We also review the gauge coupling threshold corrections in these
models and attempt to quantify both types of correction terms.
These effects are applied in the context of gauge
and third generation Yukawa unification in minimal SU(5), and are used
to probe the superheavy spectrum and the top quark mass.
For definiteness, we consider the scenario in which these unifications,
modulo grand unified scale thresholds, are used to solve for
the physical top mass $M_t$, the colored Higgs triplet mass $\Mhs$
and an effective GUT scale $\Mx$. These solutions are
considered for experimental ranges of values of $\sin\theta_W$,
the strong coupling $\alpha_s$ and
the physical bottom quark mass $M_b$ and central values of
the remaining weak scale parameters. We also give, especially for
the $M_t$ solution, the dependence on $\tan\beta$, the
ratio of the VEVs of the two Higgs doublets in the MSSM (see Appendix
A for conventions). We highlight the importance
of threshold effects in these solutions by displaying their
quantitative dependence on the sparticle and GUT scale spectrum.

We find significant variation of the solutions for $\Mhs$ and $\Mx$
with top masses in the range $130 < M_t < 200$ GeV. The corresponding
weak threshold corrections are between $1$ and $40\%$ for the $\Mx$
solution and between $6$ and $-95\%$ for the $\Mhs$ solution.
The latter solution is more sensitive to the
large $\alpha_s$ error bars, two loop effects and
boundary corrections at $M_Z$ and $\Ms$, particularly since
$\Mhs$ itself only appears in threshold matching functions.
The decline of the
$\Mhs$ solution with $M_t$ is to be contrasted with the increase
in the lower bound on $\Mhs$ from proton decay in which a large top
Yukawa enhances the corresponding effective dimension 5 operator
through renormalization group effects. We allude to the possibilities
for constraining the minimal SUSY-SU(5) model using proton
decay limits without introducing theoretical prejudices about
the GUT scale spectrum.
We also re-emphasize the sensitivities of $\Mx$ and $\Mhs$ on
the gluino and Higgsino masses respectively.

The effect of the Yukawa thresholds is seen in the $M_t$ solution
or, equivalently, $m_b(M_Z)$ for fixed $M_t$. We discuss the robustness
of the former solution compared to the latter in the context
of both Yukawa threshold corrections at $\Ms$ and $\Mx$. We
find that these GUT scale corrections to $M_t$ typically range from
$+10$ to $-5\%$ for extreme ranges of the ratio of the superheavy
vector and adjoint scalar masses.
We dissect the various supersymmetric Yukawa coupling
threshold corrections, in particular the various sources
of enhancements in the large $\tan\beta$ region coming
from certain vertex diagrams involving gluinos and charginos.
Quantitative effects for $M_t$ are
given for sample sparticle spectra for low, intermediate and
large $\tan\beta$. Characteristically, we find positive corrections
to $M_t$ of order $2$ - $7\%$ for low $\tan\beta = 1.5$ and for both
signs of the supersymmetric Higgs mass parameter $\muh$.
For higher $\tan\beta$ and $\muh > 0$, we find typically
negative corrections of up to $10\%$ for $\tan\beta = 15$.
For $\tan\beta \gsim 40$, the corrections are larger and highly
spectrum dependent. The inclusion of the Yukawa
coupling thresholds can also significantly modify
perturbativity constraints on third generation Yukawa couplings.

The importance of
Yukawa coupling thresholds has come to the
attention of several authors
recently\rlap.\refmark{\HRS - \CarenaSOten}
Loop corrections to low energy
mass parameters have been treated in certain cases such as
the well-known case of radiative corrections to the Higgs boson
masses in the MSSM\refmark{\Hempfling} although
usually not explicitly in the form of threshold conditions on
running mass parameters except in the case of the
Standard Model\refmark{\template} and for $y_b$ and $y_\tau$ in
ordinary GUTs\rlap.\refmark{\FO} However, in the context
of renormalization group analyses of high scale predictions
they have not been treated generally.

This work is organized as follows.
We first consider in Section 2 the general treatment of
Yukawa and mass parameter thresholds. This analysis provides a
framework which can be applied to more complicated
mass and mixing angle textures.
In Section 3 we give a complete calculation of the GUT scale matching
functions for the Yukawa couplings in minimal SUSY-SU(5).
In the spirit of quantifying uncertainties in SUSY-GUT predictions,
we consider in particular the one loop threshold corrections
to the tree-level relation $y_b(\Mx) = y_{\tau}(\Mx)$ and
highlight its sensitivity to the top Yukawa coupling $y_t(\Mx)$
and the possible splitting
of the colored Higgs triplet superfields from the other superheavy
fields.
This is one of many threshold corrections to various mass relations
discussed recently in Refs.~{\DHR\ - \ADHRS} when a
specific GUT or string model
is taken to generate certain textures in the fermion mass matrices
at the unification scale.
We also review the GUT scale matching functions for the gauge
couplngs.

In Section 4 we give the most important contributions
to the Yukawa coupling matching conditions at $\Ms$ in the
MSSM in the case of one light Higgs doublet in the effective SM
below $\Ms$. We show the origin of the
large enhancements which occur for large $\tan\beta$.
We discuss the sensitivity of these corrections to the soft
supersymmetry breaking parameters of the model.
A complete calculation of these thresholds for a general
sparticle spectrum is given in Appendix A.
We also review the gauge coupling thresholds at $\Ms$ as well as
the top mass dependent gauge and Yukawa thresholds
when the top quark together with the light Higgs and
$W$ and $Z$ bosons are integrated out at $M_Z$.
Finally we describe our procedure
for extracting the heavy quark pole masses.

In Section 5 we apply these threshold corrections to
analytic predictions for $\Mx = (\Mv^2\Msig)^{\scrscr{1\over 3}}$,
$\Mhs$ and $m_t(M_Z)$, where $\Mv$ and $\Msig$ are the superheavy
vector and the superheavy adjoint scalar masses, respectively.
The first two predictions are essentially determined
by gauge coupling unification. Generally gauge unification
can be used to determine the unification scale $\Mx$, the value
of the couplings at $\Mx$, $\ag$, and either $\alpha_s(M_Z)$,
$\sin\theta_W(M_Z)$ or $\Mhs$. In the latter case the
mismatch of the gauge couplings at the GUT scale determines
the colored Higgs triplet mass through threshold corrections.
We choose this somewhat unconventional case as it is more useful
in discussing proton decay constraints. We quantify the
importance of threshold corrections at $M_Z$ and $\Ms$ on the
$\Mx$ and $\Mhs$ solutions. In particular, we describe the
importance of the top
mass dependence of $\sin^2\theta_W(M_Z)$ and the gluino and
Higgsino masses, respectively, on these two solutions.
For the last prediction, we give a semi-analytic solution
for $m_t$ which allows one to identify the relative
importance of the various correction terms. We note the robustness of
the $m_t$ solution in all but the extreme large $\tan\beta$ region
and relate this to the attraction of the infrared quasi-fixed point
in $y_t$. For all the solutions we show their dependence
on the experimental uncertainties in low scale parameters.

In Section 6 we briefly review the proton decay bound on $\Mhs$
and indicate how conservative bounds can be improved when
applied to the context considered here. We also review how
perturbativity arguments are used to put theoretical bounds on
the GUT scale spectrum. We apply these, together with the condition
that all masses lie below the Planck mass $\Mpl$, only to
delineate the cases in which a perturbative analysis
is appropriate.

We give the full two loop numerical solutions for
$\Mx$, $\Mhs$ and $M_t$ in Section 7. We give
the extreme ranges of $\Mx$ and $\Mhs$ consistent with
gauge coupling unification and low energy data.
We display the effects of the weak and
supersymmetry scale threshold corrections as well as the
error bars on $\alpha_s$ of these solutions in graphical form.
The effects of the thresholds at $\Ms$ and $\Mx$ on the $M_t$
solution are separately discussed. In the former case we quantify
the effects in the low, intermediate and high $\tan\beta$ regions for
sample sparticle spectra. In the latter case we show how the
GUT scale spectrum can modify the prediction for $M_t$ as a function
of $\tan\beta$. We also show the dominant effects of the uncertainties
in $M_b$ and $\alpha_s$ on $M_t$.

Finally we make two general remarks.
The first concerns our restriction to minimal SUSY-SU(5).
This restriction is chosen merely for its relative simplicity.
The minimal case may also be indicative of what one may expect
in more general GUTs. Thus we do not address the less appealing
features of minimal SU(5) such as the doublet-triplet
splitting problem and the unsuccessful mass relations for
the first and second generation fermions. These
difficulties can be resolved by complicating the
GUT scale superpotential\rlap.\refmark{\missdoublet} Instead we regard
minimal SUSY-SU(5) as a toy model for investigation.
In general, one expects the threshold corrections in
more complicated models to be at least as large as those
in the minimal case.

Second, we would like to advocate a particular philosophy for
the treatment of threshold corrections in minimal subtraction (MS) schemes.
{}From the Appelquist-Carrazzone decoupling theorem\refmark{\apple}
we expect the physics at energies below a given mass scale to be independent
of the particles with masses higher than this threshold.
However, MS schemes are not physical
in the sense that they are scale dependent and  mass independent so that
the  decoupling theorem is not manifest.
As described in Refs.~\weinberg\ - \Schnitzrut ,
one implements the decoupling in MS schemes by formulating a low energy
effective theory obtained by integrating out the heavy fields to one loop.
The effect of this procedure is to
give relations between renormalized parameters just above and
below the particle mass(es),
the so-called matching functions, and to modify the various $\beta$
function coefficients so that in the lower scale theory the contribution
of the particle(s) to these coefficients is removed. The running parameters
are typically discontinuous at the boundary at which the matching
function is applied unless one tunes the boundary scale. In either case
reliable values for the parameters of the theory are obtained asymptotically
away from the boundary.

In the treatment of gauge coupling thresholds,
particularly in the MSSM and in GUTs, many authors have identified a threshold
with each sparticle (often assuming small mixings or large degeneracies)
or superheavy particle mass, and modified the beta function coefficients
between each threshold. Here the couplings are usually taken to be
continuous at the boundary, which is justified since the gauge
matching functions in supersymmetric theories in the dimensional
reduction (DR) scheme
vanish at the mass of the particle defining the boundary.
However this procedure can be complicated when the effect of integrating
out the particle breaks the gauge symmetry of the higher scale theory.
In the case of split multiplets (such as the Higgs {\bf 5} of SU(5)),
one typically must use different sets of $\beta$ functions depending on
different mass orderings of the spectrum.  Also, when Yukawa thresholds are
included the Yukawa and mass parameters are in general not continuous
at threshold boundaries even in DR schemes.

Instead we use the arguably simpler scheme advocated by Hall\refmark{\hall}
in which one integrates out together all particles with similar masses
at a single scale. If one uses one loop matching functions then this
is justified as long as one loop $\beta$ functions can be used between the
different particle masses\rlap.\refmark{\hall} This is generally
the case. This is simpler both from
the analytic and numerical standpoint when it is applied to the MSSM
and its grand unified extensions. For example, in the analysis below in which
one has a light SM Higgs doublet below $M_{{\scrscr SUSY}}$,
the top quark as well as the $W$, $Z$, and SM Higgs are integrated out
at $M_Z$, all sparticles including the heavier Higgs doublet can be integrated
out at a fixed scale, $M_{{\scrscr SUSY}}$ (with complicated matching
functions incorporating the details of the spectrum) and all superheavy
particles at $M_{{\scrscr GUT}}$. With more generality and without a
loss of accuracy, the proliferation of scales and sets of $\beta$
functions is reduced to at most three.

\chapter{Yukawa Thresholds: The General Case}

The treatment of Yukawa coupling and mass parameter thresholds
in minimal subtraction (MS) schemes is analogous to the treatment of gauge
thresholds given by Weinberg\rlap,\refmark{\weinberg} Hall\refmark{\hall} and
Ovrut and Schnitzer\rlap.\refmark{\Schnitzrut}
For completeness we outline the procedure for Yukawa coupling and mass
thresholds
below.

Consider a generic gauge theory in which the kinetic and Yukawa contributions
to the bare Lagrangian are
$$ \eqalign{\Lagr_{kin} = \ &\Lagr_{gauge}
		   + i\ovl{\psi}_{Li}\!\dslash{\psi}^{}_{Li}
                   + i\ovl{\psi}_{Ri}\!\dslash{\psi}^{}_{Ri}
		   + \partial_{\mu}\phi^{\dag}_a\partial^{\mu}\phi^{}_a
		   + \dots \cr
	    \Lagr_{Y} = \ &-\ovl{\psi}_{Li}M_{ij}\psi^{}_{Rj}
                   - \ovl{\psi}_{Li}Y_{ij}^a\psi^{}_{Rj}\phi^{}_a
	           + {\rm h.c.}~,\cr}\eqn\lagrfull
$$
where $\psi_i$ is a generic fermion of type $i$, $\phi_a$ is a generic
scalar field with vacuum expectation value $v_a$, and $M_{ij} = Y_{ij}^av_a$.
We assume that it is possible to decompose the fields $\psi_i$ and $\phi_a$
into light components $\hat{\psi}_i$, $\hat{\phi}_a$ and heavy
components $\Psi_I$ and $\Phi_\alpha$
When the heavy fields are integrated out to one loop we generate the following
low energy effective Lagrangian,
$$ \eqalign{\Lagr^{eff}_{kin} = \ &\Lagr^{eff}_{gauge}
                   + i\ovl{\hat{\psi}}_{Li}\!\dslash Z^{ij}_L
			{\hat{\psi}}^{}_{Lj}
                   + i\ovl{\hat{\psi}}_{Ri}\!\dslash Z^{ij}_R
			{\hat{\psi}}^{}_{Rj}
 + \partial_{\mu}\hat{\phi}_a^{\dag}Z^{ab}_{\phi}\partial^{\mu}\hat{\phi}^{}_b
 + \dots \cr
\Lagr^{eff}_{Y} = \ &-\ovl{\hat{\psi}}_{Li}\hat{M}_{ij}\hat{\psi}^{}_{Rj}
       - \ovl{\hat{\psi}}_{Li}\hat{Y}_{ij}^a\hat{\psi}^{}_{Rj}\hat{\phi}_a
                   + {\rm h.c.} + {\rm NR}~,\cr}\eqn\lagreff
$$
where NR indicates
induced nonrenormalizable interactions. The matrix parameters $Z_{(L,R)}$
can be written as
$$Z_{(L,R)} = {1} + K_{(L,R)}~, \eqn\zlr $$
where $K_{(L,R)}$
come from light fermion wavefunction renormalization diagrams with heavy
fields in the loop. The parameters $K_{\phi}$ in
$$Z_{\phi} = {1} + K_{\phi}~, \eqn\zphi $$
comes from analogous light scalar
wavefunction renormalization. The matrices $\hat{M}$ and $\hat{Y}$ can
be written as
$$\eqalign{\hat{M} =\ & M + \delta M \equiv M(1 + K_M)~,\cr
\hat{Y} =\ & Y + \delta Y \equiv Y(1 + K_Y)~, \cr}\eqn\my
$$
where $\delta M$ and $\delta Y$ come from one loop light field self energy
and three point Yukawa vertex diagrams, respectively,
with heavy fields in the loop.

To properly normalize the kinetic terms one must first
diagonalize the hermitian matrices $Z_{(L,R)}$ and then rescale each
bare fermion field by the appropriate eigenvalue. Let
$Z_L = U_L^{\dag} Z^d_L U^{}_L$ where $Z^d_L$ is diagonal and $U_L$ is unitary,
with similar definitions for $U_R$ and $U_{\phi}$.
We redefine the
fields via
$$\eqalign {\psi'_{Li} =\ & (Z^d_{Li})^{\scrtwo} U_L^{ij} \hat{\psi}^{}_{Lj}~,
\cr
\psi'_{Ri} =\ & (Z^d_{Ri})^{\scrtwo} U_R^{ij} \hat{\psi}^{}_{Rj}~, \cr
\phi'_a =\ & (Z^d_{\phi a})^{\scrtwo} U_{\phi}^{ab}\hat{\phi}^{}_b~. \cr }
\eqn\rescale
$$
In terms of the primed fields, the kinetic terms are canonical
and the bare effective mass and Yukawa matrices are
$$ M^{eff} = (Z^d_L)^{-\scrtwo} U^{}_L (M + \delta M) U_R^{\dag}
(Z^d_R)^{-\scrtwo}~,
\eqn\meff $$
$$ Y^{eff\, a} = (Z^d_L)^{-\scrtwo} U^{}_L (Y + \delta Y)^b
U_R^{\dag} (Z^d_R)^{-\scrtwo}
U_{\phi}^{\dag ba}(Z^d_{\phi a})^{-\scrtwo}~,\eqn\yeff $$
To get the one loop matching functions one must turn these bare relations
into renormalized ones. The divergent parts in $M^{eff}$ and $Y^{eff\, a}$
are just the difference between the self energy counterterms of the effective
theory and those of the full theory. So by replacing all parameters in
\meff\ and \yeff\ by their finite parts (indicated by a bar over the quantity)
as defined in $\ovl{{\rm MS}}$ and neglecting higher order corrections,
one obtains the renormalized relations:
$$ M^{eff}_{ij}(\mu ) = ( U^{}_L M(\mu ) U_R^{\dag})_{ij}
	+ M_{ij}(\mu)(\ovl{K}_{M ij} - \ttwo( \ovl{K}^d_{Li}
+ \ovl{K}^d_{Rj} ))~,
\eqn\meffrunning $$
$$ Y^{eff\, a}_{ij}(\mu ) = (U^{}_L Y^b(\mu ) U_R^{\dag})^{ij} U_{\phi}^{\dag
ba}
+ Y^a_{ij}(\mu)(\ovl{K}_{Y ij}^a - \ttwo( \ovl{K}^d_{Li}
+ \ovl{K}^d_{Rj}  + \ovl{K}^d_{\phi a}))~,
\eqn\yeffrunning $$
In practice $M^{eff}$ must be diagonalized by a biunitary transformation
even if the original theory is written in terms of mass eigenstates.
In the case of the MSSM, for example, this leads to threshold corrections
in the renormalized Cabibbo-Kobayashi-Maskawa matrix at the effective
scale $M_{{\scriptscriptstyle SUSY}}$ when the superparticles
are integrated out.
The effect of such thresholds, of potential relevance to RG
analyses of GUT scale mass textures, will be discussed in a
future work\rlap.\refmark{\inprep}

\chapter{Yukawa and Gauge Thresholds in Minimal SUSY-SU(5)}

Next we apply the preceding formalism to the one loop GUT
threshold corrections to the tree-level
relation $y_b(\Mx) = y_{\tau}(\Mx)$
in minimal supersymmetric SU(5).
This is of particular interest since this condition strongly
constrains the allowed range of the top quark mass in an
experimentally accessible region\rlap.\refmark{\us - \GHS}
Here we find two contributions, one
proportional to the square of the top Yukawa coupling coming
from integrating out heavy color triplet Higgs
and Higgsino fields, the other proportional to the GUT scale gauge
coupling coming from integrating out superheavy vector bosons.
Since the soft supersymmetry breaking mass
parameters are much smaller than typical GUT scale masses we can
work in an approximately supersymmetric formalism in which both light fields
along with their superpartners are treated as massless in loops involving
superheavy fields. We therefore use supergraph methods to simplify the
calculations.

We start with a superpotential of the form
$$
\eqalign{ P =\ &  \sqrt{2} \Psi_a {Y}^{(d)} \chi^{ab} H^{(1)}_b -
{\tfour}\epsilon_{abcde} \chi^{ab} {Y}^{(u)} \chi^{cd} H^{(2)e} \cr
&\quad + M_2 H^{(2)a} H^{(1)}_a + \lambda_2 H^{(2)a} \Sigma^b_a H^{(1)}_b
+ {\lambda_1\over 3}\Tr \Sigma^3 + {M_1\over 2}\Tr \Sigma^2~, \cr}\eqn\supot$$
where $\Psi$, $H^{(1)}$, $H^{(2)}$, $\chi$ and $\Sigma$ are SU(5) superfields
transforming as the ${\bf 5}$, ${\bf 5}$, ${\bf\bar{5}}$, ${\bf 10}$ and ${\bf
24}$
dimensional representations, respectively. In the usual way we associate these
superfields with the ${\rm SU(3)}\times{\rm SU(2)}\times{\rm U(1)}$
decomposition by
$$
\chi = {1\over\sqrt{2}}\pmatrix{0&U_3^c&-U_2^c&-U^1&-D^1\cr
				-U_3^c&0&U_1^c&-U^2&-D^2\cr
				U_2^c&-U_1^c&0&-U^3&-D^3\cr
				U^1&U^2&U^3&0&-E^c\cr
				D^1&D^2&D^3&E^c&0\cr}~,\eqn\tenplet
$$
and $\Psi^{T} = (D_{\alpha}^c, E, -\nu )$, where lowercase Greek/Latin indices
denote SU(3) and SU(5) indices, respectively. In component fields the charge
conjugate Weyl fermion fields with be related to the right-handed component
of Dirac fermions.
The five dimensional Higgs supermultiplets
are likewise split into color triplet and SU(2) doublet components
according to
$$
H^{(1)} = (H^{(1)}_{3\alpha}, H^{(1)-}, -H^{(1)}_0)\ \ {\rm and}\ \
H^{(2)} = (H^{(2)\alpha}_{3}, H^{(2)+}, H^{(2)}_0)~,\eqn\triplets
$$
where the $3$ denotes
the strongly interacting triplets. These superfields couple to the SU(5) vector
supermultiplet
in the usual way. In addition to the component field interactions obtained from
\supot, one
also can add the most general set of SU(5) invariant soft supersymmetry
breaking terms.
In the threshold calculations that follow, however supersymmetry as
well as electroweak symmetry
breaking effects are unimportant as the corresponding mass scales are much
smaller than
$M_{{\scrscr GUT}}$. The SU(5) symmetry is broken without breaking
supersymmetry when
the adjoint Higgs field gets the vacuum expectation
value $\VEV{\Sigma} = V {\rm diag}(1,1,1,-3/2,-3/2)$, where $V =
2M_1/\lambda_1$.
This gives the following mass spectrum in the minimal model: two degenerate
Higgs triplet
superfields of mass $M_{{\scrscr H_3}}$, degenerate $X$ and $Y$ leptoquark
gauge superfields
of mass $M_{{\scrscr V}}$, color octet and SU(2) triplet superfields of mass
$M_{{\scrscr\Sigma}}$
along with a singlet of mass $0.2 M_{{\scrscr\Sigma}}$. The electroweak
doublets remain light
as long as the fine-tuning constraint
$\lambda_1M_2 - 3\lambda_2M_1 \approx M_{{\scrscr EW}}$ is satisfied,
where $M_{\scrscr EW}$ is a typical electroweak scale mass. This can
of course be relaxed in the missing doublet model
for example\rlap,\refmark{\missdoublet} but we limit our discussion to the
minimal model
for simplicity.

The GUT scale threshold corrections are obtained by integrating
out the superheavy $X$ and $Y$ gauge supermultiplets, the Higgs
triplet, and the Higgs adjoint superfields
to obtain an effective MSSM. These were obtained
for the gauge couplings long ago in Refs.~\EJ\ and \IKT, and have
recently been reanalyzed
in the present context in Refs.~\HY\ and \HMY. However, explicit
model dependent threshold corrections for
the SU(5) Yukawa couplings have not been computed and studied.
To properly generalize
the analysis of Weinberg and Hall\refmark{\weinberg,\hall} to
construct an effective
MSSM, one must gauge-fix the superfield Lagrangian in a
so-called supersymmetric S-covariant gauge so that the low energy gauge
symmetry S is ${\rm SU(3)}\times{\rm SU(2)}\times{\rm U(1)}$.
One essentially replaces derivatives by S-covariant derivatives in the
$R_\xi$ gauge-fixing functional of
the high energy theory.
Details in the superfield formalism can be found
in Refs.~\Rxi\ and \SRxi , but the
corresponding vertices are not involved in the Yukawa threshold calculation.
As mentioned above, at the high scale we can use
the supergraph formalism\refmark{\GSR}
to construct the effective action in Feynman gauge ($\xi = 1$)
obtained when the aforementioned
heavy fields are integrated out. The one loop divergences
are naturally regulated by
dimensional reduction ($\ovl{{\rm DR}}$)\rlap,\refmark{\DR} which
preserves supersymmetry at least up to one loop.

Due to the supersymmetric nonrenormalization
theorems\rlap,\refmark{\GSR,\nonrenorm}
the only modifications
to the parameters of the superpotential in the effective action
arise through superfield wavefunction renormalizations.
Hence if the original action has the form
$$
S = \int d^4x \int d^4\theta (\Phi_a^{\dag}{\rm e}^{{\scrscr
2gV}}\Phi^{}_a + P[\Phi ]\delta^2(\bar\theta )
+ P^{\dag}[\Phi ]\delta^2(\theta ) + \dots )~,\eqn\action
$$
with $\Phi_a = (\Phi_I,\phi_i)$ being the decomposition into heavy/light
superfields
respectively,
then the effective action (with an unconventional normalization of the kinetic
term),
suppressing gauge kinetic and interaction terms, is
$$
S^{eff} = \int d^4x \int d^4\theta ( Z_{ij}\phi_i^{\dag}{\rm e}^{{\scrscr 2gv}}
\phi^{}_j + P'[\phi ]\delta^2(\bar\theta ) + P^{\prime\dag}[\phi
]\delta^2(\theta ) + {\rm NR} + \dots )~.
\eqn\effaction
$$
As before, NR denotes effective nonrenormalizable interactions
suppressed by inverse powers of superheavy masses which, in the GUT
case, will include operators responsible for proton decay.
To evaluate Yukawa coupling corrections
one need only consider supergraph two point functions with external light MSSM
matter (s)fermion and
Higgs doublet superfields and at least one superheavy superfield in the loop.
Using the methods of Ref.~\GSR\ we find that the effective action contributions
to the light electron family, down quark family, up quark family and
neutral Higgs superfield kinetic terms
relevant for the Yukawa thresholds are
$$
\eqalignno{S^{eff} = &\int\! d^4x \int\! d^4\theta\,
(E^\dag{Z}_{{\scrscr E}}E + E^{c\dag}{Z}_{{\scrscr E^c}}E^c +
D^\dag{Z}_{{\scrscr D}}D
+ D^{c\dag}{Z}_{{\scrscr D^c}}D^c &\cr
&\quad + U^\dag{Z}_{{\scrscr U}}U
+ U^{c\dag}{Z}_{{\scrscr U^c}}U^c +
Z_{{\scrscr H^{(1)}_0}}H^{(1)\dag}_0H^{(1)}_0 +
Z_{{\scrscr H^{(2)}_0}}H^{(2)\dag}_0H^{(2)}_0 + \dots &\eqnalign\EDeffaction\cr
&\quad + E{Y}^{(e)}E^cH^{(1)}_0 + D^c{Y}^{(d)}DH^{(1)}_0 +
U^c{Y}^{(u)}UH^{(2)}_0 &\cr
&\quad + \hc + {\rm NR} + \dots )~,&\cr}
$$
where $Z_{{\scrscr E,D,U}}$ are matrices in family and color space.
Writing $Z = 1 + K$ we find
$$
\eqalign{
{K}_{{\scrscr E}} =\ & -3\bar{g}^2 A(0,M_{{\scrscr V}}^2,0){1} +
3({Y}^{(d)}{Y}^{(d)\dag})^T A(0,M_{{\scrscr H_3}}^2,0)~,\cr
{K}_{{\scrscr E^c}} =\ & -6\bar{g}^2 A(0,M_{{\scrscr V}}^2,0){1} +
3{Y}^{(u)\dag}{Y}^{(u)} A(0,M_{{\scrscr H_3}}^2,0)~,\cr
{K}_{{\scrscr D}} =\ & -3\bar{g}^2 A(0,M_{{\scrscr V}}^2,0){1} +
(2{Y}^{(u)\dag}{Y}^{(u)} + {Y}^{(d)\dag}{Y}^{(d)}) A(0,M_{{\scrscr
H_3}}^2,0)~,\cr
{K}_{{\scrscr D^c}} =\ & -2\bar{g}^2 A(0,M_{{\scrscr V}}^2,0){1} +
2({Y}^{(d)}{Y}^{(d)\dag})^T A(0,M_{{\scrscr H_3}}^2,0)~,\cr
{K}_{{\scrscr U}} =\ & -3\bar{g}^2 A(0,M_{{\scrscr V}}^2,0){1} +
(2{Y}^{(u)\dag}{Y}^{(u)} + {Y}^{(d)\dag}{Y}^{(d)}) A(0,M_{{\scrscr
H_3}}^2,0)~,\cr
{K}_{{\scrscr U^c}} =\ & -2\bar{g}^2 A(0,M_{{\scrscr V}}^2,0){1} +
({Y}^{(u)\dag}{Y}^{(u)} + 2{Y}^{(d)\dag}{Y}^{(d)}) A(0,M_{{\scrscr
H_3}}^2,0)~,\cr
{\rm K}_{{\scrscr H^{(1)}_0}} =\ & {\rm K}_{{\scrscr H^{(2)}_0}} =
-3\bar{g}^2 A(0,M_{{\scrscr H_3}}^2,M_{{\scrscr V}}^2) +
|\lambda_2 |^2 \Bigl( 3 A(0,M_{{\scrscr H_3}}^2,M_{{\scrscr V}}^2)\cr
&\qquad + {\textstyle {3\over 2}} A(0,0,M_{{\scrscr\Sigma}}^2)
+ {\textstyle{3\over 10}} A(0,0,{\textstyle{1\over 5}}M_{{\scrscr\Sigma}}^2)
\Bigr)~,\cr}
\eqn\wavefcnrenorm
$$
where $\bar{g}$ is the GUT scale gauge coupling.
Here the universal function $A$ arising from the $d$ dimensional
loop integrals is
$$
A(p,M_A^2,M_B^2) = -i\mu^{2\epsilon}\int {d^dk\over (2\pi )^d} {1\over (p+k)^2
- M_A^2 }
{1\over k^2 - M_B^2}~,
\eqn\loopint
$$
where $\epsilon = 2 - {d\over 2}$.
In the case $p = 0$ we get
$$
A(0,M_A^2,M_B^2) = {1\over (4\pi )^2}\Bigl( {1\over\eta} - F_1(M_A^2,M_B^2)
\Bigr)~,
\eqn\limloopint
$$
where ${1\over\eta} = {1\over\epsilon} + \ln{4\pi} - \gamma_{{\scrscr E}}$ and
$F_1$
is defined in Appendix B.

To determine the Yukawa threshold corrections we simply follow the generic
component
field analysis. First redefine the bare low scale effective light superfields
so that they
have a conventionally normalized kinetic term. This gives a relation between
the bare high
and low scale Yukawa matrices. The running relation is obtained by absorbing
the divergent
parts from the wavefunction renormalizations using the definition of the high
and low scale
$\beta$ functions. The resulting threshold matching functions relating
the renormalized couplings in the region of the unification scale then
depend only on the finite parts of the $Z$s. We obtain $y_\alpha (\mu) =
\bar{y}_\alpha (\mu)
(1 + \Delta^{\scrscr GUT}_{y_\alpha})$, where\goodbreak
$$
\eqalignno{
16\pi^2\Delta^{\scrscr GUT}_{y_t} =& -{\bar{g}^2\over 2}
\bigl(5 F_1(\Mx^2,0) + 3 F_1(\Mhs^2,\Mx^2)\bigr)&\cr
&\quad + \tfrac32 (\bar{y}^2_t + \bar{y}^2_b)F_1(\Mhs^2,0)&\cr
&\quad + {\lambda_2\over 2}\bigl(3 F_1(\Mhs^2,\Mx^2)
+ \tfrac32 F_1(\Msig^2,0) + {\textstyle{3\over 10}}
F_1({\textstyle{1\over 5}}\Msig^2,0)\bigr)~,&\cr
16\pi^2\Delta^{\scrscr GUT}_{y_b} =& -{\bar{g}^2\over 2}
\bigl(5 F_1(\Mx^2,0) + 3 F_1(\Mhs^2,\Mx^2)\bigr)&\cr
&\quad + (\bar{y}^2_t + \tfrac32\bar{y}^2_b)F_1(\Mhs^2,0)&\eqnalign\ytybytau\cr
&\quad + {\lambda_2\over 2}\bigl(3 F_1(\Mhs^2,\Mx^2)
+ \tfrac32 F_1(\Msig^2,0) + {\textstyle{3\over 10}}
F_1({\textstyle{1\over 5}}\Msig^2,0)\bigr)~,&\cr
16\pi^2\Delta^{\scrscr GUT}_{y_\tau} =& -{\bar{g}^2\over 2}
\bigl(9 F_1(\Mx^2,0) + 3 F_1(\Mhs^2,\Mx^2)\bigr)&\cr
&\quad + \tfrac32 (\bar{y}^2_t + \bar{y}^2_b)F_1(\Mhs^2,0)&\cr
&\quad + {\lambda_2\over 2}\bigl(3 F_1(\Mhs^2,\Mx^2)
+ \tfrac32 F_1(\Msig^2,0) + {\textstyle{3\over 10}}
F_1({\textstyle{1\over 5}}\Msig^2,0)\bigr)~,&\cr
}
$$
and where the (un)barred couplings are (low) GUT scale
parameters ($\bar{y}_b = \bar{y}_{\tau}$).
The $y_b/y_{\tau}$ threshold matching function is therefore
$$
{y_b\over y_{\tau}}(\mu ) = 1 + {1\over 16\pi^2}\bigl( 2\bar{g}^2(\mu )
(2 \ln{M_{{\scrscr V}}\over\mu} - 1)
- \ttwo \bar{y}_t^2(\mu ) (2 \ln{M_{{\scrscr H_3}}\over\mu} - 1) \bigr)~,
\eqn\Rbtauthreshold
$$
and can be applied for $\mu \simeq M_{{\scrscr GUT}}$.

We also quote the gauge matching conditions for minimal SUSY-SU(5) in
the $\ovl{{\rm DR}}$ scheme\refmark{\EJ,\IKT}:
$$
{1\over \alpha_i(\mu )} = {1\over \alpha_{{\scrscr G}}(\mu )} -
\Delta_i^{{\scrscr GUT}}(\mu )~,
\eqn\gutgaugethresh
$$
where $\ag = \bar{g}^2/4\pi$ and
$$
\eqalign{\Delta_1^{{\scrscr GUT}}(\mu ) =& -{5\over\pi}\ln {\Mv\over\mu} +
{1\over 5\pi}
\ln {\Mhs\over\mu}~,\cr
\Delta_2^{{\scrscr GUT}}(\mu ) =& -{3\over\pi}\ln {\Mv\over\mu} +
{1\over\pi}\ln{\Msig\over\mu}~,\cr
\Delta_3^{{\scrscr GUT}}(\mu ) =& -{2\over\pi}\ln {\Mv\over\mu} +
{3\over 2\pi}\ln{\Msig\over\mu} + {1\over 2\pi}\ln{\Mhs\over\mu}~.\cr}
\eqn\deltagut
$$
In the interest of simplicity and minimality we do not include
the effects of gravity-induced nonrenormalizable operators\refmark{\NRO} here.
As noted in Ref.~\HMY , differences of the $\Delta_i$ depend on
$\Mv$ and $\Msig$ in the combination $\Mx = (\Mv^2\Msig )^{\scrscr{1\over 3}}$
which
we will take as GUT scale at which the matching functions are applied
in the analysis to follow.

\chapter{Electroweak and Supersymmetric Thresholds}

In addition to the GUT scale thresholds there are also thresholds
at the electroweak scale and the effective
supersymmetry scale $\Ms$.
At the electroweak scale, $M_Z$, we integrate out the top quark, the weak gauge
bosons
and the SM Higgs. The definition of
the inverse electromagnetic coupling $\alpha(M_Z ) = 127.9 \pm 0.1$ and the
Weinberg angle
$s^2_0(M_Z ) = 0.2324 \pm 0.0003$ includes one loop
corrections from electroweak gauge
bosons and the top quark\refmark{\DFS} for a pole mass $M_{t0} = 143$ GeV.
This must be corrected for different top masses above $M_Z$.  The result
is\refmark{\LangB}
$$
\eqalign{
{1\over \alpha_1(M_Z)} =& {3\over 5}(1 - s^2(M_Z))\Bigl({1\over\alpha(M_Z)} +
{8\over 9\pi}\ln {M_t\over M_{t0}}\Bigr) -
{3\over 5}{\Delta_{{\scrscr s^2}}^{top}\over\alpha(M_Z )}~,\cr
{1\over \alpha_2(M_Z)} =& s^2(M_Z)\Bigl({1\over\alpha(M_Z)} +
{8\over 9\pi}\ln {M_t\over M_{t0}}\Bigr)
+ {\Delta_{{\scrscr s^2}}^{top}\over\alpha(M_Z )}~,\cr
{1\over \alpha_3(M_Z)} =& {1\over \alpha_s(M_Z)} + {1\over 3\pi}\ln {M_t\over
M_Z}~,\cr}
\eqn\alphasmzplus
$$
where $\Delta_{{\scrscr s^2}}^{top} \approx -0.92\times 10^{-7}
{\rm GeV}^{-2} (M_t^2 - M_{t0}^2)$\refmark{\LangC}
and accounts for the quadratic top mass dependence of
$\sin^2\theta_W(M_Z)\equiv s^2(M_Z)$ in Ref.~\DFS .
Below $M_Z$ the ``true'' decoupled QED coupling is related to $\alpha(M_Z )$
via
$$
{1\over \alpha^-(M_Z )} = {1\over \alpha(M_Z )} - {8\over 9\pi}\ln {M_{t0}\over
M_Z}
 - {1\over 6\pi}(1 + 21\ln {M_Z\over M_W})~.\eqn\alphaemlow
$$

We also incorporate the mass thresholds of Ref.~\template\ to determine the
relation
between the running fermion masses defined in the effective ${\rm SU(3)}\times
{\rm U(1)}$
low energy theory and those above $M_Z$. The effect of these thresholds is
quite small
given the smallness of $\alpha_2(M_Z ) \approx 0.03 $ and the Yukawa couplings
of the light
fermions. However these results did not account for a heavy top quark.
Essentially the only
modification of the results is the matching function for the bottom quark mass.
By properly integrating out the charged Nambu-Goldstone bosons in Feynman gauge
one obtains a dependence on the SM top Yukawa coupling,
$y^{\scrscr SM}_t$ ($y^{\scrscr SM}_i(\mu) = m_i(\mu)/v$ where $v =
174.104$ GeV and $i$ runs over all SM fermions).
The general result for all
massive fermions lighter than the top quark is\refmark{\template}
$$
m_i^{low}(\mu ) = m_i^{\scrscr SM}(\mu )(1 + \Delta_{m_i}^{{\scrscr
SM}})~,\eqn\lowmassthresh
$$
where
$$
\Delta_{m_i}^{{\scrscr SM}} =  {\alpha_2(\mu )\over 8\pi c^2}\Bigl(
(-{\textstyle {3\over 4}}(g_V^i)^2 + {\textstyle {5\over 4}}(g_A^i)^2)
(\ln {M_Z\over\mu} - \tfour ) + c^2(\ln {M_W\over\mu} - \tfour
)\Bigr)~,\eqn\deltasmmass
$$
for all but the bottom quark mass and $c = \cos\theta_W$.
Here  $g_{V,A}^i = 2 ( g_L^i \pm g_R^i)$, where $g_{L,R}^i= T_{3\, L,R}^i - s^2
Q^i $
and $T_{3\, L,R}^i$ and $Q^i$ are the third component of weak isospin and the
electric charge, respectively, for a given handedness of the $i^{\rm th}$
fermion.
For the different quark and lepton charge sectors one has
$$\eqalign{ g_A^{\nu} = \hphantom{-}1,\ \ \qquad g_V^{\nu} &=\  1,\cr
g_A^{e} = -1,\ \ \qquad g_V^{e}&= -1 + 4s^2,\cr
g_A^{u} = \hphantom{-}1,\ \ \qquad g_V^{u} &=\  1 - {\textstyle {8\over
3}}s^2,\cr
g_A^{d} = -1,\ \ \qquad g_V^{d} &= -1 + {\textstyle {4\over 3}}s^2.\cr}
\eqn\qnumbers $$
For the case of the bottom mass one obtains
$$
\eqalign{
\Delta_b^{{\scrscr SM}} =\ & {\alpha_2(\mu )\over 8\pi c^2}\Bigl(
(-{\textstyle {3\over 4}}g_V^{d2} + {\textstyle {5\over 4}})
(\ln {M_Z\over\mu} - \tfour ) + c^2 |V_{tb}|^2 F_2(M_W^2,M_t^2)\Bigr)\cr
&\quad - {(y_t^{\scrscr SM})^2(\mu )\over 16\pi^2}|V_{tb}|^2
\Bigl(F_1(M_W^2,M_t^2) -
\tfour F_2(M_W^2,M_t^2)\Bigr)~,\cr}\eqn\deltabsmmass
$$
where the CKM matrix element $|V_{tb}| \approx 1$ and $F_{1,2}$ are defined in
Appendix B.
The term proportional to $y^{\scrscr SM}_t$ is dominant. For pole
masses $M_t$ up to $200$ GeV
this effect corresponds to a shift of $m_b(M_Z)$ downwards by at most $0.6\% $
from its
value below $M_Z$.

The pole masses for the heavier quarks, $M_q (q = c,b,t)$, are determined by
finding a consistent solution to the three loop relation
$$
M_q = m_q(M_q)\biggl(1+{4\over 3}{\alpha_s(M_q)\over\pi}
+ K_q\big({\alpha_s(M_q)\over\pi}\big)^2\biggr)~,\eqn\runpol
$$
where
$$
K_q = 16.11 - 1.04\sum_{M_i< M_q}(1-{M_i\over M_q})~,\eqn\bigk
$$
with $K_c=13.3$ and $K_b=12.4$\rlap.\refmark{\gray}
In this analysis we will take the pole mass $M_b$ in the range
from $4.7$ to $5.3$ GeV. Using the three loop running of the strong
coupling, $\alpha_s$,\footnote{1}{We also include two loop QCD
threshold effects\refmark{\BW} which, however, are negligible in the
mass range of interest.} we find $m_b(M_b) = 4.06,\ 4.24,\ 4.43,\ 4.61 \pm
0.08$ GeV
and $m_b(M_Z) = 2.81,\ 2.96,\ 3.11,\ 3.26 \pm 0.16$ GeV
for $M_b = 4.7,\ 4.9,\ 5.1,\ 5.3$ GeV, respectively, and we have included the
uncertainty for the strong coupling $\alpha_s(M_Z) = 0.120\pm 0.007$.
The pole mass for the $\tau$ lepton is given experimentally by
$M_{\tau} = 1.7771 \pm 0.0005$ GeV\rlap,\refmark{\Marsiske}
which, using the QED relation between the pole and running masses
gives $m_{\tau}(M_Z) = 1.7476 \pm 0.0006$ just below the electroweak
threshold. With the loop effects of $W$ and $Z$ bosons included as in
Eq.~\deltasmmass\ one
has $m_{\tau}^+(M_Z) = 1.7494 \pm 0.0006$ just above the threshold.

The supersymmetric threshold is a potentially more important correction to
to RG evolution of the couplings. It is sensitive to the
details of the sparticle spectrum and in the case of the Yukawa
coupling thresholds
has quite different effects in different regions of parameter space.
We shall first review the gauge thresholds and then give a simplified form
for the Yukawa thresholds. The exact Yukawa coupling matching
functions for the MSSM in the
case of one light Higgs doublet are given in Appendix A.

The exact form of the matching functions for the gauge couplings at $\Ms$ is
well known.
We will parametrize the gauge thresholds in terms of three
mass scales\rlap,\refmark{\LangA} $M_i$. The general result for the
gauge threshold matching conditions is
$$
{1\over \alpha_i^-(\mu )} = {1\over \alpha_i^+(\mu )} - \Delta_i^{{\scrscr
SUSY}}(\mu )
- \Delta_i^{{\scrscr DR}}~,
\eqn\susygaugethresh
$$
where $\alpha_i^{\pm}$ denotes the gauge couplings just above/below $\Ms$.
We convert from $\ovl{\rm MS}$ to $\ovl{\rm DR}$
couplings above $\Ms$ with the conversion factor given by
$\Delta_i^{{\scrscr DR}} = -C_2(G_i)/12\pi$,
where the quadratic Casimir $C_2(G)$ is N for SU(N) and 0 for U(1) groups.
The matching functions are
$$
\Delta_i^{{\scrscr SUSY}}(\mu ) = {1\over 2\pi}\sum_p b^{(p)}_i \ln
{M_p\over\mu}~,
\eqn\deltasusy
$$
where $p$ runs over all sparticles integrated out near $\mu\approx\Ms$.
Here $b^{(p)}_i$ is the contribution of sparticle $p$ to the one loop
coefficient $b_i = ({33\over 5},1,-3),\ (i=1,2,3)$ of the
$\beta$ function for $g_i$ in the MSSM,
$$ \beta_{g_i} = {dg_i\over dt} = {b_i\over 16\pi^2}g_i^3~, \eqn\betag $$
where $t = \ln\mu$.
The cumulative effect of these thresholds can be parametrized in terms of mass
scales $M_i$
via
$$
\Delta_i^{{\scrscr SUSY}}(\mu ) = {1\over 2\pi}(b_i - b_i^{{\scrscr SM}}) \ln
{M_i\over\mu}~,
\eqn\deltasusyb
$$
where $b_i^{{\scrscr SM}} = ({41\over 10},-{19\over 6},-7)$ are
the corresponding coefficients in the SM and
$b_i - b_i^{{\scrscr SM}} = ({\textstyle {5\over 2}, {25\over 6}}, 4)$.
To get some idea of the dependence of the $M_i$ on the sparticle spectrum
we compute them for the case of separate degeneracies among squarks, sleptons,
gauginos, higginos, and heavy Higgs particles:
$$
\eqalign{
M_1 =\ & m_{\tilde{\ell}}^{9\over 25} m_{\tilde{q}}^{11\over 25}
m_{\tilde{H}}^{4\over 25}
m_H^{1\over 25}~,\cr
M_2 =\ & m_{\tilde{\ell}}^{3\over 25} m_{\tilde{q}}^{9\over 25}
m_{\tilde{H}}^{4\over 25}
m_{\tilde{W}}^{8\over 25} m_H^{1\over 25}~,\cr
M_3 =\ & m_{\tilde{q}}^{1\over 2} \mgl^{1\over 2}~.\cr}
\eqn\misusy
$$

The complete Yukawa thresholds at $\Ms$
are quite complicated, however we can give some indication of their
form and estimate their effects.
The most tractable form useful for making estimates occurs in the
limit of small gaugino-Higgsino mixing (see Eqs.~(A.30)-(A32)) and is given by
$$
\eqalign{
y_{t}^{\scrscr SM}(\mu ) =\ & y_t(\mu )\sin\beta (1 + \Delta^{\scrscr
SUSY}_{y_t})~,\cr
y_{b}^{\scrscr SM}(\mu ) =\ & y_b(\mu )\cos\beta (1 + \Delta^{\scrscr
SUSY}_{y_b})~,\cr
y_{\tau}^{\scrscr SM}(\mu ) =\ & y_{\tau}(\mu )\cos\beta
(1 + \Delta^{\scrscr SUSY}_{y_\tau})~,\cr}
\eqn\susyyukmatcha
$$
where
$$
\eqalignno{
16\pi^2\Delta^{\scrscr SUSY}_{y_\tau} \simeq
\tfrac14 &y_\tau^2\biggl(F_2(\msnutau^2,\mhino^2)
+ \sum_i F_2(\mstaui^2,\mh0ino^2) + \sin^2\!\beta\, F_2(\mhpm^2,0)\biggr)&\cr
+ \tfrac12 &g_2^2\biggl(\muh{\cal M}_2\tan\beta
\sum_i (G_2(\mstaui^2,\mh0ino^2,\mw3ino^2)(O^{\tau}_{i1})^2
&\eqnalign\approxdytaususy\cr
&\quad + G_2(\msnutau^2,\mhino^2,\mwino^2))&\cr
& \quad - 2t^2{\cal M}_1(\muh\tan\beta + A_{\tau})
\sum_{i,j} G_2(\mbino^2,\mstaui^2,\mstauj^2)(P'_{ij}(\tau ))^2\biggr)~,&\cr
}
$$
$$
\eqalign{
16\pi^2\Delta^{\scrscr SUSY}_{y_b} \simeq \tfrac83
&g^2_3\biggl(-2\mgl(\muh\tan\beta + A_b)
\sum_{i,j} G_2(\mgl^2,\msbi^2,\msbj^2)(P'_{ij}(b))^2\cr
&\quad + \tfrac14\sum_i F_2(\msbi^2,\mgl^2)\biggr)\cr
+ &y_t^2\biggl(-2\muh\tan\beta\sum_{i,j}
\Bigl(m_t G_2(\mhino^2,\msti^2,\mstj^2)P'_{ij}(t)P^{}_{ij}(t)\cr
&\quad + (\muh\cot\beta +
A_t)G_2(\mhino^2,\msti^2,\mstj^2)(P'_{ij}(t))^2\Bigr)\cr
&\quad + \tfrac14\Bigl(\sum_i F_2(\msti^2,\mhino^2)(O^t_{i2})^2
+ \cos^2\!\beta\, F_2(\mhpm^2,m_t^2)\Bigr)\biggr)\cr
+ \tfrac14 &y_b^2\biggl(\sum_i F_2(\msti^2,\mhino^2)(O^t_{i1})^2
+ \sum_i F_2(\msbi^2,\mh0ino^2)\cr
&\quad + \sin^2\!\beta\, F_2(\mhpm^2,m_t^2)\biggr)\cr
+ \tfrac12 &g_2^2\biggl(\muh{\cal M}_2\tan\beta \sum_i
\Bigl(G_2(\msbi^2,\mh0ino^2,\mw3ino^2)(O^b_{i1})^2\cr
&\quad +
G_2(\msti^2,\mhino^2,\mwino^2)(O^t_{i1})^2\Bigr)\biggr)~,\cr}\eqn\approxdybsusy
$$
$$
\eqalign{
16\pi^2\Delta^{\scrscr SUSY}_{y_t} \simeq \tfrac83
&g^2_3\biggl(-2\mgl\sum_{i,j}\Bigl( m_t
G_2(\mgl^2,\msti^2,\mstj^2)P'_{ij}(t)P^{}_{ij}(t)\cr
&\quad + (\muh\cot\beta + A_t)G_2(\mgl^2,\msti^2,\mstj^2)(P'_{ij}(t))^2
\Bigr)\cr
&\quad + \tfrac14\sum_i F_2(\msti^2,\mgl^2)\biggr)\cr
+ \tfrac14 &y_t^2\biggl(\sum_i F_2(\msbi^2,\mhino^2)(O^b_{i1})^2
+ \sum_i F_2(\msti^2,\mh0ino^2)\cr
&\quad + \cos^2\!\beta\, F_2(\mhpm^2,m_b^2)\biggr)\cr
+ \tfrac14 &y_b^2\biggl(\sum_i F_2(\msbi^2,\mhino^2)(O^b_{i2})^2
+ \sin^2\!\beta\, F_2(\mhpm^2,m_b^2)\biggr)~,\cr}\eqn\approxdytsusy
$$
\noindent where $t = \tan\theta_W$, $\muh$ is the supersymmetric Higgs mass
parameter
defined in Eq.~(A.1) and $A_{t,b,\tau}$ and ${\cal M}_{1,2}$ are the
soft supersymmetry breaking parameters defined in Eqs.~(A.2).
All parameters as well as the function $F_2$ have an implicit
dependence on the renormalization scale ($\sim \Ms$)
The function $G_2$ is positive and is defined in Appendix B.
The $P$ matrices involve products of the squark and slepton mixing matrices
$O^{t,b,\tau}$ and are defined in Eqs.~(A.3) and (A.22).
Note that all terms proportional to the products $PP'$ drop out in the limit of
no squark or slepton mixing. To put these into perhaps more
conventional notation in terms of the left-right squark and slepton
mixing angles $\theta_t$, $\theta_b$ and $\theta_{\tau}$ defined as in
Eq.~(A.3), use the general relations
$$
\eqalignno{
\sum_{i,j}G_2(m^2,m^2_{\tilde{q}_i},m^2_{\tilde{q}_j}) P^{\prime 2}_{ij}=\ &
\tfrac14\sin^2 2\theta_q\Bigl(G_2(m^2,m^2_{\tilde{q}_1},m^2_{\tilde{q}_1})
+ G_2(m^2,m^2_{\tilde{q}_2},m^2_{\tilde{q}_2})\Bigr)&\cr
&\quad + \tfrac12\cos^2 2\theta_q
G_2(m^2,m^2_{\tilde{q}_1},m^2_{\tilde{q}_2})~,&\eqnalign\pstothetas\cr
\sum_{i,j}G_2(m^2,m^2_{\tilde{q}_i},m^2_{\tilde{q}_j}) P'_{ij}P^{}_{ij}=\ &
\tfrac12 \sin 2\theta_q \Bigl(G_2(m^2,m^2_{\tilde{q}_1},m^2_{\tilde{q}_1})
- G_2(m^2,m^2_{\tilde{q}_2},m^2_{\tilde{q}_2})\Bigr)~,&\cr}
$$
where $q = t, b, \tau$. The explicit form of these angles in terms of
the superpotential and soft supersymmetry breaking parameters of
Eqs.~(A.1-2) can be found in Ref.~\BBOspectra, for example.

For small $\tan\beta$ one may further neglect
squark and slepton mixing except for the stop mixing effects.
In this case the gluino contributions to the quark wavefunction
renormalization and to the vertex diagrams of Fig.~1 give the
dominant effect. Note that for $\muh < 0$ one typically finds
larger stop squark mixing, so that the threshold effect in $y_t$
should be largest in that case.

At large $\tan\beta$,
where the corrections to $y_b$ and $y_\tau$ tend to be more significant,
the sbottom and stau mixing must be included. In this case enhancements
proportional to $\tan\beta$ occur from vertex graphs containing
gluinos, charginos and neutralinos. The gluino contribution from Fig.~1
dominates the correction to $y_b$ for large $\tan\beta$. The chargino
and neutralino contributions of Fig.~2 are also large in this regime.
Depending on the relative signs of $\muh$ and $A_t$, the part of the finite
diagram
of Fig.~2 proportional to $\muh A_ty_t^2\tan\beta$ can
either enhance or decrease the gluino contribution. The divergent wino exchange
graph
tends to depress the gluino contribution slightly.
The enhanced $y_{\tau}$ vertex corrections come from the order $g_2^2\tan\beta$
contributions of Fig.~3. In this case the bino exchange graph can be important
due to the maximal hypercharge of the leptons. It is opposite in sign to
wino and Higgsino vertex contributions.
In analyzing the possible large $\tan\beta$ enhancements, it is
important to note that the sign of most of the relevant vertex
contributions is controlled by the sign of $\muh$ (or $\muh$ and $A_t$)
and the various large contributions can appear with opposite sign.
In addition, if we take $\Ms \simeq M_Z$ then\footnote{1}{The choice of $\Ms$
is
not important when the full threshold corrections are included since they
incorporate the effects of one loop running between different
effective supersymmetry scales.} the wavefunction
renormalization contributions tend to be positive. In light of these
facts a complete analysis is required.
We will discuss the numerical significance of these thresholds in Section~7.

\chapter{Semi-Analytic Unification Analysis}

We outline a semi-analytic analysis of gauge and Yukawa coupling unification.
In particular, we discuss the solutions for $\Mx$, $\Mhs$, and
$m_t(M_Z)$ obtained when the $\beta$ functions for the
gauge and Yukawa coupings are integrated at the two and one loop level
respectively. For the gauge couplings we must approximate the two loop
Yukawa coupling effects and for the Yukawa couplings we can
give only approximate solutions to the one loop $\beta$ functions.
We then look at the effect of the one loop threshold matching
functions on these solutions.
This of course can only give some idea of the dependencies of these
masses on other parameters since a full analysis with one loop thresholds
requires integration
of the two loop $\beta$ functions. The full numerical analysis will be given in
Section~7.

We use the condition of gauge coupling unification to determine
part of the GUT scale spectrum from deviations from the naive unification
condition.
Threshold conditions at the GUT scale relate the mismatch of the gauge
couplings to
masses of superheavy particles. At the two loop level we shall consider the
scenario, earlier considered in Ref.~\HMY, in which one predicts
$\Mhs$, $\Mx = (\Mv^2\Msig )^{\scrscr{1\over 3}}$ and the coupling
constant at the GUT scale, $\ag$,
for fixed $\alpha$, $\sin^2\theta_W$ and $\alpha_s$ at $M_Z$.
This scenario, which at one loop essentially reduces to determining
the effective scale $\Ms$ rather than $\Mhs$, is sensitive to the
large $\alpha_s$ error bars as well as boundary corrections at $M_Z$
and $\Ms$, particularly since
$\Mhs$ itself only appears in threshold matching functions.

We also consider the implication of the Yukawa coupling thresholds derived
in Sections 3 and 4 to the minimal SU(5) prediction
$y_b(\Mx ) = y_{\tau}(\Mx )$. One may either fix $M_t$ to predict
the pole mass $M_b$ (or the running mass $m_b(M_Z)$) or use the
accepted range of allowed $M_b$ values
to predict $M_t$ (or $m_t(M_Z)$). In the former case the threshold
corrections are straightforward to
estimate and constitute a significant correction to $M_b$ which should be
included
into the analysis of Ref.~{\LangA} (These Yukawa corrections
essentially belong to their correction
parameter $\rho_{\scrscr F}$.). In the latter case the analysis is more
difficult
and must be done numerically, however we will see how both the gauge and Yukawa
coupling
thresholds filter into a semi-analytic expression for $m_t(M_Z)$.

In the first approximation $\Mx$, $\Mhs$ and $\ag$
are determined entirely by gauge coupling unification.
They depend on $M_t$ only through two loop effects in the gauge running and
through
threshold effects, primarily at the weak scale. The solutions to the two loop
gauge coupling
$\beta$ functions generalizing Eq.~\betag\ are
$$
{1\over \alpha_i(M_Z)} = {1\over \ag} + {(b^{\scrscr SM}_i - b_i)\over
2\pi}\ln{\Ms\over M_Z} + {b_i\over
2\pi}\ln{\Mx\over M_Z} + \theta_i - \Delta_i~, \eqn\gaugesol
$$
where $\Delta_i = \Delta_i^{\scrscr SM} + \Delta_i^{\scrscr SUSY} +
\Delta_i^{\scrscr GUT}$ and the $\theta_i$ are the two loop
contributions, including the effects of the top Yukawa coupling.
By utilizing the threshold corrections
given in Eqs.~\gutgaugethresh, \deltagut,
\alphasmzplus, \susygaugethresh\ and \deltasusyb, one obtains the solutions
$$
\eqalign{
t_{\scrscr G} =\ & \ln{\Mx\over M_Z} = t^0_{\scrscr G} +
\Delta_{t_{\scrscr G}}~,\cr
t_{\scrscr H} =\ & \ln{\Mhs\over M_Z} = t^0_{\scrscr H} +
\Delta_{t_{\scrscr H}}~,\cr
{1\over\ag} =\ & {1\over\ag^0} + \Delta_{\ag}~,\cr}
\eqn\thagut
$$
where the naive one loop results are
$$
\eqalign{
t^0_{\scrscr G} =\ & {\pi\over 6}\biggl({1-2s^2\over\alpha}
- {2\over 3}{1\over\alpha_s}\biggr)
- {2\over 9}\ln{\Ms\over M_Z}~,\cr
t^0_{\scrscr H} =\ & {\pi\over 2}\biggl({-1+6s^2\over\alpha} - {10\over
3\alpha_s}
\biggr) + {5\over 6}\ln{\Ms\over M_Z}~,\cr
{1\over\ag^0} =\ & {-1+14s^2\over 12\alpha} + {1\over 18\alpha_s}
 + {79\over 36\pi}\ln{\Ms\over M_Z}~,\cr}\eqn\thagutonelp
$$
and the threshold and two loop correction terms are
$$
\eqalign{
\Delta_{t_{\scrscr G}} =\ & -{2\over 9}\ln{U_{\scrscr SUSY}\over \Ms}
+ \delta_{t_{\scrscr G}}^{\scrscr SM} + \delta_{t_{\scrscr G}}^{\ovl{\scrscr
DR}}
- {\pi\over 18}(5\theta_1-3\theta_2 -2\theta_3)~,\cr
\Delta_{t_{\scrscr H}} =\ & {5\over 6}\ln{V_{\scrscr SUSY}\over \Ms}
+ \delta_{t_{\scrscr H}}^{\scrscr SM} - {5\pi\over 6}(-\theta_1 +3\theta_2
-2\theta_3)~,\cr
\Delta_{\ag} =\ & {79\over 36\pi}\ln{W_{\scrscr SUSY}\over \Ms}
+ \delta_{{\ag^{-1}}}^{\scrscr SM} + \delta_{{\ag^{-1}}}^{\ovl{\scrscr DR}}
+ \delta_{{\ag^{-1}}}^{\scrscr GUT}
+ {1\over 36}(-5\theta_1 + 39\theta_2 + 2\theta_3)~,\cr}
\eqn\deltathagut
$$
where $s = \sin\theta_W$, $\delta_{t_{\scrscr G}}^{\ovl{\scrscr DR}} =
{1\over 18}$ and $\delta_{\ag^{-1}}^{\ovl{\scrscr DR}} = -{7\over 36\pi}$.
We will use superscript 0's to denote the one loop results
which use a naive step approximation to the thresholds.
In the above all low scale parameters are implicitly evaluated
at $M_Z$.
Note that the two loop contribution of the top Yukawa drops out of
the solution for $\Mx$ while if we estimate constant
$y_t \approx 1$ then this effect
changes $t_{\scrscr H}$ by of order
${1\over 8\pi^2}y_t^2 t_{\scrscr G} \approx 0.4 $ and $1/\ag$ by
${3\over 16\pi^3}y_t^2 t_{\scrscr G} \approx 0.2 $, respectively.

The weak scale threshold corrections are
$$ \eqalign{
\delta_{t_{\scrscr G}}^{\scrscr SM} =\ & \delta_{t_{\scrscr G}}^{top;s^2}
+ \delta_{t_{\scrscr G}}^{top;\alpha} + \delta_{t_{\scrscr
G}}^{top;\alpha_s}\cr
=\ & -{\pi\over 3}{\Delta^{top}_{s^2}\over\alpha} +
{4\over 27}(1-2s^2)\ln{M_t\over M_{t0}} - {1\over 27}\ln{M_t\over M_Z}~,\cr
\delta_{t_{\scrscr H}}^{\scrscr SM} =\ & \delta_{t_{\scrscr H}}^{top;s^2}
+ \delta_{t_{\scrscr H}}^{top;\alpha} + \delta_{t_{\scrscr
H}}^{top;\alpha_s}\cr
=\ & 3\pi{\Delta^{top}_{s^2}\over\alpha} +
{4\over 9}(-1+6s^2)\ln{M_t\over M_{t0}} - {5\over 9}\ln{M_t\over M_Z}~,\cr
\delta_{{\ag^{-1}}}^{\scrscr SM} =\ & \delta_{{\ag^{-1}}}^{top;s^2}
+ \delta_{{\ag^{-1}}}^{top;\alpha} + \delta_{{\ag^{-1}}}^{top;\alpha_s}\cr
=\ & {7\over 6}{\Delta^{top}_{s^2}\over\alpha} +
{2\over 27\pi}(-1+14s^2)\ln{M_t\over M_{t0}} + {1\over 54\pi}\ln{M_t\over
M_Z}~.\cr}
\eqn\txhsmthresh
$$
The residual GUT threshold due to the undetermined value of $\Mv$ is
$$\delta_{{\ag^{-1}}}^{\scrscr GUT} = -{5\over\pi}\ln{\Mv\over \Mx}
= -{5\over 3\pi}\ln{\Mv\over \Msig}~.\eqn\aggutthresh $$
The three effective scales replacing the naive scale $\Ms$ are
$$
\eqalign{
U_{\scrscr SUSY} =\ & M_1^{-{25\over 16}}M_2^{25\over 16}M_3
\simeq \Bigl({m_{\tilde{q}}\over m_{\tilde{\ell}}}\Bigr)^{3\over
8}m_{\tilde{W}}^{1\over 2}
m_{\tilde{g}}^{1\over 2}~,\cr
V_{\scrscr SUSY} =\ & M_1^{-{5\over 4}}M_2^{25\over 4}M_3^{-4}
\simeq \Bigl({m_{\tilde{\ell}}\over m_{\tilde{q}}}\Bigr)^{3\over 10}
\Bigl({m_{\tilde{W}}\over m_{\tilde{g}}}\Bigr)^2
m_{\vphantom{\tilde{H}}H}^{1\over 5} m_{\tilde{H}}^{4\over 5}~,\cr
W_{\scrscr SUSY} =\ & M_1^{-{25\over 316}}M_2^{325\over 316}M_3^{4\over 79}
\simeq \Bigl({m_{\tilde{\ell}}\over m_{\tilde{q}}}\Bigr)^{15\over 158}
m_{\tilde{q}}^{36\over 79} m_{\tilde{W}}^{26\over 79} m_{\tilde{H}}^{12\over
79}
m_{\vphantom{\tilde{H}}H}^{3\over 79}m_{\tilde{g}}^{2\over 79}~.\cr}
\eqn\uvwsusy
$$
We can approximate a range for these effective scales in the
context of the canonical soft breaking of supersymmetry induced from a hidden
sector
of N$=1$ supergravity. In this case there is a common SU(5) invariant gaugino
mass
$M_{\scrscr {1\over 2}}$, a common soft supersymmetry breaking scalar
mass $M_0$ and a common trilinear scalar coupling $A$ (see Eq.~(A.2))
at $\Mx$. Then there is a simple one loop
relation between the gaugino masses:
$$ {m_{\tilde{W}}\over m_{\tilde{g}}} = {\alpha_2(\Ms )\over\alpha_3(\Ms )}
\approx 0.28~.
\eqn\wgl
$$
Assuming $m_{\tilde{\ell}}\simeq m_{\tilde{q}}$ we obtain the approximate
relations
$$
U_{\scrscr SUSY}\approx 0.53\, m_{\tilde{g}}~, \qquad
V_{\scrscr SUSY} \approx 0.08\, m_{\vphantom{\tilde{H}}H}^{1\over 5}
m_{\tilde{H}}^{4\over 5}~, \qquad
W_{\scrscr SUSY} \approx m_{\tilde{q}}^{1\over 2}m_{\tilde{W}}^{1\over 3}
m_{\tilde{H}}^{1\over 6}~.\eqn\uvwapprox
$$
Alternatively, in the no-scale supergravity case ($M_0 = A = 0$) one
has $m_{\tilde{q}}\simeq m_{\tilde{g}}\simeq 3m_{\tilde{W}}\simeq
3m_{\tilde{\ell}}$. We then have instead the approximate relations
$$
U_{\scrscr SUSY}\approx 0.80\, m_{\tilde{g}}~, \qquad
V_{\scrscr SUSY} \approx 0.06\, m_H^{1\over 5} m_{\tilde{H}}^{4\over 5}~,
\qquad
W_{\scrscr SUSY} \approx 0.70\, m_{\tilde{g}}^{5\over 6}
m_{\tilde{H}}^{1\over 6}~.\eqn\uvwapproxns
$$
As pointed out in Ref.~\HMY, the supersymmetric threshold corrections to $\Mx$
depend mainly on the wino and gluino masses while $\Mhs$ is
most sensitive to the Higgsino mass. The corrections to $\ag$ depend mainly on
$M_2$
and are most sensitive to the squark masses.
If we allow a range of $100 - 1000$ GeV for
$m_H$ and $m_{\tilde{H}}$ while the gluino mass is taken to range
between $120 - 1000$ GeV\rlap,\footnote{1}{The lower limit on the
gluino mass is near the experimental
lower bound\refmark{\glexp} if the light gluino window is closed.}
then for degenerate squarks and sleptons
$64 \lsim U_{\scrscr SUSY} \lsim 530$ GeV and
$8 \lsim V_{\scrscr SUSY} \lsim 80$ GeV, and bear little relation to the
naive $M_Z \lsim \Ms \lsim 1$ TeV. We also take $100 \lsim W_{\scrscr
SUSY}\sim M_2 \lsim 700$ GeV in this case. Note that the approximate
upper limit on the sparticle masses of $1$ TeV is the generic
order of magnitude expectation if supersymmetry is to
avoid introducing a new low scale hierarchy problem.
In the no-scale case the corresponding limits are
$96 \lsim U_{\scrscr SUSY} \lsim 800$ GeV,
$6 \lsim V_{\scrscr SUSY} \lsim 60$ GeV, and
$73 \lsim W_{\scrscr SUSY}\sim M_2 \lsim 630$ GeV.

Before turning to the $M_t$ solution we give an
estimate of the relative importance of the
different corrections to $t_{\scrscr G}$ and $t_{\scrscr H}$.
The relevant low energy parameters are\refmark{\DFS,\LEPupdate}
$$
\eqalign{
M_Z =\ & 91.187 \pm 0.007\ {\rm GeV}~,\cr
\alpha^{-1}(M_Z) =\ & 127.9 \pm 0.1~,\cr
\alpha_s(M_Z) =\ & 0.120 \pm 0.007~,\cr
s^2_0 =\ & 0.2324 \pm 0.0003~,\cr}
\eqn\inputs
$$
where, as in Ref.~\LangB, we denote by $s_0$ the sine of the Weinberg
angle for a central value of $M_{t0} = 143$ GeV with the
quadratic top mass uncertainty removed and treated as a threshold correction as
discussed in the previous section. The current best fit value of
$\alpha_s$ from LEP and all collider and neutrino experiments is
$\alpha_s = 0.120\pm 0.006 \pm 0.002$\refmark{\LEPupdate} where the
second set of error bars corresponds to a SM Higgs mass in the range
$60< m_h < 1000$ GeV and we have included a conservative error in \inputs.

For central values of the parameters the one loop solutions are $t^{0}_{\scrscr
G} = 32.9$
corresponding to $\Mx^{0} = 1.83\times 10^{16}$ GeV and $1/\ag^{0} = 24.5$ for
$\Ms = M_Z$.
We should note that the one loop solution, $\Mhs^{0}$ given in Eq.~\thagut, is
only
a definition and is related to the scenario is which the effective
scale $\Ms$ is determined by the requirement of gauge coupling
unification. We may write
$$\eqalign{
t_{\scrscr HX} =& \ln{\Mhs\over\Mx} = \ln{\Mhs^{0}\over\Mx^{0}}
+ {19\over 18}\ln{T_{\scrscr SUSY}\over\Ms}\cr
&\quad + {2\pi\over 9}(5\theta_1 - 12\theta_2 + 7\theta_3) + \delta_{t_{\scrscr
HX}}^{\scrscr SM}
+ \delta_{t_{\scrscr HX}}^{\ovl{\scrscr DR}}~,\cr}\eqn\thx
$$
where
$$
\ln{\Mhs^{0}\over\Mx^{0}} = {19\over 18}\ln{\Ms\over\Ms^{0}} = -{2\pi\over 9}
\Bigl({3(1-5s^2)\over\alpha} + {7\over\alpha_s}\Bigr) + {19\over
18}\ln{\Ms\over M_Z}~,
\eqn\honelp
$$
and $\Ms^{0}$, the naive one loop value required for gauge unification,
is $79.2$, $7.3$, $0.9$ GeV for $\alpha_s(M_Z) = 0.113,\ 0.120,\ 0.127$,
respectively.
The scale $T_{\scrscr SUSY} = M_1^{-{25\over 19}}M_2^{100\over
19}M_3^{-{56\over 19}}$,
using the notation of  Ref.~{\Carena}.
For $\Ms = M_Z$ this gives $\Mhs^{0} \approx 2.6\times 10^{17}$ GeV
for $\alpha_s(M_Z) = 0.120$, however,
$\Mhs$ can only be reliably determined at the two loop level.

Next we quantify the correction terms to these naive estimates.
The current experimental limit on the top quark mass from fits
to electroweak data is
$M_t = 164^{+16+18}_{-17-21}$ GeV\rlap,\refmark{\LEPupdate} where
the central value is for a Higgs mass of $300$ GeV and the
second set of error bars corresponds to $60 < m_h < 1000$ GeV.
In the MSSM a range $50 < m_h < 150$ GeV is more appropriate,
giving $M_t = 143^{+17+6}_{-19-8}$ GeV\rlap.\refmark{\LangC}
The current lower limit from the D0\llap/ experiment
is $M_t > 131$ GeV (95\% C.L.)\rlap.\refmark{\Dzerotop}
We will allow $M_t$ to range from $130$ to $200$ GeV, and obtain the
following ranges for the top dependent electroweak corrections:
$$
\eqalignno{
\delta_{t_{\scrscr G}}^{top;s^2} =& (-0.04,0.24) \quad
\delta_{t_{\scrscr G}}^{top;\alpha} = (-0.008,0.03)\quad
\delta_{t_{\scrscr G}}^{top;\alpha_s} = (-0.013,-0.03)~,&\cr
\delta_{t_{\scrscr H}}^{top;s^2} =& (.39,-2.2) \quad
\delta_{t_{\scrscr H}}^{top;\alpha} = (-0.02,0.06)\quad
\delta_{t_{\scrscr H}}^{top;\alpha_s} = (-0.20,-0.44)~,&\eqnalign\dtgaugesm\cr
\delta_{{\ag^{-1}}}^{top;s^2} =& (0.05,-0.27) \quad
\delta_{{\ag^{-1}}}^{top;\alpha} = (-0.005,0.02)\quad
\delta_{{\ag^{-1}}}^{top;\alpha_s} = (0.002,0.005)~,&\cr}
$$
where the first(second) entries are for the lower(upper) limit for $M_t$.
The influences of corrections from thresholds, two loop terms and the
experimental
error bars in \inputs\ are summarized in Table~1. The top dependent thresholds
in \dtgaugesm\
are summed together with $\ovl{DR}$ conversion factors in the SM threshold
entries
for the above range of $M_t$ values. The MSSM threshold entries include the
supersymmmetric
threshold as well as the effect of running SM parameters
from $M_Z$ to $\Ms$ for the previously
mentioned ranges taken for $U_{\scrscr SUSY}$, $V_{\scrscr SUSY}$ and
$W_{\scrscr SUSY}$. The upper range
corresponds to the degenerate squark-slepton case while the lower
entry is for the no-scale case.
The $\alpha_s$ error bar entries are given first for the lower, then the upper
limit on $\alpha_s$. The percent deviations given are relative to the
naive one loop predictions given above. We see that $\Mhs$ decreases
with increasing $M_t$ and increases with increasing $\alpha_s$ and
Higgsino mass ($V_{\scrscr SUSY}$). The
effective GUT scale increases with increasing $M_t$ and $\alpha_s$ and
decreases with increasing gluino mass ($U_{\scrscr SUSY}$). The magnitude of
the
effect on  $\Mx$ due to each of these parameters is comparable, ranging up to
$40\%$.
\bigskip
{\Tenpoint
\begintable
Correction |\multispan 3 \hfil Parameter Variations \hfil |
\multispan 3 \hfil \% Deviations \hfil\nr
\omit\hfil |\omit\hrulefill
|\omit\hrulefill |\omit\hrulefill |\omit\hrulefill |
\omit\hrulefill |\omit\hrulefill \nr
	\hfill		| $\delta_{t_{\scrscr G}}$ |
$\delta_{t_{\scrscr H}}$ | $\delta{\ag^{-1}}$ |
$\Mx$ | $\Mhs$ | ${1\over\ag}$ \crthick
SM (top) threshold 	| $(+0.008,+0.35)$ | $(+.06,-3.0)$ | $(-0.03,-0.36)$ |
$(0.8,41)$ |$(6,-95)$|$(-0.1,-1.5)$    \cr
\omit\bigtstrut\hfil MSSM threshold \hfil |
\omit\bigtstrut\hfil\vbox{\hbox{$(+0.08,-0.39)$}\hbox{$(-0.01,-0.48)$}}\hfil
|\omit\bigtstrut\hfil\vbox{\hbox{$(-2.0,-.11)$}\hbox{$(-2.3,-.35)$}}\hfil
|\omit\bigtstrut\hfil\vbox{\hbox{$(+0.07,+1.6)$}\hbox{$(-0.17,+1.5)$}}\hfil
|\omit\bigtstrut\hfil\vbox{\hbox{$(\hphantom{-}8,-32)$}\hbox{$(-1,-38)$}}\hfil
|\omit\bigtstrut\hfil\vbox{\hbox{$(-86,-10)$}\hbox{$(-90,-30)$}}\hfil
|\omit\bigtstrut\hfil\vbox{\hbox{$(\hphantom{-}0.3,6.5)$}
\hbox{$(-0.7,6.1)$}}\hfil\cr
$s^2_0$ error bar	| $\pm 0.04$ | $\pm 0.36$ | $\pm 0.05$ |
$\pm 4$ |$(-30,43)$ | $\pm 0.2$    \cr
$\alpha$ error bar	| $\pm 0.03$ | $\pm 0.06$ | $\pm 0.02$ |
$\pm 3$ |$\pm 6$ | $\pm 0.1$   \cr
$\alpha_s$ error bar	| $(-0.19,+0.17)$ | $(-2.8,+2.5)$  | $(+0.03,-0.03)$ |
$(-17,18.5)$ |$(-94,+1120)$ | $\pm 0.1$    \cr
two loop ($y_t=0$)	| $0.18$ | $-3.8$ | $1.1$ |
$20$    |$-98$    |$4.4$    \cr
two loop ($y_t$ only)   |  $0$   | $0.4$ | $0.2$  |
$0$    | $49$   | $1$
\endtable\nobreak
}
\smallskip\nobreak
\centerline{\singlespace Table~1.\ \vtop{\parindent=0pt\hsize=4truein
Corrections to $\Mx$, $\Mhs$ and $1/\ag$.}}
\bigskip

Note that the $\Delta^{top}_{s^2}$ correction dominates the SM threshold
effects
and, as expected, the ``correction'' terms for $\Mhs$ can be large.
The $\alpha_s$ error bar
correction is the largest, changing $\Mhs$ over three orders of
magnitude. The MSSM threshold
correction to $\Mhs$ was included in the analysis of Ref.~\HMY, but the top
dependent SM threshold was not discussed. These two effects can be comparable
in magnitude, with the SM effects changing $\Mhs$ by up to two
orders of magnitude, corresponding
to a change in the proton decay lifetime of four orders of magnitude. Clearly,
when we
also solve for $M_t$ from the condition of Yukawa unification,
the solution for $\Mhs$ will be substantially correlated with
the $M_t$ solution. The tendency to predict large values of $M_t$
means that the effect of the SM threshold will generally be to depress $\Mhs$.
As we shall see in Section~6, the opposite is true for the lower
bound on $\Mhs$ from proton decay. The combination of these results
may be useful in strengthening proton decay constraints
on the parameter space\rlap.\refmark{\inpreptwo}

We next turn to the Yukawa unification predictions for either $M_t$ or
$m_b(M_Z)$.
We discuss both cases with emphasis on the former.
The one loop $\beta$ functions for the Yukawa
couplings are not solvable analytically, however we shall give some
semi-analytic
solutions. The one loop Yukawa $\beta$ functions for the third generation in
the MSSM are
$$
\beta_{y_{\alpha}} = {dy_{\alpha}\over dt} = {y_{\alpha}\over 16\pi^2}
( -c_{\alpha i}g_i^2 + b_{\alpha\beta}y_{\beta}^2 )~,\eqn\betayuk
$$
where $\alpha = (t,b,\tau)$ and $b_{\alpha\beta}$ and $c_{\alpha i}$ are given
by the matrices
$$
\pmatrix{6&1&0\cr 1&6&1\cr 0&3&4\cr}{\rm\ and\ }\pmatrix{{13\over
15}&{3}&{16\over 3}\cr
{7\over 15}&3&{16\over 3}\cr {9\over 5}&3&0\cr}~,\eqn\bcmat
$$
respectively.
The analogous matrices in the SM, $b^{\scrscr SM}_{\alpha\beta}$ and
$c^{\scrscr SM}_{\alpha i}$, are
given by
$$
\pmatrix{{9\over 2}&{3\over 2}&1\cr {3\over 2}&{9\over 2}&1\cr
3&3&{9\over 2}\cr}{\rm\ and\ }
\pmatrix{{17\over 20}&{9\over 4}&{8}\cr
{1\over 4}&{9\over 4}&{8}\cr {9\over 4}&{9\over 4}&0\cr}~,\eqn\bcmatsm
$$
respectively. At one loop, in the naive step approximation, the SM Yukawa
couplings are
matched to those in the MSSM at $\Ms$ via $y_t = y_t^{\scrscr SM}/\sin\beta$,
$y_b = y_b^{\scrscr SM}/\cos\beta$ and
$y_{\tau} = y_{\tau}^{\scrscr SM}/\cos\beta$, where
$\tan\beta = {v_2\over v_1}$ is the ratio
of the VEVs of the two Higgs doublets $H_1$ and $H_2$ in the MSSM (see
Appendix~A).
Using Eqs.~\betayuk, the one loop solution for $m_b/m_{\tau}$ at $M_Z$ can be
written as
$$
{m_b^{0}(M_Z)\over m_{\tau}^{0}(M_Z)} = A^{0\,{\scrscr
SM}}_{b/\tau}A^{0}_{b/\tau}B_t^{0\,{\scrscr SM}\,-{3\over 2}}
B_b^{0\,{\scrscr SM}\,{3\over 2}}B_{\tau}^{0\,{\scrscr SM}\,-{7\over
2}}B_t^{0}\biggl({B_b^0\over B_{\tau}^0}\biggr)^3~,
\eqn\mbmtauonelp
$$
where
$$
\eqalign{
A^{0\,{\scrscr SM}}_{b/\tau} =\ & \Bigl(
{\alpha_1^{0}(\Ms)\over\alpha_1^{0}(M_Z)}\Bigr)^{-{10\over 41}}
\Bigl({\alpha_3^{0}(\Ms)\over\alpha_3^{0}(M_Z)}\Bigr)^{-{4\over 7}}~,\cr
A^{0}_{b/\tau} =\ & \Bigl({\ag^{0}\over\alpha_1^{0}(\Ms)}\Bigr)^{-{10\over 99}}
\Bigl({\ag^{0}\over\alpha_3^{0}(\Ms)}\Bigr)^{-{8\over 9}}~,\cr}\eqn\AA
$$
and
$$
\eqalign{
B^{0\,{\scrscr SM}}_{\alpha}(M_Z,\Ms) =\ & {\rm exp}\Biggl(
-{1\over 16\pi^2}\int^{t_s}_{t_z}\! dt\, (y_{\alpha}^{0\,{\scrscr SM}})^2
\Biggr)~,\cr
B^{0}_{\alpha}(\Ms,\Mx^{0}) =\ & {\rm exp}\Biggl(
-{1\over 16\pi^2}\int^{t_{\scrscr G}^{\scrscr 0}}_{t_s}\! dt\,
(y_{\alpha}^{0})^2\Biggr)~.\cr}
\eqn\BB
$$
Here the parameters $t_s = \ln{\Ms\over M_Z}$, $t_z = 1$ and
$t^0_{\scrscr G}$ is given in Eq.~\thagutonelp.
On the other hand the full two loop solution including the
Yukawa thresholds of Sections~3 and 4 is
$$
\eqalign{
{m_b(M_Z)\over m_{\tau}(M_Z)} =\ & A^{{\scrscr SM}}_{b/\tau}
A_{b/\tau}B_t^{{\scrscr SM}\,-{3\over 2}}
B_b^{{\scrscr SM}\,{3\over 2}}B_{\tau}^{{\scrscr SM}\,-{7\over 2}}
B_t\biggl({B_b\over B_{\tau}}\biggr)^3\Theta_{b/\tau}\cr
&\quad \times(1 + \Delta_{b/\tau}^{\scrscr SM} + \Delta_{b/\tau}^{\scrscr SUSY}
+ \Delta_{b/\tau}^{\scrscr GUT})~,\cr}
\eqn\mbmtaufull
$$
where
$$
\eqalign{
A^{{\scrscr SM}}_{b/\tau} =\ &
\Bigl({\alpha_1^-(\Ms)\over\alpha_1^+(M_Z)}\Bigr)^{-{10\over 41}}
\Bigl({\alpha_3^-(\Ms)\over\alpha_3^+(M_Z)}\Bigr)^{-{4\over 7}}~,\cr
A_{b/\tau} =\ & \Bigl({\alpha_1^-(\Mx)\over\alpha_1^+(\Ms)}\Bigr)^{-{10\over
99}}
\Bigl({\alpha_3^-(\Mx)\over\alpha_3^+(\Ms)}\Bigr)^{-{8\over
9}}~,\cr}\eqn\AAfull
$$
and
$$
\eqalign{
B^{{\scrscr SM}}_{\alpha}(M_Z^+,\Ms^-) =\ & {\rm exp}\Biggl(
{-{1\over 16\pi^2}\int^{t_s^-}_{t_z^+}\! dt\, (y_{\alpha}^{\scrscr
SM})^2}\Biggr)~,\cr
B_{\alpha}(\Ms^+,\Mx^-) =\ & {\rm exp}\Biggl(
{-{1\over 16\pi^2}\int^{t_{\scrscr G}^-}_{t_s^+}\! dt\,
(y_{\alpha})^2}\Biggr)~.\cr}
\eqn\BBfull
$$
The thresholds corrections $\Delta_{b/\tau}$ are contained in
Eqs.~\Rbtauthreshold, \deltabsmmass\
and \approxdytaususy\ - \approxdytsusy.
The two loop corrections are contained in $\Theta_{b/\tau}$ which we will take
to be unity in this
section. The $\pm$ superscripts indicate parameters just above/below the scale
at which they are
evaluated, \ie\ on either side of the threshold boundary. We call the threshold
corrections
$\Delta_{b/\tau}$ in \mbmtaufull\ the direct threshold corrections.
There are also indirect threshold corrections which arise in rewriting the the
quantities
in \AAfull\ and \BBfull\ in terms of the naive one loop results. The
discontinuities in the
other parameters at the threshold boundaries indirectly filter into the result
for $m_b/m_{\tau}$.
These indirect threshold effects will
primarily involve the gauge coupling thresholds.

For the indirect threshold effects in the $A_{b/\tau}$ parameters we find
$$
\eqalign{
A^{{\scrscr SM}}_{b/\tau} =&\ A^{0\,{\scrscr SM}}_{b/\tau}
(1 + \Delta^{\alpha\, {\scrscr SM}}_{b/\tau})~,\cr
A_{b/\tau} =&\ A^{0}_{b/\tau}(1 + \Delta^{\alpha}_{b/\tau})~,\cr}
\eqn\indthra
$$
where
$$
\eqalign {
1 + \Delta^{\alpha\, {\scrscr SM}}_{b/\tau} =&\ \biggl(
{1 + \alpha_1^0(M_Z)\Delta_1^{\scrscr SM}\over 1 +
\alpha_1^0(\Ms)\Delta_1^{\scrscr SM}}
\biggr)^{-{10\over 41}}
\biggl(
{1 + \alpha_s^0(M_Z)\Delta_3^{\scrscr SM}\over 1 +
\alpha_3^0(\Ms)\Delta_3^{\scrscr SM}}
\biggr)^{-{4\over 7}}~,\cr
1 + \Delta^{\alpha}_{b/\tau} =&\ \biggl(
{1 + \alpha_1^0(\Ms)(\Delta_1^{\scrscr SM} + \Delta_1^{\scrscr SUSY})\over
1 + \ag^0(\Delta_{\ag} - \Delta_1^{\scrscr GUT}(\Mx))}
\biggr)^{-{10\over 99}}\cr
\quad &\times\biggl(
{1 + \alpha_3^0(\Ms)(\Delta_3^{\scrscr SM} + \Delta_3^{\scrscr SUSY} +
\Delta_3^{\scrscr DR})\over
1 + \ag^0(\Delta_{\ag} - \Delta_3^{\scrscr GUT}(\Mx))}
\biggr)^{-{8\over 9}}~.\cr }
\eqn\indthrb
$$
Later we will also need
$$
A^{\scrscr SM}_t = A^{0\, \scrscr SM}_t(1 +
\Delta^{\alpha\, \scrscr SM}_t)~,\eqn\atsm
$$
where
$$
A^{0\, \scrscr SM}_t =
\Bigl({\alpha_1^0(\Ms)\over\alpha_1^0(M_Z)}\Bigr)^{17\over 164}
\Bigl({\alpha_2^0(\Ms)\over\alpha_2^0(M_Z)}\Bigr)^{-{27\over 76}}
\Bigl({\alpha_3^0(\Ms)\over\alpha_3^0(M_Z)}\Bigr)^{-{4\over 7}}
{}~,\eqn\atzerosm
$$
and
$$
\eqalign{
1 + \Delta^{\alpha\, \scrscr SM}_t =\ &
\Bigl({1 + \alpha_1^0(M_Z)\Delta^{\scrscr SM}_1\over
1 + \alpha_1^0(\Ms)\Delta^{\scrscr SM}_1}\Bigr)^{17\over 164}
\Bigl({1 + \alpha_2^0(M_Z)\Delta^{\scrscr SM}_1\over
1 + \alpha_2^0(\Ms)\Delta^{\scrscr SM}_1}\Bigr)^{-{27\over 76}}\cr
&\quad\times\Bigl({1 + \alpha_3^0(M_Z)\Delta^{\scrscr SM}_1\over
1 + \alpha_3^0(\Ms)\Delta^{\scrscr SM}_1}\Bigr)^{-{4\over 7}}
{}~.\cr}\eqn\deltaat
$$
The threshold correction to $\ag$ is given in Eq.~\deltathagut.
In the expressions for $\Delta_i^{\scrscr GUT}$ of Eq.~\deltagut\ we
shall insert the one loop expressions for $\ln \Mhs/\Mx$ given in
Eq.~\honelp, consistent with the scenario considered here.
Note that terms proportional to $\ln \Mv/\Mx$ cancel in the
differences $\Delta_{\ag} - \Delta_i^{\scrscr GUT}$ so that these
threshold corrections are computable given the sparticle spectrum.
In all cases we find that
$\Delta^{\alpha\, {\scrscr SM}}_{b/\tau}$ and
$\Delta^{\alpha\, {\scrscr SM}}_t$ are negligible ($\lsim 0.1\%$
for $\Ms$ up to $1$ TeV) due to the absence of large logarithms.

The indirect threshold corrections to the $B$ parameters are
primarily of second order compared to the direct thresholds in
Eq.~\mbmtaufull\ and are more difficult to estimate analytically.
However, one can account simply for the threshold effects in
$t_{\scrscr G}$ given in Eq.~\deltathagut.
We may write
$$ B_{\alpha} \approx B_{\alpha}^0 \Bigr(1 - {1\over 16\pi^2}
\Delta_{t_{\scrscr G}}\ovl{y}^2_{\alpha} \Bigl)~,\eqn\bthresh
$$
although the full corrections should be computed numerically.
The threshold corrections to the $B^{\scrscr SM}$ can be neglected.
We can now rewrite Eq.~\mbmtaufull\ in terms of the naive one loop
result
$$
\eqalign{
{m_b(M_Z)\over m_{\tau}(M_Z)} \approx &\ {m_b^0(M_Z)\over m_{\tau}^0(M_Z)}
\Theta_{b/\tau}(1 + \Delta_{b/\tau}^{\scrscr SM} +
\Delta_{b/\tau}^{\scrscr SUSY} + \Delta_{b/\tau}^{\scrscr GUT}\cr
\quad &+ \Delta^{\alpha\, {\scrscr SM}}_{b/\tau} + \Delta^{\alpha}_{b/\tau}
- {1\over 16\pi^2}\Delta_{t_{\scrscr G}}(\ovl{y}^2_t + 3\ovl{y}^2_b -
3\ovl{y}^2_{\tau}))~.\cr}\eqn\mbtautotthresh
$$
This may be used to determine $m_b(M_Z)$ using the Yukawa unification
condition. However it is important to note that such a solution will
be sensitive to potentially large threshold corrections both at $\Ms$
and $\Mx$, the former occurring for large $\tan\beta$ and the latter
occuring for large allowed splittings $\Mv \gg \Msig$
or $\Mhs \gg \Mx$ and large $\ovl{y}_t$ (see Sections 3 and 4).
We shall discuss these corrections further in Section 7.

We focus instead on using Eq.~\mbmtaufull\ to solve for
$m_t(M_Z)$. Here we use a good approximate analytic solution for
$y_t$ for small to moderate $\tan\beta$ (see also Ref.~\Naculich).
$$
B_t(\Ms^+,\Mx^-) \simeq (1 - y_t^2(\Ms^+)K_t(\Ms^+,\Mx^-))^{1\over
12}~,\eqn\btapprox
$$
where
$$
K_t(\Ms^+,\Mx^-) = {3\over 4\pi^2}\int^{t_{\scrscr G}^-}_{t_s^+}\!
dt\, A_t^{-2}(t_s,t)~,\eqn\ktdef
$$
and
$$
A_t(t_s^+,t_{\scrscr G}^-) = \Bigl({\alpha_1^-(\Mx)\over
\alpha_1^+(\Ms)}\Bigr)^{13\over 198}
\Bigl({\alpha_2^-(\Mx)\over \alpha_2^+(\Ms)}\Bigr)^{3\over 2}
\Bigl({\alpha_3^-(\Mx)\over \alpha_3^+(\Ms)}\Bigr)^{-{8\over 9}}~.\eqn\atsusy
$$
We then solve \mbmtaufull\ for $y_t(M_Z)$ using $B_t = k({B_{\tau}\over
B_b})^3$ where $k = k^0(1 - \delta k)$ and
$$
\eqalign{
k^0 =&\ {m_b(M_Z)\over m_{\tau}(M_Z)}
(A^{0\,{\scrscr SM}}_{b/\tau}A^{0}_{b/\tau})^{-1}B_t^{0\,{\scrscr SM}\,{3\over
2}}
B_b^{0\,{\scrscr SM}\,-{3\over 2}}B_{\tau}^{0\,{\scrscr SM}\,{7\over 2}}~,\cr
\delta k \simeq &\ \Delta_{b/\tau}^{\scrscr SM} +
\Delta_{b/\tau}^{\scrscr SUSY} + \Delta_{b/\tau}^{\scrscr GUT}
+ \Delta^{\alpha\, {\scrscr SM}}_{b/\tau} + \Delta^{\alpha}_{b/\tau}~.\cr}
\eqn\deltak
$$
The solution for $m_t(M_Z)$ is
$$
\eqalignno{
m_t(M_Z) =&\  m^0_t(M_Z)\Bigl(1 + \Delta_{y_t}^{\scrscr SUSY} +
\Delta^{\alpha\, {\scrscr SM}}_t - \tfrac12\Delta_{K_t}&\cr
&\quad + C\bigl(\delta k + 3(\Delta_{B_b}
- \Delta_{B_{\tau}})\bigr)\Bigr)~, &\eqnalign\mtsol\cr
\noalign{\hbox{where}}
m^0_t(M_Z) =&\ v\sin\beta A_t^{0\scrscr SM}B_t^{0{\scrscr SM}\, {9\over 2}}
B_b^{0{\scrscr SM}\, {3\over 2}}B_{\tau}^{0{\scrscr SM}}
{\sqrt{1 - (k^0)^{12}\Bigl({B^0_{\tau}\over B^0_b}\Bigr)^{36}}
\over \sqrt{K_t^0}}~,&\eqnalign\mtzero\cr}
$$
$v = 174.104$ GeV and we have inserted the appropriate
threshold corrections to $y_t$.
The factor $C$ is given by
$$
C = {6(k^0)^{12}\Bigl({B^0_{\tau}\over B^0_b}\Bigr)^{36}\over
1 - (k^0)^{12}\Bigl({B^0_{\tau}\over B^0_b}\Bigr)^{36}}~. \eqn\bigc
$$
Also, we have introduced a threshold correction factor $\Delta_{K_t}$,
defined by $K_t = K_t^0(1 + \Delta_{K_t})$, which must be determined
numerically.

Next we evaluate some of the parameters appearing in Eq.~\mtsol.
In particular we estimate the corrections arising from the gauge
coupling thresholds. We will evaluate the Yukawa thresholds from
supersymmetric particles of the MSSM and SUSY-SU(5) in the numerical
section to follow. Table~2 gives values for the corrections to
$y_b/y_{\tau}$ and $K_t$ from gauge coupling thresholds which
affect $\Mx$ and $\ag$. Results are given for
$\tan\beta = 1.5$ for explicit two loop $M_t$ solutions.
The results are insensitive to $M_t$ and $\tan\beta$ (a two loop
effect in the gauge coupling evolution).
The gauge coupling sparticle spectrum parameters are taken
to be $M_1 = 368$ GeV, $M_2 = 256$ GeV and $M_3 = 375$ GeV.
\vskip 2em
\begintable
$\Ms$ | $M_t$ | $\Delta^{\alpha}_{b/\tau}$|
$\Delta_{K_t}$\crthick
$91.2$ | $157.4$ | $-0.06$ | $-0.03$ \cr
$500$ | $173.9$ | $0.02$ | $0.003$
\endtable\nobreak
\smallskip\nobreak
\centerline{\singlespace Table~2.\ \vtop{\parindent=0pt\hsize=4truein
Corrections to $M_t$ from Gauge Coupling Thresholds in the MSSM
($\alpha_s(M_Z) = 0.120$, $M_b = 4.9$ GeV, $\tan\beta = 1.5$).}}
\bigskip
\noindent The correction factors $\Delta_{K_t}$ are for central values
$K^0_t = 0.77,\ 0.85$ for $\Ms = M_Z,\ 500$ GeV, respectively.
The correction term $\Delta_{B_b} - \Delta_{B_{\tau}}$
in Eq.~\mtsol\ is negligible except for large $\tan\beta$
in which case it has a value of $-0.01$ for $\Ms = M_Z$.
For low $\tan\beta$ we also find $m^0_t(M_Z) = (195.7\, {\rm GeV})\sin\beta$,
consistent with the infrared
quasi-fixed point solution\refmark{\BBOfixpt,\CarenaIRone} for $m_t$.
This behaviour is modified for larger $\tan\beta$ since the factor
of $B^0_{\tau}/B^0_b$ is proportional to $\exp(\sec^2\beta)$.
For low $\tan\beta$ the fixed point solution implies that this factor is very
close to $1$. Thus the factor $C$ multiplying the $y_b/y_{\tau}$
threshold corrections in Eq.~\mtsol\ is determined by $k^0$ only.
For the $\Ms = M_Z$ and $\tan\beta = 1.5$ case in Table~2 we find
$k^0 = 0.72$ and $C = 0.12$. Hence we see explicitly how the
fixed point solution inhibits the influence of potentially
large GUT or MSSM threshold corrections to $y_b/y_{\tau}$ on
the $M_t$ solution. In this sense
the solution for $m_t(M_Z)$ tends to be more robust than the
alternative $m_b(M_Z)$ solution entertained by some authors.
Of course, away from the fixed point $B^0_{\tau}/B^0_b$ tends to be
larger than $1$, and the factor $C$ need not be small. Still, for
moderate values of $\tan\beta$, one finds $C \lsim 0.3$, and to the
extent that the approximation of Eq.~\btapprox\ is still valid,
the robustness of the $M_t$ solution is maintained except for
extremely large $\tan\beta$.

\chapter{GUT Scale Constraints}

In order the extract any meaningful information concerning the parameter space
we must inpose some constraints on the superheavy sector beyond that of
unification. Proton decay is the strongest constraint on the Higgsino color
triplet mass
due to the dimension 5 operators which it induces\rlap.\refmark{\pdecayorig}
A recent thorough analysis by the authors of Ref.~\HMY\ suggests that a
conservative
lower bound is
$$\Mhs \ge 5.3\times 10^{15}\ {\rm GeV}~.\eqn\mhclimpdecay $$
This corresponds to
extreme limits on the superparticles, \ie\ heavy squarks of order $1$
TeV in mass and a maximal ratio of squark to charged wino masses, low
$\tan\beta$ and a top mass around $100$ GeV. The proton lifetime
for the expected dominant decay mode $p\rightarrow K^+\bar{\nu}_{\mu}$
is given by\refmark{\HMY}
$$
\eqalign{
\tau(p\rightarrow K^+\bar{\nu}_{\mu}) =\ & 6.9\times 10^{31}{\rm yr}
\Bigg|{0.003\ {\rm GeV}^3\over \beta_n}{0.67\over A_{\scrscr S}}
{\sin 2\beta\over 1+y^{tK}}{\Mhs\over 10^{17}\ {\rm GeV}}\cr
&\quad \times {{\rm TeV}^{-1}\over
\mwino (G_2(\mwino^2,m^2_{\tilde{u}},m^2_{\tilde{d}}) +
G_2(\mwino^2,m^2_{\tilde{u}},m^2_{\tilde{e}}))}\Bigg|^2~,\cr}\eqn\pknu
$$
where $\beta_n$ is the relevant nuclear matrix element and
lies between $0.003$ and $0.03$ GeV${}^3$. The parameter $y^{tK}$
gives the ratio of the third to the second family contributions.
Since it involves undetermined complex phases from CKM matrix elements
one must allow for possible cancellations between different families.
A conservative value of $|1+y^{tK}| \gsim 0.4$ can be taken to get
conservative limits on $\Mhs$. The quantity $A_{\scrscr S}$ is a
short-distance factor incorporating the anomalous dimensions
of the important dimension $5$ operators as well as the
renormalization of the strange and charm masses from $M_Z$ to $\Mx$.
The current experimental limit is $\tau(p\rightarrow
K^+\bar{\nu}_{\mu})\gsim 1\times 10^{32}$ yr\rlap,\refmark{\Kam}
implying that the squared factor in \pknu\ must be $\gsim O(1)$.

We would like to note that the limit \mhclimpdecay\ can be
substantially strengthened when combined with sparticle spectrum
predictions and the large $M_t$ solutions required by bottom-tau
Yukawa unification. Note that $A_{\scrscr S}$ is enhanced by up
to a factor of $3$ when such large values of $y_t$ are included
in the running charm Yukawa coupling, decreasing the predicted
lifetime by nearly an order of magnitude.
In particular, we have an additional motivation to
compute $\Mhs$ rather than $\alpha_s(M_Z)$ as a gauge unification
scenario. By determining $\Mhs$ for each sparticle spectrum
determined by one's favorite set of theoretical criterion, \eg\
universal soft SUSY breaking, radiative electroweak breaking, \etc,
one can immediately apply strengthened and correlated limits from
proton decay to this spectrum. The well-known figure
of merit is typically the factor
$R \equiv m_{\tilde{q}}^2/(m_{\tilde{W}^\pm}\times 1\,{\rm TeV})$,
where $m_{\tilde{q}}$ is a first generation squark mass. This factor
obtains a lower bound from proton decay, while typical sparticle
spectra generated with the standard theoretical assumptions require
$R \lsim 25$ ($\lsim 4$ in the no-scale case assuming $\mgl \gsim 120$ GeV)
for sparticle masses below $1$ TeV. Even more of the parameter space
of the MSSM can be excluded by this procedure\rlap.\refmark{\inpreptwo}
For the present analysis however, we will consider the conservative
limit of Eq.~\mhclimpdecay.

One may also bound the adjoint and color
triplet Higgs masses from above by the requirement that the superpotential
couplings
$\lambda_{1,2}$ be perturbative ($ < 2\sqrt{\pi}$) below the
Planck scale\rlap.\refmark{\arnnath, \HMY}
Of course, this is purely theoretical prejudice as there may indeed be new
physics between $\Mx$ and $\Mpl \simeq 1.2\times 10^{19}$ GeV. We therefore
use the perturbativity constraint only to give an indication of a realistic GUT
spectrum
and not a strict requirement. We do not impose fine tuning constraints which
require that
these couplings are not too small\refmark{\arnnath} as such small
couplings in the superpotential
are technically (though not aesthetically) natural due to the
nonrenormalization theorem\rlap.\refmark{\nonrenorm}

In the one loop approximation the GUT scale renormalization
group equations for the Yukawa-like couplings, taking into account only third
generation effects, are
$$ \eqalign{
\beta_g =& -{3\over 16\pi^2}g^3~,\cr
\beta_{\lambda_1} =& {1\over 16\pi^2}\lambda_1 ( {63\over 5}\lambda_1^2 +
3\lambda_2^2 - 30 g^2 )~,\cr
\beta_{\lambda_2} =& {1\over 16\pi^2}\lambda_2 ( {21\over 5}\lambda_1^2 +
{53\over 5}\lambda_2^2
- {98\over 5} g^2 + 3 y_t^2 + 4y_b^2 )~,\cr
\beta_{y_t} =& {1\over 16\pi^2}y_t ( {24\over 5}\lambda_2^2 + 9 y_t^2 + 4 y_b^2
- {96\over5} g^2 )~,\cr
\beta_{y_b} =& {1\over 16\pi^2}y_b ( {24\over 5}\lambda_2^2 + 3 y_t^2 + 10
y_b^2
- {84\over5} g^2 )~,\cr}\eqn\gutrge
$$
where we have corrected an error in the $\beta_{y_t}$ of Ref.~{\HMY}.
Using the minimal SU(5) relations
$$
{\Mhs\over\Mv} = {\lambda_2\over g}~, \qquad {\Msig\over\Mv} = {2\lambda_1\over
g}~,\eqn\gutmassrel
$$
we find the following results for one loop perturbativity up to
$\Mpl/\sqrt{8\pi}$ for $\Mx = 2\times 10^{16}$ GeV and $\alpha_{\scrscr G} =
1/24.5$:
$$
\Mhs \lsim 1.9 \Mv~, \qquad \Msig \lsim 3.8\Mv~.\eqn\mhmsiglimit
$$
If we require perturbativity up to $10^{17}$ GeV instead, then the
factors becomes $2.7$ and $5.3$, respectively.
The limits are obtained for each Yukawa coupling by taking all other Yukawa
couplings to zero
in its $\beta$ function.
The same constraints applied to the ordinary Yukawa couplings gives $y_t(\Mx)
\lsim 1.5 (2.1)$
and $y_b(\Mx) \lsim 1.4 (2.0)$, where the values in parenthesis
correspond to perturbativity up to $10^{17}$ GeV.
However in the analysis below we choose to be somewhat
inconsistent by requiring only that these couplings remain perturbative up to
$\Mx$.

\chapter{Full Renormalization Group Analysis}

A complete two loop numerical analysis is necessary to properly determine the
influence of threshold effects on the solutions for $M_t$, $\Mhs$ and $\Mx$.
We will consider the effect of turning on and off various thresholds
as well as the effect of varying various parameters appearing in the threshold
formulae.

{}From Table~1 we saw that $\Mhs$ decreases with
$m_t$ and $\alpha$ and increases with $s^2$, $\alpha_s$ and
$V_{\scrscr SUSY}\ (m_{\tilde{H}})$ while $\Mx$ increases with $m_t$ and
$\alpha_s$ and
decreases with $s^2$, $\alpha$ and $U_{\scrscr SUSY}\
(m_{\tilde{W}},m_{\tilde{g}})$.
Just considering gauge unification for the moment, we can quantify
these dependencies numerically, incorporating all the gauge coupling threshold
corrections. We numerically integrate the relevant two loop $\beta$
functions (see Refs.~\template, \MV\ and \BBORGE\ and references
therein) and
use a modified globally convergent nonlinear Newton method to find
the solutions for $\Mhs$ and $\Mx$ consistent with the mismatch of the
gauge couplings at $\Mx$. The results are indicated in
Figs.~4-7, and highlight the various dependencies mentioned above.
In all the figures we indicate the dominant effect of the
uncertainty in $\alpha_s(M_Z)$ as well as the influence of
large and small $\tan\beta$ for central values of $\sin\theta_W(M_Z)$
and $\alpha(M_Z)$.
In Fig.~4 we depict the logarithmic dependence of $\Mx$ on the
gluino mass for $120 {\rm\ GeV}\le \mgl \le 1$ TeV for fixed
$M_t = 165.4$ GeV. For fixed $\alpha_s$ we see that the variation
of $\Mx$ over the entire range corresponds approximately to a
$50\%$ deviation.
Similarly, Fig.~5 shows the variation of $\Mx$ with $M_t$ for
$130 {\rm\ GeV}\le m_t(M_Z) \le 200$ GeV for fixed $\mgl = 400$ GeV.
Here the rise in $\Mx$ is determined by the quadratic dependence
of $\sin\theta_W$ on $M_t$ and gives a $20$-$30\%$ deviation
over the entire range of $M_t$. Note that the curves cut off at
different values of $M_t$ corresponding to the
nonperturbative limit of $y_t$.

The logarithmic variation of $\Mhs$ with $V_{\scrscr SUSY}\simeq
.08m_{\tilde{H}}$ is shown in Fig.~6 for
$100 {\rm\ GeV}\le m_{\tilde{H}} \le 1$ TeV
and $M_t = 165.4$ GeV. Note that only in the case
of large values of $\alpha_s$ do we obtain consistency
with even the most conservative proton decay
bound, $\Mhs > 10^{15.7}$ GeV. Here we get an order of magnitude
change in $\Mhs$ over the range of $V_{\scrscr SUSY}$.
The more complicated dependence of $\Mhs$ with $M_t$ for
$130 {\rm\ GeV}\le m_t(M_Z) \le 200$ GeV is given in Fig.~7
for fixed $V_{\scrscr SUSY} = 45$ GeV.
The initial fall of $\Mhs$ with $M_t$ is due to the
electroweak threshold while its rise for large $M_t$ is
due to the effect of large $y_t$ on the two loop gauge $\beta$
functions. The variations in $\Mhs$ are on the order of $30\%$.

Before imposing the condition of Yukawa unification, gauge
unification constrains the solutions for $\Mx$ and $\Mhs$.
For $M_t$ between $130$ and $200$ GeV and
$0.5 \lsim \tan\beta \lsim 60$ we find the ranges
$$
\eqalign{
1.0(0.96)\times 10^{16}< &\Mx < 3.9(3.7)\times 10^{16}\ {\rm GeV}~,\cr
1.7(1.2)\times 10^{13}< &\Mhs < 1.6(0.9)\times 10^{17}\ {\rm
GeV}~,\cr}\eqn\mhmxlims
$$
for the case of approximately degenerate squarks and sleptons, with
the no-scale limits given in parentheses. These results are comparable
with those of Ref.~\HMY.
Note that the lower limit on $\Mx$ and the upper limit on $\Mhs$ will be
sensitive to the restricted range of values of $M_t$ which result from Yukawa
unification.
The application of the most conservative proton decay bound from the
previous section can be translated into a lower bound on the strong
coupling constant. By tuning the remaining parameters so that $\Mhs$
is as large as possible we find
$$ \alpha_s(M_Z) \gsim 0.118\, (0.119)~,\eqn\asbound $$
where again the limit in parenthesis corresponds to the no-scale case.

To get some idea of the size of the GUT scale Yukawa coupling
threshold corrections, we combine the renormalization group
constraints with the perturbativity constraints of Section~6.
The perturbativity constraints of Eq.~\mhmsiglimit\ together
with Eq.~\mhmxlims\ imply
$$
\eqalign{
\Mv >& 7.0(6.1)\times 10^{15}\ {\rm GeV}~,\cr
\Msig <& 0.9(1.2)\times 10^{17}\ {\rm GeV}~,\cr}
$$
where the two values correspond to perturbativity up to
$\Mpl/\sqrt{8\pi} = 2.4\times 10^{18}$ GeV and $10^{17}$ GeV, respectively.
If we further assume that no superheavy masses lie above the relevant
Planck mass scale, we find in addition
$$ {\Mv\over\Msig} \lsim 1.2\times 10^7~.\eqn\mvsiglim $$
Although this may appear to be an extreme choice there
is nothing to rule it out phenomenologically, in fact such a case was
considered recently\refmark{\MY} where the fine-tuning of $\lambda_1$ is used
to
solve the Polonyi problem\rlap.\refmark{\Polonyi} What we regard as a
fine-tuning of parameters may indeed be fixed by some not as yet not
fully understood quantum gravitational model. In any case,
supersymmetry insures that any tuning of superpotential parameters is
at least technically natural as they are multiplicatively
renormalized. From Eqs.~\Rbtauthreshold, \deltak\ and \mtsol\ we
see that such a large splitting tends to increase the top mass
prediction. In many cases no perturbative solution can be found,
however, away from the infrared quasi-fixed point of $y_t$, sizable
enhancements of the predictions can occur. In Fig.~8 we
plot the $M_t$ solution {\it vs.} $\tan\beta$ for various values of the
ratio $\Mv/\Msig$ for $M_b = 5.2$ GeV, $\alpha_s(M_Z)= 0.120$
and $\Ms = M_Z$. The largest deviation from the naive case
is a $15\%$ increase in $M_t$ and corresponds to
$\Delta^{\scrscr GUT}_{b/\tau} \simeq 0.1$.
The sparticle spectrum parameters
are fixed at $\mgl = M_3 = 1$ TeV and $V_{\scrscr SUSY} = 80$ GeV.
The typical GUT masses are then $\Mx = 10^{16.1 \pm 0.1}$ GeV and
$\Mhs = 10^{15.7 \pm 0.4}$ GeV for these solutions.
We also plot the bound from the nonperturbative
limit on $y_t$, indicating that for this particular
example, one is not in the domain of attraction of the fixed point.

The GUT scale matching function for $y_b/y_\tau$ of
Eq.~\Rbtauthreshold\ can also be negative for large $y_t$ and
$\Mhs$. We set $\Mv/\Msig = 0.3$, near the minimum required by
the perturbativity argument. In Fig.~9 we give an example
for this case for $M_b = 4.9$ GeV, $\alpha_s(M_Z)= 0.127$
and $\Ms = M_Z$. Using the same set of sparticle spectrum
parameters as in the previous figure, we obtain GUT masses
in the ranges $\Mx = 10^{16.2\pm 0.1}$ GeV and
$\Mhs = 10^{16.8\pm 0.3}$ GeV. The deviations here
are smaller: $\Delta^{\scrscr GUT}_{b/\tau} \simeq -0.02$
since $\Mhs$ cannot be larger than $\Mx$ by more than an
order of magnitude. The important thing to note is that
the lack of knowledge of the GUT scale spectrum does not
preclude extracting reliable predictions for $M_t$ over much
of the parameter space and the
uncertainties in these predictions can be reasonably estimated.
In regions where predictions are not robust, \ie\ away from the
attraction of the infrared fixed point for $y_t$, one must worry
about $10$ - $20\%$ effects in the minimal model.

For completeness we also depict the most significant
uncertainty in the $M_t$ solution: the error bars on $\alpha_s$ and
$M_b$. Fig.~10 shows the $M_t$ {\it vs.} $\tan\beta$ contours for different
$\alpha_s$ for the generic case $\Mv = \Msig$, $M_b = 4.9$ GeV
and sparticle spectrum parameters $\mgl = 400$ GeV, $V_{\scrscr SUSY}
= 35$ GeV and $M_3 = 375$ GeV. In Fig.~11 we show these contours for different
$M_b$ ranging from $4.7$ to $5.3$ GeV for $\alpha_s(M_Z) = 0.120$ and
the same remaining inputs as in Fig.~10. Note that for small $M_b$ the
curves are cut off due to the lack of a perturbative solution
for $M_t$ for intermediate values of $\tan\beta$. Clearly, in the
absence of additional theoretical cuts such as proton decay or
radiative electroweak symmetry breaking there remains a large
parameter space. Even if one imposes the restricted $M_t$ limits from
fits to electroweak data, one can find solutions for
all regions of $\tan\beta$, at least for larger $M_b$.
A better determination of $M_b$ could
eliminate the intermediate $\tan\beta$ solutions, although this
subject is frought with theoretical uncertainties.

Next we turn to a discussion of the Yukawa threshold corrections
in the MSSM. Using the approximate matching functions
of Eq.~\approxdytaususy\ - \approxdytsusy\ we incorporate $M_t$ into
the numerical routine together with $\Mx$ and $\Mhs$ and compute the threshold
effects on the top mass solution for representative
sparticle spectra consistent with universal
soft supersymmetry breaking parameters at the GUT scale as well
as radiative electroweak symmetry breaking\rlap.\refmark{\BBOspectra}
We also consider the low, intermediate and high
$\tan\beta$ cases separately.
Table~3 gives the effect of the threshold corrections for low
$\tan\beta = 1.47$ and some sample soft breaking parameters including
the generic no-scale case.
We also consider both signs of $\muh$ and take $A = 0$ throughout.
\smallskip
\begintable
$M_0$ | $M_{\scrscr{1\over 2}}$ | $\muh$ | $\Delta^{\scrscr SUSY}_{y_t}$|
$\Delta^{\scrscr SUSY}_{y_b}$ | $\Delta^{\scrscr SUSY}_{y_{\tau}}$|
$M_t$ (naive) | $M_t$ \crthick
$100$ | $140$ | $+375$ | $0.02$ | $0.006$ | $-0.006$ | $164.1$ | $167.6$ \cr
$100$ | $140$ | $-375$ | $0.03$ | $0.03$  | $-0.01$  | $164.1$ | $170.5$ \cr
$0$   | $150$ | $+353$ | $0.02$ | $0.007$ | $-0.007$ | $164.2$ | $167.8$ \cr
$0$   | $200$ | $-466$ | $0.07$ | $0.04$  | $-0.01$  | $163.0$ | $175.4$
\endtable\nobreak
\medskip\nobreak
\centerline{\singlespace Table~3.\ \vtop{\parindent=0pt\hsize=4truein
Low $\tan\beta$ Threshold Corrections to $M_t$ in the MSSM
($\tan\beta = 1.47$, $\alpha_s(M_Z) = 0.120$, $M_b = 4.8$ GeV, $A = 0$ )
}}
\medskip
\noindent All masses in the tables are in GeV.
Note that the increased stop squark splitting in the case of
$\muh < 0$ increases the $y_t$ corrections. The SUSY thresholds
increase the top mass prediction relative to the naive result
which includes only the gauge coupling thresholds.
The dominant effect arises
from the gluino-induced vertex corrections of Fig.~1 (for $\muh < 0$)
and the gluino contribution to the top quark wavefunction renormalization.
These are also the dominant contributions to the $y_b$ threshold,
although for $\muh > 0$ the gluino-induced vertex and wavefunction
renormalization contributions almost cancel, while they add
constructively for $\muh < 0$.
Since $y_t$ is near the fixed point in this region, the order
$5\%$ corrections to $y_b/y_{\tau}$ translate into a negligible effect
on the $M_t$ solution as discussed in Section~5.

For intermediate and high $\tan\beta$ the bottom and $\tau$
Yukawa corrections become more important.
For intermediate $\tan\beta = 15$ and
$M_{\scrscr{1\over 2}} = 100$ GeV, Table~4
summarizes the typical threshold effects.
\bigskip
\begintable
$M_0$ | $\muh$ | $\Delta^{\scrscr SUSY}_{y_t}$|
$\Delta^{\scrscr SUSY}_{y_b}$ | $\Delta^{\scrscr SUSY}_{y_{\tau}}$|
$M_t$ (naive) | $M_t$ \crthick
$100$ | $164$ | $0.01$  | $-0.1$  | $0.025$ | $194.5$ | $178.3$ \cr
$200$ | $316$ | $0.03$  | $-0.03$ | $0.025$ | $190.3$ | $187.5$ \cr
$300$ | $462$ | $0.035$ | $-0.01$ | $0.02$  | $187.3$ | $188.7$
\endtable\nobreak
\smallskip\nobreak
\centerline{\singlespace Table~4.\ \vtop{\parindent=0pt\hsize=4truein
Intermediate $\tan\beta$ Threshold Corrections to $M_t$ in the MSSM
($\tan\beta = 15$, $\alpha_s(M_Z) = 0.118$,
$M_b = 4.9$ GeV, $M_{\scrscr{1\over 2}} = 100$ GeV, $A = 0$ )}}
\bigskip
\noindent It is interesting to see the source of the enhanced $y_b$ and
$y_\tau$
corrections. For $M_0 = 100$ GeV the gluino induced vertex of Fig.~1
contributes $-0.13$, while the chargino induced vertices of Fig.~2
with a stop squark internal lines contribute $0.0175$. Accounting for
other small corrections leaves a $10\%$ effect. On the other hand for
$M_0 = 300$ GeV and correspondingly larger squark and slepton masses, many
contributions come into play. Now the chargino contributions add to $0.03$,
the gluino vertex correction is $-0.09$ and the total wavefunction
renormalization contribution is $0.03$. Accounting for heavy Higgs
effects and the neutralino contributions leaves a net $1\%$ effect in $y_b$.
Clearly a complete analysis is important even for intermediate
values of $\tan\beta$. The dominant corrections
to $y_\tau$ come from the neutral and charged wino induced graphs
of Fig.~3.

For larger $\tan\beta = 40$ and
$M_{\scrscr{1\over 2}} = 100$ GeV, we give typical threshold effects
in Table~5.
\vskip 2em
\begintable
$M_{\scrscr{1\over 2}}$ | $\muh$ | $\Delta^{\scrscr SUSY}_{y_t}$|
$\Delta^{\scrscr SUSY}_{y_b}$ | $\Delta^{\scrscr SUSY}_{y_{\tau}}$|
$M_t$ (naive) | $M_t$ \crthick
$200$ | $205$ | $0.04$  | $-0.035$ | $0.04$  | $177.6$ | $154.5$ \cr
$400$ | $450$ | $0.05$  | $-0.045$ | $0.06$  | $169.0$ | $78.2$ \cr
$600$ | $660$ | $0.055$ | $-0.03$  | $0.065$ | $162.3$ | $45.2$
\endtable\nobreak
\smallskip\nobreak
\centerline{\singlespace Table~5.\ \vtop{\parindent=0pt\hsize=4truein
Large $\tan\beta$ Threshold Corrections to $M_t$ in the MSSM
($\tan\beta = 40$, $\alpha_s(M_Z) = 0.118$,
$M_b = 4.9$ GeV, $M_0 = 400$ GeV, $A = 0$ )
}}
\bigskip
\noindent Here the heavier spectrum chosen enhances all
the wavefunction renormalization contributions and the gluino vertex
contributions to $y_b$ approach $20\%$. For the positive $\muh$
chosen however, this is partially cancelled as described in the intermediate
$\tan\beta$ case. Even so, the effect on $M_t$ is dramatic.
Here one is far from the attraction of the infrared fixed point.
The typically large $M_t$ predictions in SUSY-GUTs normally arise
because the QCD running of $y_b$ typically places it below $y_\tau$
at $\Mx$. A large top Yukawa coupling counters this behaviour.
However, here the large, order $10\%$, threshold correction
increases $y_b$ relative to $y_\tau$ at $\Mx$ and smaller values
of $M_t$ are required.
In fact, for some spectra the situation becomes
unstable in the sense that no positive $y_t$ can be found such
that the $y_b(\Mx) = y_\tau(\Mx)$ unification occurs.

Finally, we use the results of Section~5 to estimate the
contribution of the gauge coupling thresholds to the $M_t$
solution which are already incorporated in the ``naive'' solution
in the above tables. Using Eq.~\indthra\ we find the ranges
$\Delta^{\alpha}_{b/\tau} \approx (-0.014,-0.03),\ (-0.03,-0.045),\
(-.044,-0.084)$ for the data of Tables~3, 4 and 5, respectively.
In the intermediate and high $\tan\beta$ regions this effect
supplements the other potentially large negative Yukawa corrections
to $y_b/y_{\tau}$. There is also a threshold correction to $K_t$
of Eq.~\ktdef\ given by $\Delta_{K_t} \approx -0.02$ for the data of
Tables~3 and 4 and in the range $\approx (-0.025,-0.046)$
for Table~5. The corresponding fractional correction to $M_t$ is
$-\tfrac12\Delta_{K_t}$ and enhances the generally positive
direct Yukawa threshold correction to $y_t$.
The generic features of the analytic solutions are
born out, at least in the low $\tan\beta$ region where they are
valid. In particular, the gauge and Yukawa corrections to
$y_b/y_{\tau}$ both at $\Ms$ and $\Mx$ do not directly feed
into the $M_t$ solution. They
are multiplied by a factor, $C$ of Eq.~\bigc, which for
the data of Table~3 is $C \approx 0.1$. The fact that the corrections
to $M_t$ are dominated by the direct threshold correction
$\Delta_{y_t}^{\scrscr SUSY}$ for low $\tan\beta$ confirms this
expectation.

\chapter{Conclusions}

We have given a complete treatment of Yukawa coupling threshold
corrections in the MSSM and minimal SUSY-SU(5). We have applied these
as well as the gauge coupling thresholds to both approximate one loop
analytic and consistent two loop numerical solutions for the
superheavy masses $\Mx$ and $\Mhs$ and the top quark mass, $M_t$ in
the context of gauge and Yukawa unification in SUSY-SU(5). We have
highlighted the sensitivities of these solutions to the low scale
parameters. The effective GUT scale was the most constrained and
variations of $\alpha_s(M_Z)$, $M_t$ and the gluino mass led to of
order $40\%$ deviations in $\Mx$.

We have also found limits on the colored Higgs triplet superfield mass,
$\Mhs$, from an RG analysis of unification. This approach avoids ad
hoc assumptions about the degeneracies of the superheavy particles.
We displayed the sensitivities of $\Mhs$ on $\alpha_s(M_Z)$, $M_t$,
$\tan\beta$ and the Higgsino masses which are much larger than for
$\Mx$. Using these results we found that a conservative analysis of
the proton decay bound could be translated into a lower bound on the
strong coupling, $\alpha_s(M_Z) \gsim 0.118$ for any sparticle
spectrum with masses less than order $1$ TeV. We also concluded
that $\Mhs$ must be lighter than $2\times 10^{17}$ GeV. This in turn
limits the size of GUT scale corrections to $y_b/y_{\tau}$ so that
these effects cannot produce too large a decrease in the
solution for $M_t$. We found also that away from the attractive
region of the infrared quasi-fixed point of the top Yukawa coupling,
there can be significant positive corrections to $M_t$ if there is a
substantial allowed splitting of the superheavy vector and adjoint
scalar masses, $\Mv \gg \Msig$. This typically occurs for larger
values of $M_b$, while for smaller values no perturbative solution can
be found. For low $\tan\beta$ these corrections to $y_b/y_{\tau}$ have
different effects on the two alternative solutions, $M_t$ for fixed
$M_b$ or $m_b(M_Z)$ for fixed $M_t$. In the former case the
corrections are suppressed, while in the latter they feed directly
into the uncertainty for $m_b(M_Z)$.

We have given a detailed calculation of the Yukawa corrections
in the MSSM and analyzed their effect on the $M_t$ solution for
various regions of $\tan\beta$ and representative sparticle spectra.
We have shown, both analytically and numerically, the robustness of
this solution for low $\tan\beta$, where the dominant corrections
come from gluino-induced effects. For larger $\tan\beta$ we discussed
the importance of the complete calculation to determining the
potentially large corrections to $y_b$ and $y_{\tau}$.
These corrections were particularly sensitive to the signs of $\muh$
and $A_t$. For all $\tan\beta$ we described the enhancing effect of
stop squark mixing on $M_t$ for both signs of $\muh$.

We emphasized the effect of the $\alpha_s$ and $M_b$ error bars on
$M_t$, by far the dominant uncertainty. With strong bounds from proton
decay in the context of canonical supergravity induced soft
supersymmetry breaking, this parameter space can be cut substantially.
A critical analysis of the theoretical uncertainty in $M_b$ would also
help reduce the parameter space, particularly the intermediate $\tan\beta$
region, which is still allowed if one accepts larger values of $M_b
\sim 5.2$ GeV. Intermediate values of $\tan\beta$ must also be
reconsidered when the supersymmetric Yukawa thresholds are included
as these can reduce $M_t$ by $10\%$. We will discuss these issues and
perform a more complete analysis of the MSSM and GUT Yukawa thresholds
under more specific assumptions for the sparticle spectrum in a
future work\rlap.\refmark{\inpreptwo}

\ACK
The author would like to thank Vernon~Barger and Pierre~Ramond
for useful comments on earlier drafts of the manuscript. He would
also like to thank Paul~Ohmann for sharing some of his sparticle
spectrum data. He is also grateful for the hospitality of the IFT at the
University of Florida where part of this work was completed.
This research was supported in part by the University of Wisconsin
Research Committee with funds granted by the Wisconsin Alumni Research
Foundation and in part by the U.~S.~Department of Energy under
contract no. DE-AC02-76ER00881.

\endpage

\APPENDIX{A}{A: Yukawa Threshold Corrections in the MSSM}

Below we give the complete supersymmetry breaking threshold corrections to
third generation Yukawa couplings in the MSSM in the case in which
intergenerational mixing is
negligible in the squark and slepton mass matrices. The relevant factors
as in Eq.~\yeffrunning\
are the finite parts of the one loop wavefunction
renormalizations for the third generation fermions and the light SM Higgs
and the Yukawa vertex corrections involving heavy sparticles. We give results
for the scenario in which there is only one light Higgs below the scale at
which the
threshold conditions are applied.

First we give some notation and conventions. Our conventions closely follow
those of Refs.~\Higgshunt\ and \HaberKane\ which the reader should consult for
more details. We define our superpotential in the MSSM as
$$
P = \epsilon_{ij}(Y_d \hat{H}^i_1\hat{Q}^j\hat{D} + Y_e
\hat{H}^i_1\hat{L}^j\hat{E}
+ Y_u \hat{H}^j_2\hat{Q}^i\hat{U} + \muh \hat{H}^i_1\hat{H}^j_2 )~,\eqn\mssmpot
$$
where the hats denote superfields, $\epsilon_{12} = -\epsilon_{21} = 1$ and
$Y_{u,d,e}$ are the conventional Yukawa matrices, with $y_t$, $y_b$
and $y_{\tau}$ the respective diagonal elements for the third family.
The superfields $Q$, $U$, $D$, $L$ and $E$ have the hypercharge assignments of
the
corresponding  quarks and leptons. The Higgs superfields $H_{1,2}$ have
hypercharge
$y = -1,+1$ and their scalar components acquire vacuum expectation values
$\VEV{H^1_1} = v_1/\sqrt{2}$ and $\VEV{H^2_2}= v_2/\sqrt{2}$, respectively.
A general set of soft supersymmetry breaking parameters is introduced without
explicit regard to their origin. These include soft trilinear scalar couplings
mimicking those of the superpotential as well as explicit mass terms for the
gauginos, squarks, sleptons and Higgs fields.
$$
\eqalignno{
V_{\rm soft} =\ & - \epsilon_{ij}(Y_d A_d H^i_1 \tilde{Q}^j\tilde{D} +
Y_e A_e H^i_1 \tilde{L}^j\tilde{E} + Y_u A_u H^j_2
\tilde{Q}^i\tilde{U} + \muh B H^i_1 H^j_2 )~,&\cr
{\cal{L}}^{\rm mass}_{\rm soft} =\ & \ttwo {\cal{M}}_1\ovl{\lambda}_B\lambda_B
+ \ttwo {\cal{M}}_2\ovl{\lambda}_W^i\lambda_W^i
+ \ttwo {\cal{M}}_3\ovl{\lambda}_g^A\lambda_g^A &\eqnalign\softbr\cr
&\quad - m_Q^2 \tilde{Q}^{i*}\tilde{Q}^i - m_U^2 \tilde{U}^*\tilde{U}
- m_D^2 \tilde{D}^*\tilde{D} &\cr
&\quad  - m_L^2 \tilde{L}^{i*}\tilde{L}^i - m_E^2 \tilde{E}^*\tilde{E}
- m_{H_1}^2 \tilde{H}_1^{i*}\tilde{H}_1^i - m_{H_2}^2
\tilde{H}_2^{i*}\tilde{H}_2^i~,&\cr}
$$
where we denote superpartners of ordinary particles with a tilde, and $\lambda$
is used to denote
the Majorana gaugino fields. We also define squark and slepton fields with the
same charge
conventions as their partners: $\tilde{Q} = (\tilde{u}_L,\tilde{d}_L )$,
$\tilde{U}= \tilde{u}_R^*$,
$\tilde{D}= \tilde{d}_R^*$, $\tilde{Q} = (\tilde{\nu}_L,\tilde{e}_L )$ and
$\tilde{E}= \tilde{e}_R^*$.

The threshold corrections will in general involve
squark and slepton mixing matrices for the third generation.
For example, denoting the stop squark mass eigentates as $\tilde{t}_{1,2}$, we
can relate
them to weak eigenstates $\tilde{t}_{L,R}$ by an orthogonal matrix $O^t$ such
that
$O^t M^2_{\tilde{t}} O^{tT} = \bar{M}^2_{\tilde{t}}$ where
$\bar{M}^2_{\tilde{t}}$ is diagonal
and
$$
\pmatrix{\tilde{t}_1\cr \tilde{t}_2\cr} = \pmatrix{\cos\theta_t &
\sin\theta_t\cr
-\sin\theta_t & \cos\theta_t \cr} \pmatrix{\tilde{t}_L\cr \tilde{t}_R\cr}
= O^t \pmatrix{\tilde{t}_L\cr \tilde{t}_R\cr}~.\eqn\stopmix
$$
We define $O^b$ and $O^{\tau}$ similarly. The neutralino mass matrix is
diagonalized by
the unitary matrix $Z$ such that $Z^{*}M_{\chi^0}Z^{-1} = \bar{M}_{\chi^0}$
and relates the weak bino, wino and Higgsino eigenstates to the mass eigenstate
Majorana fields $\chi^{0}_i$ via
$$ \eqalign{
-i\lambda_{B} =\ & Z_{i1}^*\chi^0_i~, \qquad -i\lambda_{W}^3 =
Z_{i2}^*\chi^0_i~, \cr
\tilde{H}_1^1 =\ & Z_{i3}^*\chi^0_i~, \qquad \tilde{H}_2^2 = Z_{i4}^*\chi^0_i~.
\cr}
\eqn\neutralinoegs
$$
The upper index on the Higgsino fileds is an ${\rm SU(2)}_L$ index.
Finally we define the unitary mixing matrices for the charginos by
$U^*M_{\chi^\pm} V^{-1} = \bar{M}_{\chi^\pm}$ where the charged Higgsino
and wino are related to the Dirac mass eigenstates
$\chi_{Di}^{T} = (\chi^+_i, \bar{\chi}^-_i)$
via
$$
\tilde{H}_2^1 = V_{i2}^*\chi_i^+ ~,\quad \tilde{H}_1^2 = U_{i2}^*\chi_i^- ~,
\quad -i\lambda_{W}^{+}
 = V_{i1}^*\chi_i^{+}~,\quad -i\lambda_{W}^{-} =
U_{i1}^*\chi_i^{-}~.\eqn\charginoegs
$$
Explicit forms for the mass and mixing matrices can be found in
Refs.~\BBOspectra, \HaberKane, and \Rosiek.

We first give the finite parts of the wavefunction renormalization
of the top and bottom quark and $\tau$ lepton
due to loops involving squarks, sleptons, gluinos, neutralinos, charginos, and
the heavy
Higgs doublet involving the charged Higgs, $H^{\pm}$, the pseudoscalar $A$ and
the heavy scalar
$H^0$. We sum the coefficients of $\dslash P_{L,R}$ in the effective action
obtained by
integrating out these fields and find (using the notation of Section 2)
for the top quark,
$$
\eqalign{
\bar{K}_{t(L + R)} =\ & -{1\over 32\pi^2}\Biggl(
{\textstyle {8\over 3}} g_3^2 \sum_i F_2(\mst^2,\mgl^2 ) \cr
+ &y_b^2 \biggl( \sum_{i,j} F_2(\msb^2,\mchg^2 ) (O^{b}_{i2})^2 |U_{j2}|^2
+ F_2(m_{H^{\pm}}^2,m_b^2 )\sin^2\!\beta \biggr)\cr
+ &y_t^2 \biggl( \sum_{i,j} F_2(\msb^2,\mchg^2 ) (O^{b}_{i1})^2 |V_{j2}|^2
+ \sum_{i,j} F_2(\mst^2,\mneut^2) |Z_{j4}|^2\cr
&\quad + \cos^2\!\beta \bigl(F_2(m_{H^{\pm}}^2,m_b^2 ) +
\ttwo (F_2(m_{H^0}^2,m_t^2) - F_2(m_{A}^2,m_t^2))\bigr) \biggr)\cr
+ &g_2^2 \biggl( \sum_{i,j} F_2(\msb^2,\mchg^2 ) (O^{b}_{i1})^2 |U_{j1}|^2\cr
&\quad + \ttwo\sum_{i,j} F_2(\mst^2,\mneut^2)((O^{t}_{i1})^2| Z_{j2} +\tthree
Z_{j1}t |^2\cr
&\quad + {\textstyle {16\over 9}} (O^{t}_{i2})^2 |Z_{j1}|^2 t^2 )\biggr)\cr
- &g_2 y_b \biggl(2\sum_{i,j} F_2(\msb^2,\mchg^2 ) O^{b}_{i1}O^{b}_{i2} \Re
(U_{j1}U_{j2}^*) \biggr)\cr
+ &g_2 y_t \biggl(\sqrt{2} \sum_{i,j} F_2(\mst^2,\mneut^2) O^{t}_{i1}O^{t}_{i2}
\Re ((Z_{j2} - Z_{j1}t)Z_{j4}^*)\biggr)\Biggr)~,\cr}\eqn\ktlr
$$
\noindent and for the bottom quark,
$$
\eqalign{
\bar{K}_{b(L + R)} =\ & -{1\over 32\pi^2}\Biggl(
{\textstyle {8\over 3}} g_3^2 \sum_i F_2(\msb^2,\mgl^2 ) \cr
+ &y_t^2 \biggl( \sum_{i,j} F_2(\mst^2,\mchg^2 ) (O^{t}_{i2})^2 |V_{j2}|^2 +
F_2(m_{H^{\pm}}^2,m_t^2 )
\cos^2\!\beta \biggr)\cr
+ &y_b^2 \biggl( \sum_{i,j} F_2(\mst^2,\mchg^2 ) (O^{t}_{i1})^2 |U_{j2}|^2
+ \sum_{i,j} F_2(\msb^2,\mneut^2) |Z_{j3}|^2\cr
&\quad + \sin^2\!\beta \bigl(F_2(m_{H^{\pm}}^2,m_t^2 ) +
\ttwo (F_2(m_{H^0}^2,m_b^2) - F_2(m_{A}^2,m_b^2))\bigr) \biggr)\cr
+ &g_2^2 \biggl( \sum_{i,j} F_2(\mst^2,\mchg^2 ) (O^{t}_{i1})^2 |V_{j1}|^2\cr
&\quad + \ttwo\sum_{i,j} F_2(\msb^2,\mneut^2)((O^{b}_{i1})^2| Z_{j2} - \tthree
Z_{j1}t |^2\cr
&\quad + {\textstyle {4\over 9}} (O^{b}_{i2})^2 |Z_{j1}|^2 t^2 )\biggr)\cr
- &g_2 y_t \biggl(2\sum_{i,j} F_2(\mst^2,\mchg^2 ) O^{t}_{i1}O^{t}_{i2} \Re
(V_{j1}V_{j2}^*) \biggr)\cr
- &g_2 y_b \biggl(\sqrt{2} \sum_{i,j} F_2(\msb^2,\mneut^2) O^{b}_{i1}O^{b}_{i2}
\Re ((Z_{j2} - Z_{j1}t)Z_{j3}^*)\biggr)\Biggr)~,\cr}\eqn\ktlr
$$
\noindent where $F_2$ is defined in Appendix B and $t = \tan\theta_{W}$.
The result for the $\tau$ lepton is
$$
\eqalignno{
\bar{K}_{\tau (L + R)} =\ & -{1\over 32\pi^2}\Biggl(
y_{\tau}^2 \biggl( \sum_{j} F_2(\msnutau^2,\mchg^2 ) |U_{j2}|^2
+ \sum_{i,j} F_2(\mstau^2,\mneut^2) |Z_{j3}|^2 &\cr
&\quad + \sin^2\!\beta \bigl(F_2(m_{H^{\pm}}^2,0 )
+ \ttwo (F_2(m_{H^0}^2,m_{\tau}^2) - F_2(m_{A}^2,m_{\tau}^2))\bigr) \biggr)&\cr
+ &g_2^2 \biggl( \sum_{i} F_2(\msnutau^2,\mchgi^2 ) |V_{i1}|^2&\cr
&\quad + \ttwo\sum_{i,j} F_2(\mstau^2,\mneut^2)((O^{\tau}_{i1})^2
| Z_{j2} + Z_{j1}t |^2&\eqnalign\ktaulr\cr
&\quad + 4 (O^{\tau}_{i2})^2 |Z_{j1}|^2 t^2 )\biggr)&\cr
- &g_2 y_{\tau} \biggl(\sqrt{2} \sum_{i,j} F_2(\mstau^2,\mneut^2)
O^{\tau}_{i1}O^{\tau}_{i2}
\Re (( Z_{j2} - Z_{j1}t )Z_{j3}^*)\biggr)\Biggr)~,&\cr}
$$
\noindent where $\bar{K}_{\alpha (L + R)} = \bar{K}_{\alpha L} +
\bar{K}_{\alpha R}$.

The neutral Higgs wavefunction renormalization contributions arise from
chargino
and neutralino loop diagrams in general giving matrix contributions so that,
$$
\bar{Z}_{H} = \pmatrix{1+\bar{K}_{H_0}&\bar{K}_{hH}\cr
\bar{K}_{hH}&1+\bar{K}_{h_0}\cr}~,\eqn\kbarhiggs
$$
where
$$
\eqalign{
\bar{K}_{h_0} =\ & {g_2^2\over 16\pi^2}\Biggl({\textstyle{1\over 6}}(3+t^2)
+ \sum_{i,j} (| A''_{ij}|^2  F_3(\mneuti,\mneutj )\cr
&\quad + 2 | A_{ij}|^2  F_3(\mchgi,\mchgj ) )\Biggr)~,\cr
\bar{K}_{H_0} =\ & {g_2^2\over 16\pi^2}\Biggl({\textstyle{1\over 6}}(3+t^2)
+ \sum_{i,j} (| B''_{ij}|^2  F_3(\mneuti,\mneutj )\cr
&\quad + 2 | B_{ij}|^2  F_3(\mchgi,\mchgj ) )\Biggr)~,\cr
\bar{K}_{hH} =\ & -{g_2^2\over 16\pi^2}\Biggl(
\sum_{i,j} (\Re (A''_{ij}B^{\prime\prime *}_{ij}) F_3(\mneuti,\mneutj )\cr
&\quad + 2\Re (A_{ij}B_{ij}^*) F_3(\mchgi,\mchgj ) )\Biggr)~,\cr}
\eqn\klthiggs
$$
\noindent and where
$$
\eqalign{
A''_{ij} =\ & -Q''_{ij}\cos\!\beta + S''_{ij}\sin\!\beta~,
\qquad A_{ij} = - Q_{ij}\cos\!\beta - S_{ij}\sin\!\beta~,\cr
B''_{ij} =\ & Q''_{ij}\sin\!\beta + S''_{ij}\cos\!\beta~,
\qquad B_{ij} = -Q_{ij}\sin\!\beta + S_{ij}\cos\!\beta~,\cr}\eqn\ABs
$$
with\refmark{\Higgshunt}
$$
\eqalign{
Q_{ij} =\ & {1\over\sqrt{2}}V_{i1} U_{j2}~, \qquad S_{ij} =
{1\over\sqrt{2}}V_{i2} U_{j1}~,\cr
Q''_{ij} =\ & \ttwo [ Z_{i3}(Z_{j2}- Z_{j1}t) + (i\leftrightarrow j)
]\epsilon_i~,\cr
S''_{ij} =\ & \ttwo [ Z_{i4}(Z_{j2}- Z_{j1}t) + (i\leftrightarrow j)
]\epsilon_i~.\cr}
$$
The factor $\epsilon_i = \pm 1$ is introduced\refmark{\HaberGun}
to insure positive mass neutralino
eigenstates. When $Z$ is defined without regard to the positivity of the mass
eigenvalues, the
elements of the matrix $\bar{M}_{\chi^0}$ are $\epsilon_i\mneuti\
(i=1,\dots,4)$.
The function $F_3$ is given in Appendix~B.
In general the matrix $\bar{Z}_{H}$ must be
diagonalized, $\bar{Z}^d_{H} = O_h\bar{Z}_{H}O_h^T$, giving
an effective light Higgs field:
$$
h_0^{eff} = (\bar{Z}^d_{H})_2^{1\over 2}((O_h)_{21} H_0 + (O_h)_{22}
h_0)~.\eqn\heff
$$
This results in a threshold correction to the neutral Higgs mixing
angle $\alpha$ of Ref.~\Higgshunt .
However, in the limit of small gaugino-Higgsino mixing considered later, these
off-diagonal
contributions vanish.

We next give the one particle irreducible Yukawa vertex corrections. These
contributions involve more complicated functions of the squark, slepton,
neutralino
and chargino mixing angles although many of them simplify in certain
approximations
for the sparticle mass matrices.
The greatest simplifications occur in the limit of no squark and slepton mixing
and vanishing mixing between the bino, neutral wino and the Higgsinos.
Except for the case of large $\tan\beta$
one would expect small L-R mixing in the sbottom and stau mass matrices.
However
due to the large top Yukawa coupling the mixing can be substantial. Also the
lack of mixing
in the neutralino sector is expected if the bino and neutral wino masses are
dominated by
their soft-breaking masses and if the Higgsino masses are determined by the
parameter $\muh$
in the superpotential. This approximation is certainly appropriate to the case
of
radiatively induced
electroweak symmetry breaking near the low $\tan\beta$
fixed point\refmark{\BBOspectra,\CarenaIRtwo}
and works reasonably well for larger $\tan\beta$.
We quote the entire results as well as the simplified results in the limit
of small gaugino-Higgsino mixing.
Note these terms can
give rise to substantial enhancements of the threshold corrections proportional
to ratios of sparticle masses to $m_b$ or $m_{\tau}$ if the mixing and
mass-splitting
between neutralinos or same generation squarks and sleptons is large.

The finite Yukawa corrections are denoted $\bar{K}^{\scrscr Y}_{\alpha} =
\delta y_{\alpha}/y_{\alpha}$
as in Section 2.
We separately give the results for Higgs, neutralino, chargino and gluino
corrections.
For the $h_0 \bar{t} t$ vertex we obtain
$$
\bar{K}^{\scrscr Y}_t = \bar{K}^{\scrscr Y}_t({\rm Higgs }) + \bar{K}^{\scrscr
Y}_t(\chi^0 )
+ \bar{K}^{\scrscr Y}_t(\chi^\pm ) + \bar{K}^{\scrscr Y}_t(\tilde{g})~,\eqn\kyt
$$
where the separate contributions are
$$
\eqalign{
\bar{K}^{\scrscr Y}_t({\rm Higgs}) =\ & {\cos^2\!\beta\over 16\pi^2}
\Biggl( \tfrac12 y_t^2\biggl(F_1(m_t^2,m_{H_0}^2) - F_1(m_t^2,m_{A}^2)\cr
&\quad + M_Z^2\bigl(-(\cos^2 2\beta - 2\sin^2
2\beta)G_2(\mbh^2,\mbh^2,m_t^2)\cr
&\quad + 12\sin^2\beta\cos 2\beta \, G_2(\mbh^2,\mlh^2,m_t^2)\cr
&\quad + \cos^2 2\beta \, G_2(m_A^2,m_A^2,m_t^2)\cr
&\quad + 4\cos 2\beta\sin^2\beta \, G_2(m_A^2,m_Z^2,m_t^2)\bigr) \biggr)\cr
+ &y_b^2 \biggl( F_1(m_b^2,\mhpm^2)\cr
&\quad + M_Z^2\bigl((2c^2 - \cos^2 2\beta)G_2(\mhpm^2,\mhpm^2,m_b^2)\cr
&\quad + 2\cos^2 2\beta \, G_2(\mhpm^2,m_W^2,m_b^2)\bigr)\biggr)
\Biggr)~,\cr}\eqn\kythiggs
$$
$$
\eqalign{
\bar{K}^{\scrscr Y}_t(\chi^0) =\ & {1\over 16\pi^2}\Biggl( -g_2^2{1+t^2\over 4}
- \sqrt{2}g_2\sum_{i,j,k}\biggl(G_1(\msti^2,\mneutj^2,\mneutk^2)N^{\scrscr
LR}_{ijk}(t)\cr
&\qquad + \mneutj\mneutk G_2(\msti^2,\mneutj^2,\mneutk^2)N^{\scrscr
RL}_{ijk}(t)
\biggr)\cr
&\qquad + 2m_t\sum_{i,j,k} \mneuti
G_2(\mneuti^2,\mstj^2,\mstk^2)\ovl{N}_{ijk}(t)\Biggr)~,\cr}
\eqn\kytneut
$$
$$
\eqalign{
\bar{K}^{\scrscr Y}_t(\chi^\pm) =\ & {1\over 16\pi^2}\Biggl( {-g_2^2\over 2}
- \sqrt{2}g_2\sum_{i,j,k}\biggl(G_1(\msbi^2,\mchgj^2,\mchgk^2)C^{\scrscr
LR}_{ijk}(t)\cr
&\qquad + \mchgj\mchgk G_2(\msbi^2,\mchgj^2,\mchgk^2)C^{\scrscr
RL}_{ijk}(t)\biggr)\cr
&\qquad + 2m_b\sum_{i,j,k} \mchgi
G_2(\mchgi^2,\msbj^2,\msbk^2)\ovl{C}_{ijk}(t)\Biggr)~,\cr}
\eqn\kytchg
$$
$$
\eqalign{
\bar{K}^{\scrscr Y}_t(\tilde{g}) =\ & {1\over 16\pi^2}\Biggl( -{16\over
3}g_3^2m_t\mgl
\sum_{i,j} G_2(\mgl^2,\msti^2,\mstj^2)\ovl{G}_{ij}(t)\Biggr)~,\cr}\eqn\kytglue
$$
\noindent and where the functions $F_i$ and $G_i$ are defined in Appendix B.
The quantities $N^{\scrscr LR}_{ijk}(t)$, $N^{\scrscr RL}_{ijk}(t)$
and $\ovl{N}_{ijk}(t)$ in the neutralino contribution are
$$
\eqalign{
N^{\scrscr LR}_{ijk}(t) =\ & y_t N^1_{ijk}(t) + g_2 N^2_{ijk}(t) +
g_2({g_2\over y_t}) N^3_{ijk}(t)~,\cr
N^{\scrscr RL}_{ijk}(t) =\ & y_t \tilde{N}^1_{ijk}(t) + g_2
\tilde{N}^2_{ijk}(t)
+ g_2({g_2\over y_t}) \tilde{N}^3_{ijk}(t)~,\cr
\ovl{N}_{ijk}(t) =\ & y_t^2 \ovl{N}^1_{ijk}(t) + y_t g_2 \ovl{N}^2_{ijk}(t) +
g_2^2 \ovl{N}^3_{ijk}(t)\cr
&\quad + g_2^2({g_2\over y_t})\ovl{N}^4_{ijk}(t) + g_2^2({g_2\over
y_t})^2\ovl{N}^5_{ijk}(t)~,\cr}
\eqn\nlr
$$
where
$$
\eqalign{
N^1_{ijk}(t) =\ & \Re [T^{0*}_{Lik}(t)T^0_{Rij}(t)(-{Q''}^*_{jk}\cot\beta
+ {S''}^*_{jk})]~,\cr
N^2_{ijk}(t) =\ & -\Re [(R^{0*}_{Lik}(t)T^0_{Rij}(t) +
T^{0*}_{Lik}(t)R^0_{Rij}(t))\cr
&\quad \times (-{Q''}^*_{jk}\cot\beta + {S''}^*_{jk})]~,\cr
N^3_{ijk}(t) =\ & \Re [R^{0*}_{Lik}(t)R^0_{Rij}(t)(-{Q''}^*_{jk}\cot\beta
+ {S''}^*_{jk})]~,\cr}
\eqn\Nis
$$
and
$$
\eqalign{
\ovl{N}^1_{ijk}(t) =\ & \Re [T^{0*}_{Lki}(t)T^0_{Rji}(t)(P_{jk}(t) +
P_{jk}'(t)({\muh\cot\beta + A_t\over m_t}))]~,\cr
\ovl{N}^2_{ijk}(t) =\ & -\Re [(R^{0*}_{Lki}(t)T^0_{Rji}(t) +
T^{0*}_{Lki}(t)R^0_{Rji}(t))\cr
&\quad \times (P_{jk}(t) + P_{jk}'(t)({\muh\cot\beta + A_t\over m_t}))]~,\cr
\ovl{N}^3_{ijk}(t) =\ & {1\over c^2}(\cot^2\beta - 1)\Re
[T^{0*}_{Lki}(t)T^0_{Rji}(t)P_{jk}''(t)]\cr
&\quad + \Re [R^{0*}_{Lki}(t)R^0_{Rji}(t)(P_{jk}(t) + P_{jk}'(t)({\muh\cot\beta
+ A_t\over m_t}))]~,\cr
\ovl{N}^4_{ijk}(t) =\ & -{1\over c^2}(\cot^2\beta - 1)\Re
[(R^{0*}_{Lki}(t)T^0_{Rji}(t)
+ T^{0*}_{Lki}(t)R^0_{Rji}(t))P_{jk}''(t)]~,\cr
\ovl{N}^5_{ijk}(t) =\ & {1\over c^2}(\cot^2\beta - 1)\Re
[R^{0*}_{Lki}(t)R^0_{Rji}(t)P_{jk}''(t)]~.\cr}
\eqn\Nbaris
$$
The quantities $\tilde{N}_{ijk}(t)$ are obtained by interchanging the $L$ and
$R$
subscripts in \Nis.
The matrices $R$ and $T$ involve products of squark and neutralino mixing
matrices while
the $P$ matrices involve products of squark mixing matrices only. They are
defined by
$$
\eqalign{
R^0_{Lij}(f) =\ & \sqrt{2}O^f_{i1}((T_{3f} - e_f)t Z_{j1}^* -
T_{3f}Z_{j2}^*)~,\cr
R^0_{Rij}(f) =\ & \sqrt{2}O^f_{i2}e_f t Z_{j1}\epsilon_j~,\cr
T^0_{Lij}(f) =\ & O^f_{i2}Z_{j{4\choose3}}^*~,\cr
T^0_{Rij}(f) =\ & O^f_{i1}Z_{j{4\choose3}}\epsilon_j~,\cr}\eqn\RTmat
$$
and
$$
\eqalign{
P_{ij}(f) =\ & O^f_{i1}O^f_{j1} + O^f_{i2}O^f_{j2}~,\cr
P_{ij}'(f) =\ & \ttwo( O^f_{i1}O^f_{j2} + O^f_{i2}O^f_{j1})~,\cr
P_{ij}''(f) =\ & O^f_{i1}O^f_{j1}(T_{3f} - e_f s^2) + O^f_{i2}O^f_{j2}e_f
s^2~,\cr}
\eqn\Pmat
$$
where $f = (t,b,\tau)$ with charge $e_f$ and weak isospin $T_{3f}$.
In $T^0_{L,R}$ the ${4\choose 3}$ indices
correspond to $f= {t\choose b,\tau}$. For the sneutrino case there
is no $\tilde{\nu}_{\tau R}$ so one must make the replacements
$P_{\nu_\tau} = 1$, $P_{\nu_\tau}' = 0$ and $P_{\nu_\tau}'' = \ttwo$.
In all these quantities $c = \cos\theta_W$ and $s$ and $t$ are the sine and
tangent,
respectively.

The quantities $C^{\scrscr LR}_{ijk}(t)$, $C^{\scrscr RL}_{ijk}(t)$
and $\ovl{C}_{ijk}(t)$ in the chargino
contribution are
$$
\eqalign{
C^{\scrscr LR}_{ijk}(t) =\ & y_b C^1_{ijk}(t) + g_2 C^2_{ijk}(t)~,\cr
C^{\scrscr RL}_{ijk}(t) =\ & y_b \tilde{C}^1_{ijk}(t) + g_2
\tilde{C}^2_{ijk}(t)~,\cr
\ovl{C}_{ijk}(t) =\ & y_b^2 \ovl{C}^1_{ijk}(t) + y_b g_2
\ovl{C}^2_{ijk}(t)\cr
&\quad + g_2^2 \ovl{C}^3_{ijk}(t) + g_2^2({g_2\over
y_b})\ovl{C}^4_{ijk}(t)~,\cr}
\eqn\clr
$$
where
$$
\eqalign{
C^1_{ijk}(t) =\ & \Re [T^{\pm *}_{Lik}(t)T^{\pm}_{Rij}(t)({Q}^*_{jk}\cot\beta
+ {S}^*_{jk})]~,\cr
C^2_{ijk}(t) =\ & \Re [R^{\pm *}_{Lik}(t)T^{\pm}_{Rij}(t)({Q}^*_{jk}\cot\beta
+ {S}^*_{jk})]~,\cr
\tilde{C}^1_{ijk}(t) =\ & \Re [T^{\pm
*}_{Lik}(t)T^{\pm}_{Rij}(t)({Q}_{kj}\cot\beta
+ {S}_{kj})]~,\cr
\tilde{C}^2_{ijk}(t) =\ & \Re [R^{\pm
*}_{Lik}(t)T^{\pm}_{Rij}(t)({Q}_{kj}\cot\beta + {S}_{kj})]~,\cr}
\eqn\Cis
$$
and
$$
\eqalign{
\ovl{C}^1_{ijk}(t) =\ & \cot\beta \,\Re [T^{\pm
*}_{Lki}(t)T^{\pm}_{Rji}(t)(P_{jk}(b) +
P_{jk}'(b)({\muh\tan\beta + A_b\over m_b}))]~,\cr
\ovl{C}^2_{ijk}(t) =\ & -\cot\beta \,\Re [R^{\pm *}_{Lki}(t)T^{\pm}_{Rji}(t)
(P_{jk}(b) + P_{jk}'(b)({\muh\tan\beta + A_b\over m_b}))]~,\cr
\ovl{C}^3_{ijk}(t) =\ & \cot 2\beta\,\Re [T^{\pm
*}_{Lki}(t)T^{\pm}_{Rji}(t)P_{jk}''(b)]~,\cr
\ovl{C}^4_{ijk}(t) =\ & -\cot 2\beta\,\Re [R^{\pm
*}_{Lki}(t)T^{\pm}_{Rji}(t)P_{jk}''(b)]~.\cr}
\eqn\Cbaris
$$
The matrices $R^\pm$ and $T^\pm$ involve products of squark and chargino mixing
matrices
$$
\eqalign{
R^{\pm}_{Lij}(t) =\ & O^b_{i1}U_{j1}^*~,\cr
R^{\pm}_{Lij}(b) =\ & O^t_{i1}V_{j1}^*~,\cr
R^{\pm}_{Li}(\tau ) =\ & V_{i1}^*~,\cr}\qquad
\eqalign{T^{\pm}_{Lij}(t) =\ & O^b_{i2}U_{j2}^*~,\cr
T^{\pm}_{Lij}(b) =\ & O^t_{i2}V_{j2}^*~,\cr
T^{\pm}_{Li}(\tau ) =\ & 0~,\cr}\qquad
\eqalign{T^{\pm}_{Rij}(t) =\ & O^b_{i1}V_{j2}~,\cr T^{\pm}_{Rij}(b) =\ &
O^t_{i1}U_{j2}~,\cr
T^{\pm}_{Ri}(\tau ) =\ & U_{i2}~.\cr}\eqn\RTpmmat
$$
Finally the gluino contribution depends on $\ovl{G}_{ij}(t)$, defined by
$$
\eqalign{
\ovl{G}_{ij}(t) =\ & P'_{ij}(t)\Biggl(P_{ij}(t) +
P'_{ij}(t)\biggl({\muh\cot\beta + A_t\over m_t}\biggr)\Biggl)\cr
& \quad + {g^2\over y_t^2}{\cot^2\beta - 1\over 2c^2}
P'_{ij}(t)P''_{ij}(t)~.\cr}\eqn\Gbar
$$

The finite corrections to the $h_0\bar{b} b$ vertex are obtained from those for
the top
vertex with the following substitutions. First interchange $b \leftrightarrow
t$
and $\cos\beta \leftrightarrow \sin\beta$ everywhere in Eqs.~\kythiggs\ -
\kytglue\ and
in the definitions of $N^i$, $C^i$ (and their $L \leftrightarrow R$
counterparts),
$\ovl{N}^i$, $\ovl{C}^i$ and $\ovl{G}$. Then in the $N^i$, substitute
$-Q''\leftrightarrow S''$
and in the $C^i$ substitute $Q_{jk} \rightarrow S_{kj},\ S_{jk} \rightarrow
Q_{kj}$.
Also in $\bar{N}^{3,4,5}_{ijk}(b)$ and $\bar{G}_{ij}(b)$ one must change the
signs of the terms with
$\tan^2\beta - 1$ factors and one must change the overall signs of
$\bar{C}^{3,4}_{ijk}(b)$.
The finite corrections to the $h_0\bar{\tau} \tau$ vertex are obtained from
those for the bottom
vertex by substituting $b\rightarrow \tau$, $y_t, m_t \rightarrow 0$ and
$t\rightarrow \nu_{\tau}$
everywhere (omitting the appropriate sums). The charged Higgs and gluino
contributions
vanish in this vertex, while the only chargino contributions come from wino
induced contributions involving $g_2^2C^2$ and $g^2_2\ovl{C}^4$.

We may now construct the threshold matching functions for the third generation
Yukawa
couplings. Using the results of Section 2,
$$
\eqalign{
y_{t}^{\scrscr SM}(\mu ) =\ & y_t(\mu )\sin\beta (1 + \Delta^{\scrscr
SUSY}_{y_t})~,\cr
y_{b}^{\scrscr SM}(\mu ) =\ & y_b(\mu )\cos\beta (1 + \Delta^{\scrscr
SUSY}_{y_b})~,\cr
y_{\tau}^{\scrscr SM}(\mu ) =\ & y_{\tau}(\mu )\cos\beta (1 + \Delta^{\scrscr
SUSY}_{y_\tau})~,\cr}
\eqn\susyyukmatch
$$
where
$$
\Delta^{\scrscr SUSY}_{y_\alpha} = \bar{K}^Y_\alpha - \ttwo(\bar{K}_{\alpha L}
+
\bar{K}_{\alpha R} + \bar{K}_{h_0})~.\eqn\deltayalpha
$$
The explicit form for the complete matching functions is rather unwieldly
however
simplified forms can be constructed in certain limits. For example, in the low
to
intermediate $\tan\beta$ range it is appropriate to include order $y_t^2$ and
$g_3^2$
effects as the dominant contribution. We shall consider a particular limit in
which
the neutralino and chargino mass matrices have a definite form, determined by
assuming
values for the Higgs potential parameter $\muh$ and the gaugino masses ${\cal
M}_{1,2}$
which are significantly larger than $M_Z$. The the neutralino mass matrix has
the approximate form
$$ M_{\chi^0} \approx \pmatrix{{\cal M}_1 &0&0&0\cr 0&{\cal M}_2 &0&0\cr
0&0&0&\muh\cr 0&0&\muh&0\cr}~,\eqn\mneutapprox
$$
so that $Z_{ij} = \delta_{ij}$, $Z_{i3} = Z_{i4} = 0$ and $Z_{i+2,j+2} =
Z_{2ij}$, where
$$
Z_2 = {1\over\sqrt{2}}\pmatrix{1&\hphantom{-}1\cr 1&-1\cr}~.\eqn\Ztwo
$$
Due to the appearance of a negative Majorana mass eigenvalue
we have $\epsilon_3 = -\epsilon_4 = \sgn(\muh )$ (note also $\epsilon_{1,2} =
1$).
The chargino mass matrix is approximately diagonal,
$$
M_{\chi^\pm} \approx \pmatrix{{\cal M}_2&0\cr 0&-\muh\cr}~,
\eqn\mchgapprox
$$
with $U_{ij} \approx \delta_{ij}$ and $V \approx {\rm diag}(1,-\sgn(\muh ))$,
where $\sgn(\muh ) = \muh/|\muh|$.
Thus in this limit the neutralino eigenstates are the bino, neutral wino and a
Dirac Higgsino
with masses $m_{\tilde B}\approx {\cal M}_1$, $m_{\tilde{W}^3}\approx {\cal
M}_2$
and $m_{\tilde{H^0}}\approx |\muh |$. The chargino eigenstates
are just the charged wino and Higgsino with masses $m_{W}\approx {\cal M}_2$
and
$m_{\tilde{H}^\pm} \approx |\muh |$.

In this limit, the explicit form of the vertex corrections can be obtained
using
simplified forms of the $N^i$, $\tilde{N}^i$, $\bar{N}^i$, $C^i$, $\tilde{C}^i$
and $\bar{C}^i$.
For the neutralino contributions we have
$$
\eqalign{
\sum_{j=3,4} N^2_{ijk}(t) =\ & {1\over 2\sqrt{2}}\Bigl((O^t_{i1})^2\delta_{k2}
+ {t^2\over 3}\delta_{k1}
(-(O^t_{i1})^2 + 4(O^t_{i2})^2)\Bigr)~,\cr
\sum_{j=3,4} N^2_{ijk}(b) =\ & {1\over 2\sqrt{2}}\Bigl((O^b_{i1})^2\delta_{k2}
+ {t^2\over 3}\delta_{k1}
((O^b_{i1})^2 + 2(O^b_{i2})^2)\Bigr)~,\cr
\sum_{j=3,4} N^2_{ijk}(\tau) =\ & {1\over
2\sqrt{2}}\Bigl((O^{\tau}_{i1})^2\delta_{k2}
+ t^2\delta_{k1}(-(O^{\tau}_{i1})^2 + 2(O^{\tau}_{i2})^2)\Bigr)~,\cr
\bar{N}^3_{ijk}(t) =\ & -\tfrac29 t^2\delta_{i1}P'_{jk}(t)\Bigl(P_{jk}(t) +
P'_{jk}(t)
\bigl({\muh\cot\beta + A_t\over m_t}\bigr)\Bigr)~,\cr
\bar{N}^3_{ijk}(b) =\ & \tfrac19 t^2\delta_{i1}P'_{jk}(b)\Bigl(P_{jk}(b) +
P'_{jk}(b)
\bigl({\muh\tan\beta + A_b\over m_b}\bigr)\Bigr)~,\cr
\bar{N}^3_{ijk}(\tau) =\ & - t^2\delta_{i1}P'_{jk}(\tau)\Bigl(P_{jk}(\tau) +
P'_{jk}(\tau)
\bigl({\muh\tan\beta + A_\tau\over m_\tau}\bigr)\Bigr)~.\cr}
\eqn\neutsimp
$$
The simplified chargino contributions can be obtained using
$$
\eqalign{
C^2_{ijk}(t) =\ & {1\over\sqrt{2}}(O^b_{i1})^2\delta_{k1}\delta_{j2}~,\cr
C^2_{ijk}(b) =\ & {1\over\sqrt{2}}(O^t_{i1})^2\delta_{k1}\delta_{j2}~,\cr
C^2_{ijk}(\tau) =\ & {1\over\sqrt{2}}\delta_{k1}\delta_{j2}~,\cr
\bar{C}^1_{ijk}(t) =\ & -\sgn(\muh)\delta_{i2}\cot\beta P'_{jk}(b)
\Bigl(P_{jk}(b) + P'_{jk}(b)
\bigl({\muh\tan\beta + A_b\over m_b}\bigr)\Bigr)~,\cr
\bar{C}^1_{ijk}(b) =\ & -\sgn(\muh)\delta_{i2}\tan\beta P'_{jk}(t)
\Bigl(P_{jk}(t) + P'_{jk}(t)\bigl({\muh\cot\beta + A_t\over
m_t}\bigr)\Bigr)~,\cr
\bar{C}^3_{ijk}(t) =\ & -\sgn(\muh)\delta_{i2}\cot2\beta
P'_{jk}(b)P''_{jk}(b)~,\cr
\bar{C}^3_{ijk}(b) =\ & -\sgn(\muh)\delta_{i2}\cot2\beta
P'_{jk}(t)P''_{jk}(t)~.\cr}
\eqn\chgsimp
$$
The contributions of $\tilde{N}^2$ and $\tilde{C}^2$ are determined by
multiplying
the results for $N^2$ and $C^2$ by $-\sgn(\muh)\cot\beta$ for the top case
and by $-\sgn(\muh)\tan\beta$ for the bottom, $\tau$ cases.
The other neutralino and chargino contributions vanish in this limit.
The wavefunction renormalization contributions also simplify in an obvious way;
in particular, the terms proportional to $g_2y_\alpha$ vanish in this limit.

Below we give the dominant contributions to the matching functions and
ignore those contributions
of less than $1\%$ to the Yukawa couplings. Some of the $y_b$ and $y_\tau$
corrections
will be suppressed in the low $\tan\beta$ region while receiving large
enhancements for high $\tan\beta$. The converse is true for the corrections to
$y_t$.
The gluino contributions dominate the top and bottom Yukawa corrections.
For large $\tan\beta$ the finite vertex contributions from graphs with chargino
and neutralino
exchange are enhanced for the bottom and $\tau$ cases and give large
contributions.
For high(low) $\tan\beta$ the finite parts of wavefunction renormalization
contributions depending on $y_t$($y_b$, $y_\tau$) can be important as well.
The matching functions are
$$
\eqalign{
16\pi^2\Delta^{\scrscr SUSY}_{y_t} \simeq \tfrac83
&g^2_3\biggl(-2\mgl\sum_{i,j}\Bigl( m_t
G_2(\mgl^2,\msti^2,\mstj^2)P'_{ij}(t)P^{}_{ij}(t)\cr
&\quad + (\muh\cot\beta + A_t)G_2(\mgl^2,\msti^2,\mstj^2)(P'_{ij}(t))^2
\Bigr)\cr
&\quad + \tfrac14\sum_i F_2(\msti^2,\mgl^2)\biggr)\cr
+ \tfrac14 &y_t^2\biggl(\sum_i F_2(\msbi^2,\mhino^2)(O^b_{i1})^2
+ \sum_i F_2(\msti^2,\mh0ino^2)\cr
&\quad + \cos^2\!\beta\, F_2(\mhpm^2,m_b^2)\biggr)\cr
+ \tfrac14 &y_b^2\biggl(\sum_i F_2(\msbi^2,\mhino^2)(O^b_{i2})^2
+ \sin^2\!\beta\, F_2(\mhpm^2,m_b^2)\biggr)~,\cr}\eqn\dytsusy
$$
$$
\eqalign{
16\pi^2\Delta^{\scrscr SUSY}_{y_b} \simeq \tfrac83
&g^2_3\biggl(-2\mgl(\muh\tan\beta + A_b)
\sum_{i,j} G_2(\mgl^2,\msbi^2,\msbj^2)(P'_{ij}(b))^2\cr
&\quad + \tfrac14\sum_i F_2(\msbi^2,\mgl^2)\biggr)\cr
+ &y_t^2\biggl(-2\muh\tan\beta\sum_{i,j}
\Bigl(m_t G_2(\mhino^2,\msti^2,\mstj^2)P'_{ij}(t)P^{}_{ij}(t)\cr
&\quad + (\muh\cot\beta +
A_t)G_2(\mhino^2,\msti^2,\mstj^2)(P'_{ij}(t))^2\Bigr)\cr
&\quad + \tfrac14\Bigl(\sum_i F_2(\msti^2,\mhino^2)(O^t_{i2})^2
+ \cos^2\!\beta\, F_2(\mhpm^2,m_t^2)\Bigr)\biggr)\cr
+ \tfrac14 &y_b^2\biggl(\sum_i F_2(\msti^2,\mhino^2)(O^t_{i1})^2
+ \sum_i F_2(\msbi^2,\mh0ino^2)\cr
&\quad + \sin^2\!\beta\, F_2(\mhpm^2,m_t^2)\biggr)\cr
+ \tfrac12 &g_2^2\biggl(\muh{\cal M}_2\tan\beta \sum_i
\Bigl(G_2(\msbi^2,\mh0ino^2,\mw3ino^2)(O^b_{i1})^2\cr
&\quad +
G_2(\msti^2,\mhino^2,\mwino^2)(O^t_{i1})^2\Bigr)\biggr)~,\cr}\eqn\dybsusy
$$
$$
\eqalignno{
16\pi^2\Delta^{\scrscr SUSY}_{y_\tau} \simeq
\tfrac14 &y_\tau^2\biggl(F_2(\msnutau^2,\mhino^2)
+ \sum_i F_2(\mstaui^2,\mh0ino^2) + \sin^2\!\beta\, F_2(\mhpm^2,0)\biggr)&\cr
+ \tfrac12 &g_2^2\biggl(\muh{\cal M}_2\tan\beta
\sum_i
(G_2(\mstaui^2,\mh0ino^2,\mw3ino^2)(O^{\tau}_{i1})^2&\eqnalign\dytaususy\cr
&\quad + G_2(\msnutau^2,\mhino^2,\mwino^2))&\cr
& \quad - 2t^2{\cal M}_1(\muh\tan\beta + A_{\tau})
\sum_{i,j} G_2(\mbino^2,\mstaui^2,\mstauj^2)(P'_{ij}(\tau ))^2\biggr)~,&\cr
}
$$
\noindent where we have used Eqs.~\mneutapprox\ - \mchgapprox.

\endpage

\APPENDIX{B}{B: Function Definitions}

In the appendix we define the various functions arising from parameter
integrals
in the evaluation of the finite parts of one loop diagrams.
The functions $F_i$ and $G_i$ come from
two and three point functions, respectively.
The $F_i$ are
$$
\eqalign{
F_1(M_A^2,M_B^2) =\ & \int^1_0 d\alpha
\ln\biggl({M_A^2\alpha + M_B^2(1-\alpha)\over\mu^2}\biggr)\cr
=\ & {1\over M_A^2 - M_B^2}\Bigl( M_A^2\ln{M_A^2\over\mu^2}
- M_B^2\ln{M_B^2\over\mu^2}\Bigr) - 1~,\cr
F_2(M_A^2,M_B^2) =\ & 2\int^1_0 d\alpha \alpha
\ln\biggl({M_A^2\alpha + M_B^2(1-\alpha)\over\mu^2}\biggr)\cr
=\ & {1\over (M_A^2 - M_B^2)^2}\Bigl( (M_A^4 - 2M_A^2
M_B^2)\ln{M_A^2\over\mu^2}
+ M_B^4\ln{M_B^2\over\mu^2}\Bigr)\cr
+ & {M_B^2\over M_A^2 - M_B^2} - \ttwo~, \cr
F_3(M_A^2,M_B^2) =\ & 6\int_0^1 d\alpha \alpha(1-\alpha)
\ln\biggl({M_A^2\alpha + M_B^2(1-\alpha)\over\mu^2}\biggr)\cr
=\ & {1\over (M_A^2 - M_B^2)^3}\Bigl(
M_A^4(M_A^2-3M_B^2)\ln{M_A^2\over\mu^2}\cr
- & M_B^4(M_B^2-3M_A^2)\ln{M_B^2\over\mu^2} \Bigr )
+ {2M_A^2M_B^2\over (M_A^2 - M_B^2)^2} -{5\over 6}~.\cr}
\eqn\Fis
$$
Both $F_1$ and $F_3$ are symmetric. The explicit forms of the functions
for the limits $M_A\gg (\ll) M_B$ are
$$
\eqalign{
F_1(M_A^2,0) =\ & \ln{M_A^2\over\mu^2} - 1~,\cr
F_2(M_A^2,0) =\ & \ln{M_A^2\over\mu^2} - \ttwo ~,\qquad
F_2(0,M_A^2) = \ln{M_A^2\over\mu^2} - {\textstyle{3\over 2}}~,\cr
F_3(M_A^2,0) =\ & \ln{M_A^2\over\mu^2} - {\textstyle{5\over 6}}~,\cr}\eqn\Flims
$$
and for $M_A = M_B$, $F_i = \ln{M_A^2/\mu^2}$.
The $G_i$ are
$$
\eqalignno{
G_1(M_A^2,M_B^2,M_C^2) =\ & 2\int^1_0 d\alpha\int^{1-\alpha}_0 d\beta
\ln\biggl({M_A^2\alpha + M_B^2\beta + M_C^2(1-\alpha -\beta
)\over\mu^2}\biggr)&\cr
=\ & {M_A^4\over (M_A^2 - M_B^2)(M_A^2 - M_C^2)}(\ln{M_A^2\over\mu^2} - \ttwo
)&\cr
&\quad + (A\leftrightarrow B) + (A\leftrightarrow C) - 1~,&\eqnalign\Gis\cr
G_2(M_A^2,M_B^2,M_C^2) =\ & \int^1_0 d\alpha\int^{1-\alpha}_0 d\beta
{1\over M_A^2\alpha + M_B^2\beta + M_C^2(1-\alpha -\beta )}&\cr
=\ & {1\over M_A^2 - M_B^2 }\Biggl(
{M_A^2\over M_A^2 - M_C^2 }\ln{M_A^2\over M_C^2}
- {M_B^2\over M_B^2 - M_C^2 }\ln{M_B^2\over M_C^2}\Biggr)~.&\cr}
$$
We also consider the limits below
$$
\eqalign{
G_1(M_A^2,M_B^2,M_C^2) =\ & \cases{
\eqalign{&{\textstyle {2M_A^2\over M_A^2-M_C^2}(\ln{M_A^2\over\mu^2} - \ttwo) -
1}\cr
&\quad {\textstyle +{M_C^4\over(M_A^2-M_C^2)^2}(\ln{M_C^2\over\mu^2} -
\ttwo)}\cr
\noalign{\vskip 10pt}}&$M_A=M_B$\cr
(\ln{M_A^2\over\mu^2} - \ttwo)&$M_A=M_B\gg M_C$\cr\noalign{\vskip 10pt}
{M_A^2\ln{M_A^2\over\mu^2}-M_B^2\ln{M_B^2\over\mu^2}\over M_A^2-M_B^2}
- {3\over 2}&$M_A,M_B \gg M_C$\cr\noalign{\vskip 10pt}
\ln{M_A^2\over\mu^2} - {\textstyle{3\over 2}}&$M_A\gg M_B, M_C$\cr}\cr
G_2(M_A^2,M_B^2,M_C^2) =\ & \cases{
{M_C^2\over (M_A^2 - M_C^2)^2}\ln{M_C^2\over M_A^2} +
{1\over M_A^2 - M_C^2}&$M_A=M_B$\cr
{1\over M_A^2}&$M_A=M_B\gg M_C$\cr
{1\over M_A^2 - M_B^2}\ln{M_A^2\over M_B^2}&$M_A,M_B \gg M_C$,\cr}\cr}
\eqn\Glims
$$
where we have included the particular cases that occur in the threshold
corrections.

\endpage
\refout
\endpage

\FIG\fone{Vertex diagrams contributing to dominant gluino
induced threshold corrections to $y_t$ and $y_b$.}
\FIG\ftwo{Vertex diagrams involving squarks, Higgsinos and winos
giving the dominant contributions to $y_b$ threshold
correction which are enhanced in the large $\tan\beta$ region.}
\FIG\fthree{Vertex diagrams involving sleptons, Higgsinos and gauginos
giving the dominant contributions to $y_\tau$ threshold
correction which are enhanced in the large $\tan\beta$ region.
The bino exchange diagram can be important due to the large
hypercharge of the $\tau$ lepton.}
\FIG\ffour{Variation of $\Mx$ with gluino mass for
$120 {\rm\ GeV}\le \mgl \le 1$ TeV and central values
of $\sin\theta_W(M_Z)$ and $\alpha(M_Z)$.
Contours are given for the allowed range of $\alpha_s(M_Z)$ and
two values of $\tan\beta$ for fixed $M_t = 165.4$ GeV.}
\FIG\ffive{Variation of $\Mx$ with $M_t$ for
$130 {\rm\ GeV}\le m_t(M_Z) \le 200$ GeV
and central values of $\sin\theta_W(M_Z)$ and $\alpha(M_Z)$.
Contours are given for the allowed range of $\alpha_s(M_Z)$ and
two values of $\tan\beta$ for fixed $\mgl = 400$ GeV.}
\FIG\fsix{Variation of $\Mhs$ with $V_{\scrscr SUSY}\simeq
.08m_{\tilde{H}}$ for $100 {\rm\ GeV}\le m_{\tilde{H}} \le 1$ TeV
and central values of $\sin\theta_W(M_Z)$ and $\alpha(M_Z)$.
Contours are given for the allowed range of $\alpha_s(M_Z)$ and
two values of $\tan\beta$ for fixed $M_t = 165.4$ GeV.}
\FIG\fseven{Variation of $\Mhs$ with $M_t$ for
$130 {\rm\ GeV}\le m_t(M_Z) \le 200$ GeV
and central values of $\sin\theta_W(M_Z)$ and $\alpha(M_Z)$.
Contours are given for the allowed range of $\alpha_s(M_Z)$ and
two values of $\tan\beta$ for fixed $V_{\scrscr SUSY} = 45$ GeV.
Note that the initial fall of $\Mhs$ with $M_t$ is due to the
electroweak threshold while its rise for large $M_t$ is
due to the effect of large $y_t$ on the two loop gauge $\beta$
functions.}
\FIG\feight{The dependence of the $M_t$ solution
on $\tan\beta$ showing the
effect of a large splitting of $\Mv$ and $\Msig$
for $M_b = 5.2$ GeV, $\alpha_s(M_Z)= 0.120$ and $\Ms = M_Z$.
The sparticle spectrum parameters are fixed at $\mgl = M_3 = 1$ TeV
and $V_{\scrscr SUSY} = 80$ GeV.
The typical GUT masses are then $\Mx = 10^{16.1 \pm 0.1}$ GeV and
$\Mhs = 10^{15.7 \pm 0.4}$ GeV for these solutions.
We also indicate the bound from the nonperturbative
limit on $y_t$.}
\FIG\fnine{The dependence of the $M_t$ solution
on $\tan\beta$ showing the
effect of a splitting of $\Mhs$ above $\Mx$
for $M_b = 4.9$ GeV, $\alpha_s(M_Z)= 0.127$ and $\Ms = M_Z$.
The sparticle spectrum parameters are fixed at $\mgl = M_3 = 1$ TeV
and $V_{\scrscr SUSY} = 80$ GeV.
The typical GUT masses are then $\Mx = 10^{16.2\pm 0.1}$ GeV and
$\Mhs = 10^{16.8\pm 0.3}$ GeV for these solutions.
We again indicate the bound from the nonperturbative
limit on $y_t$.}
\FIG\ften{The dependence of the $M_t$ solution from
Yukawa unification on $\tan\beta$ for the allowed
range of $\alpha_s(M_Z)$ for $M_b = 4.9$ GeV and
$\Ms = M_Z$. We give the solutions with and without
GUT scale Yukawa thresholds
for the generic case $\Mv = \Msig$.}
\FIG\feleven{The dependence of the $M_t$ solution
on $\tan\beta$ for bottom pole masses
in the range $4.7 {\rm\ GeV}\le M_b \le 5.3$ GeV for
$\alpha_s(M_Z)= 0.120$ and $\Ms = M_Z$. Again,
the solutions with and without GUT scale Yukawa thresholds
correspond to the generic case $\Mv = \Msig$. The curves
for $M_b = 4.7$ and $4.8$ GeV with GUT thresholds are
cut off due to the absence of perturbative solutions.}

\figout

\endpage
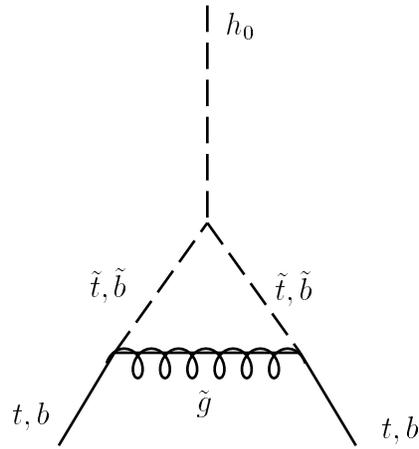
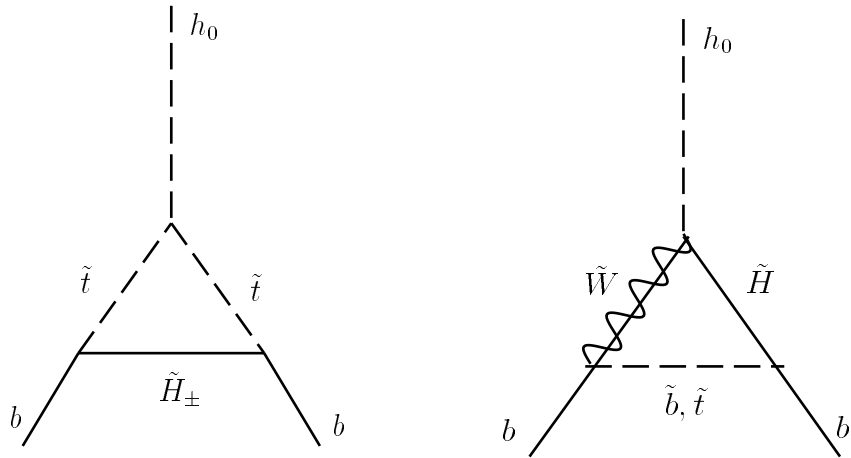
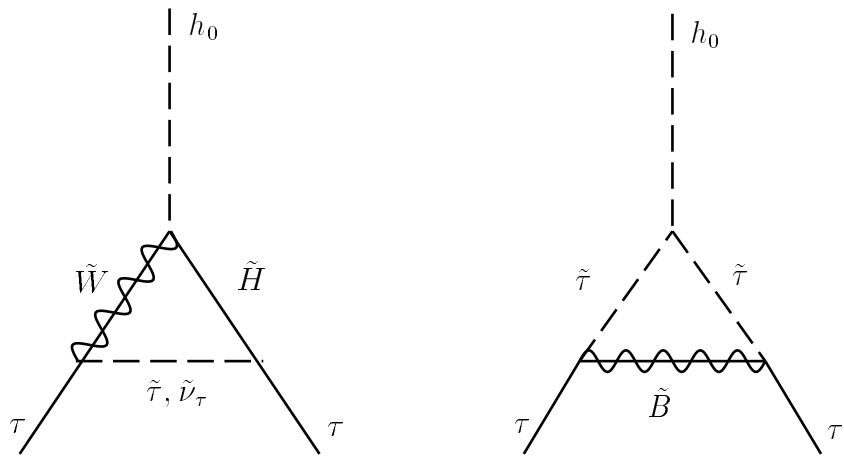
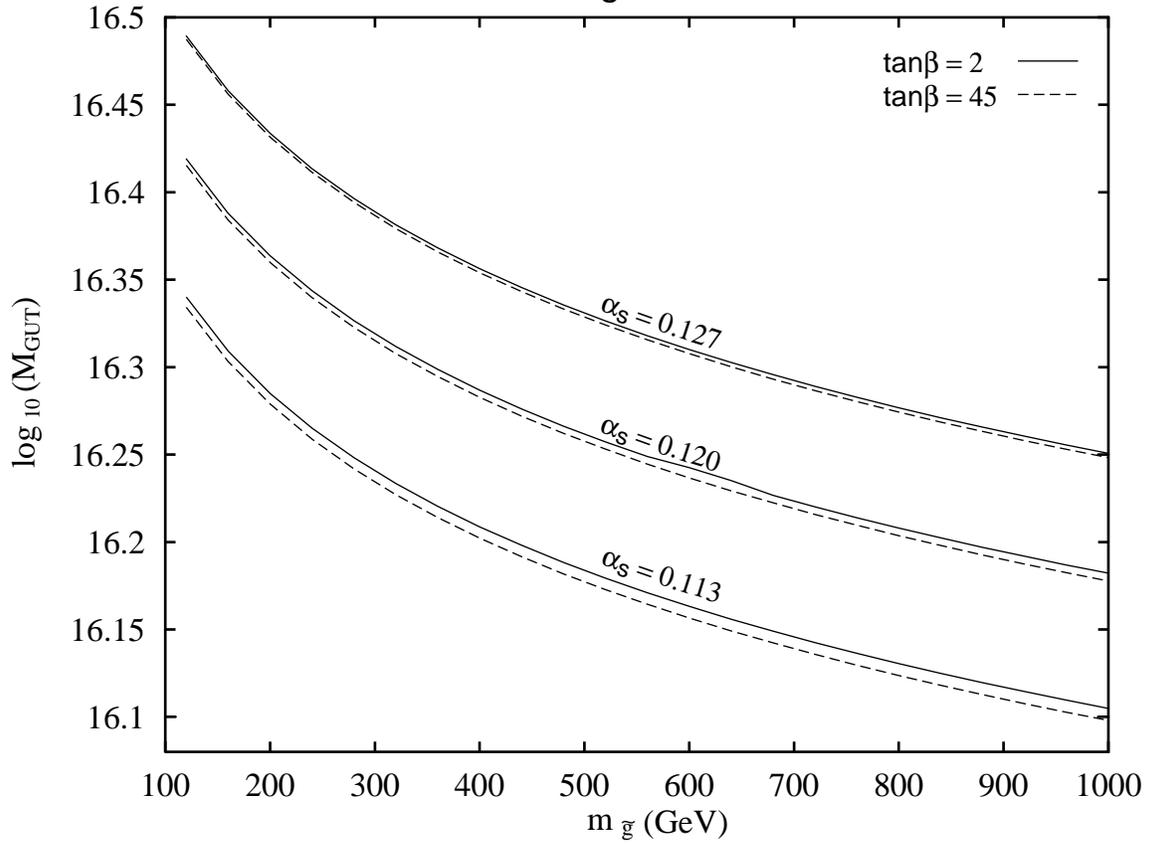
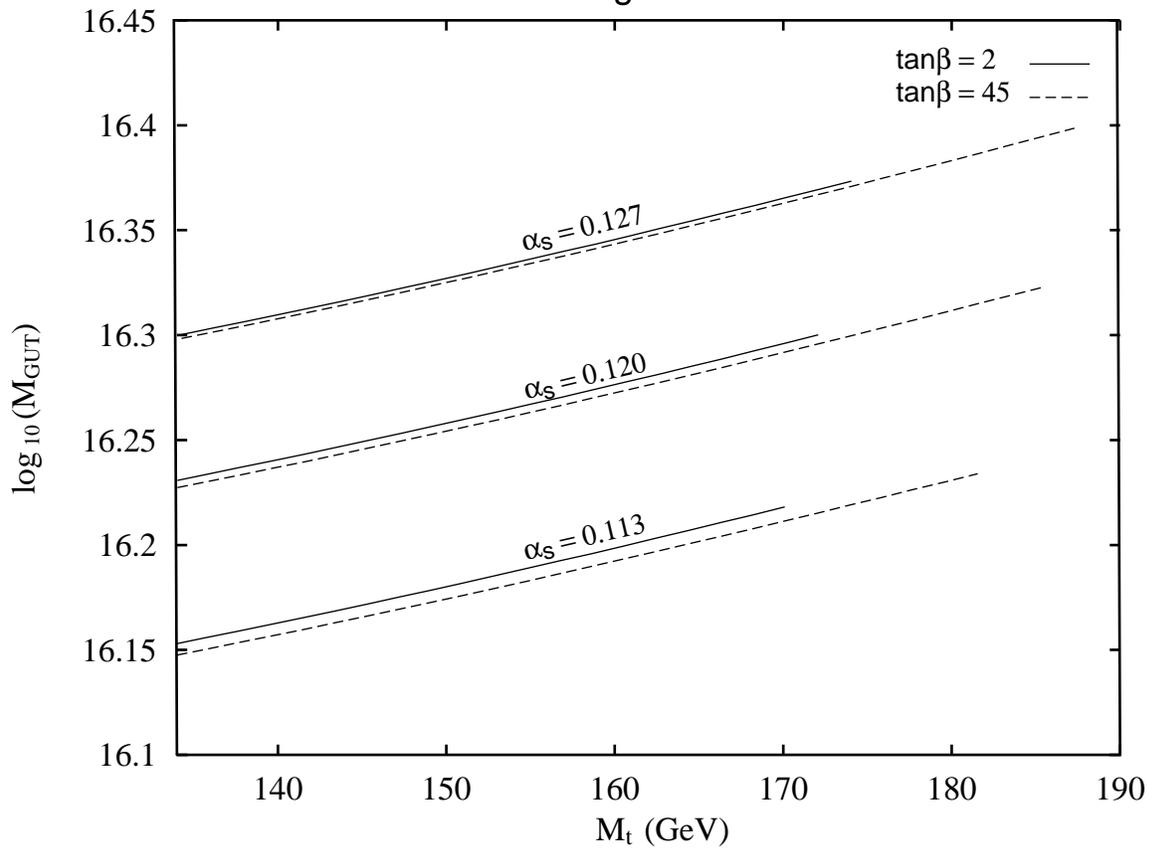
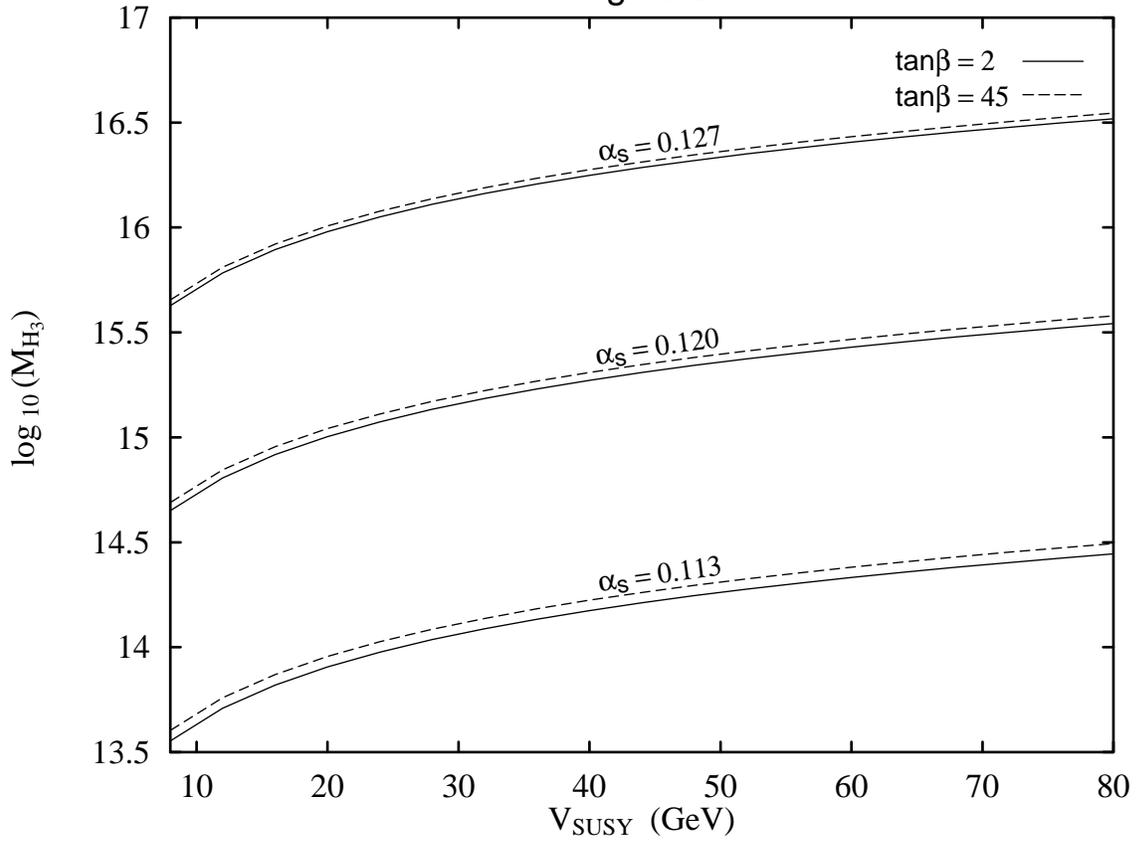
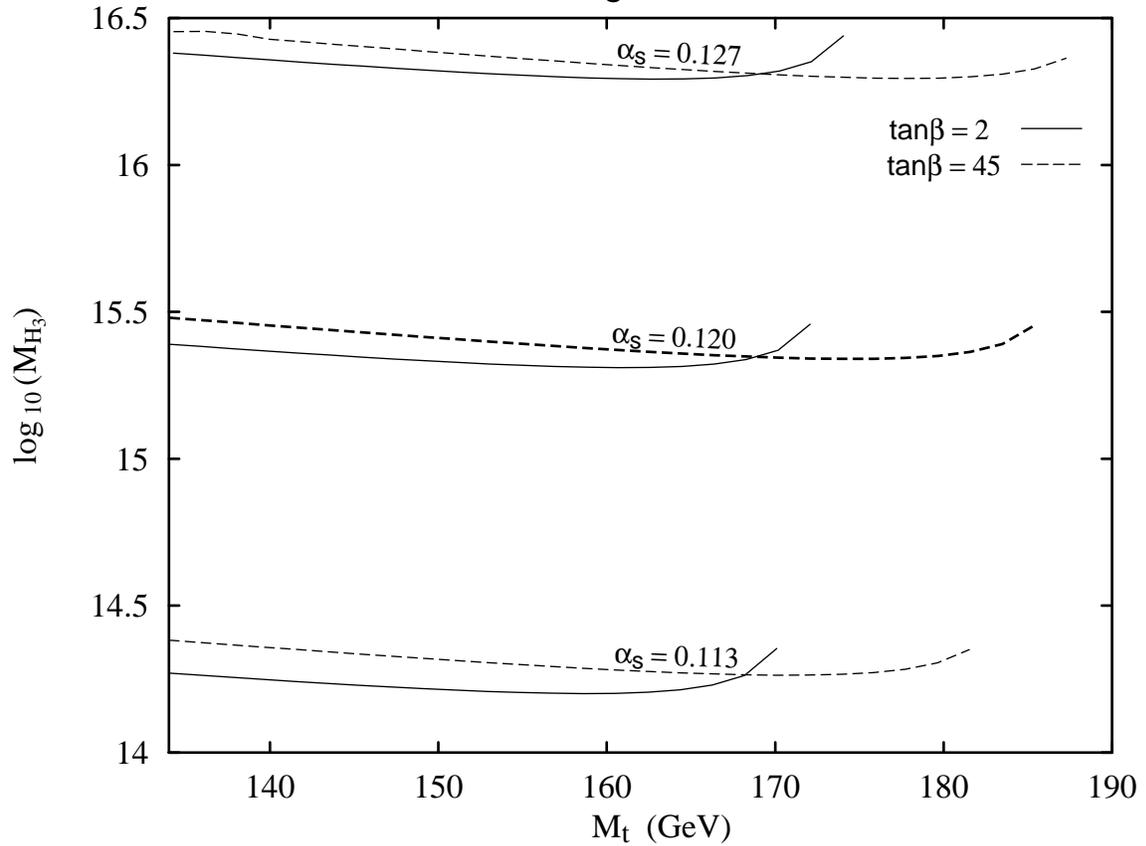
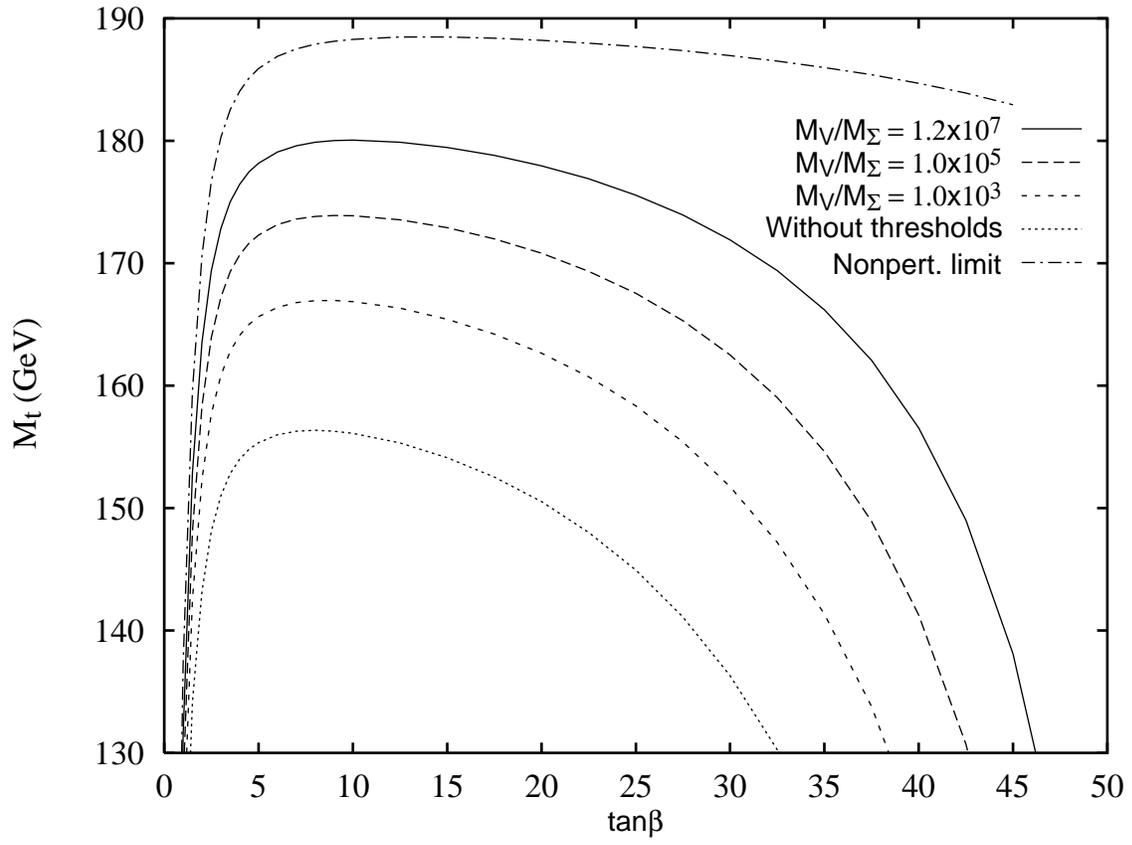
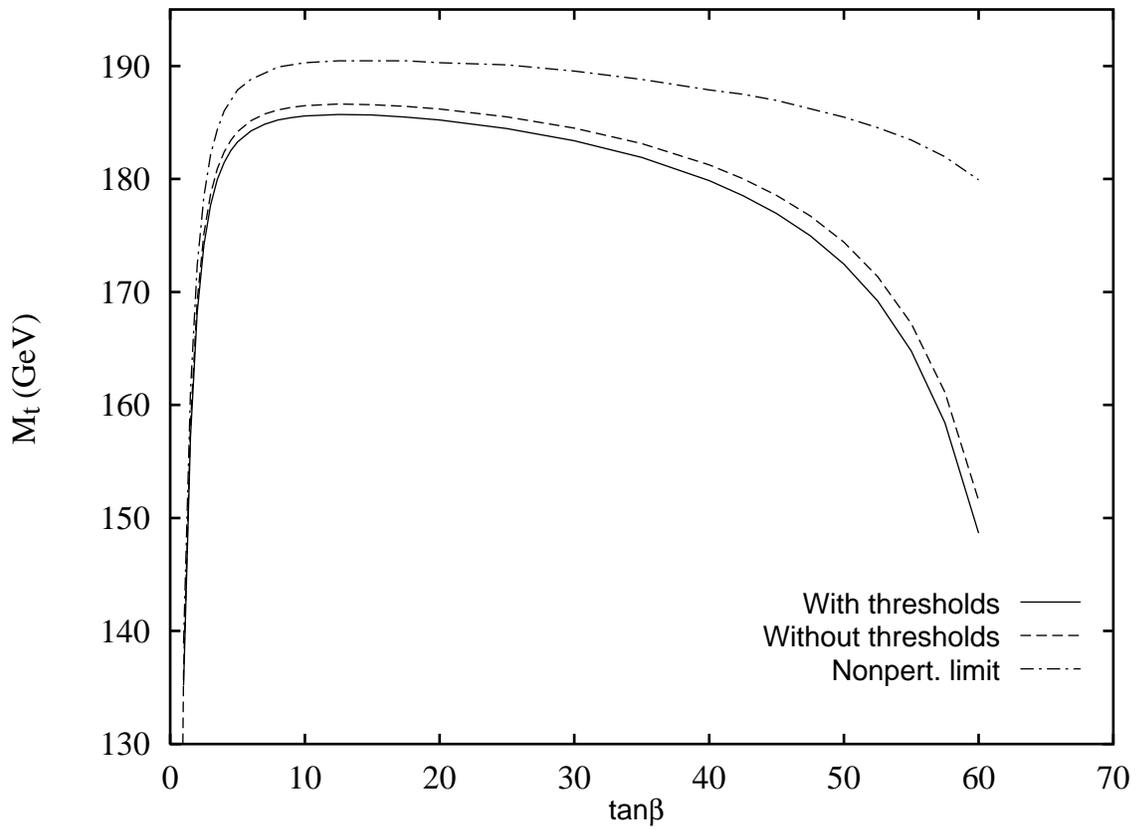
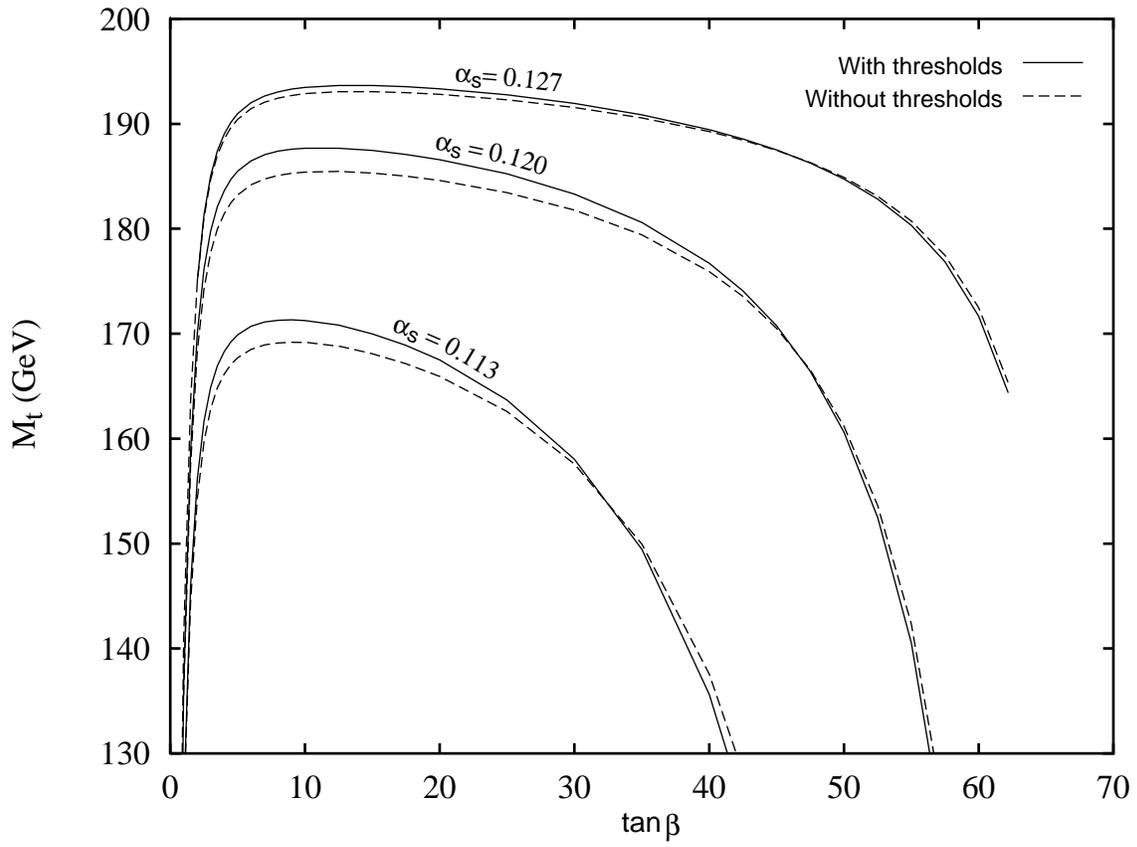
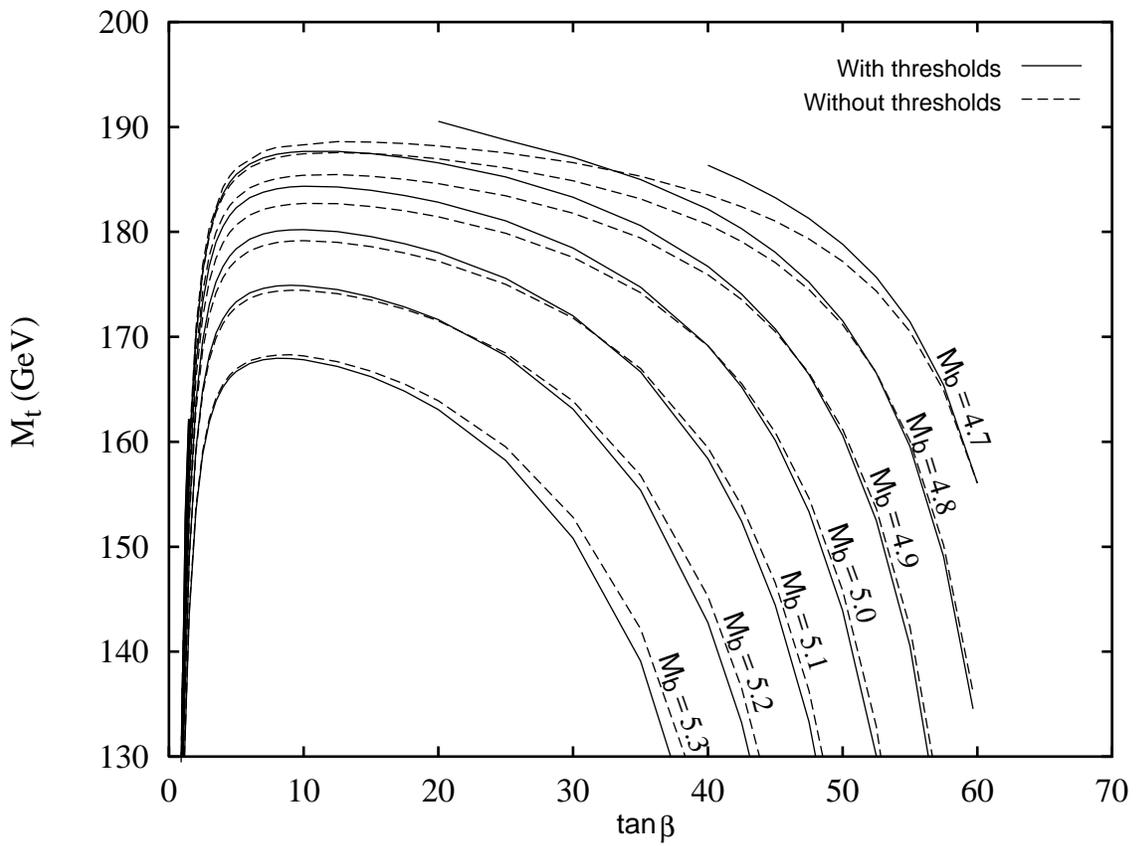
\end